\documentclass[floatfix,%
 reprint,%
 amssymb, amsmath,%
 aip,jcp,
frontmatterverbose,
]{revtex4-2}

\usepackage{amsmath,amssymb,amsfonts,mathtools}
\usepackage{multirow}
\usepackage{float}
\usepackage{csquotes}
\usepackage{dsfont}
\usepackage[mathscr]{euscript}
\usepackage{natbib}
\usepackage[english]{babel}

\usepackage{enumitem}
\setlist{nosep} 

\usepackage{graphicx}
\usepackage{bm}%
\usepackage{longtable}
\usepackage{supertabular}
\usepackage{rotating}
\usepackage[normalem]{ulem}
\usepackage{ifthen}
\newboolean{includeMQDT}
\setboolean{includeMQDT}{false}  

\usepackage[dvipsnames]{xcolor}

\def\measurehat#1{%
   \setbox0=\vbox{$\hat{#1}\hfil\break$\null\par
      \setbox0=\lastbox\unskip\unpenalty\global\setbox1=\lastbox}%
   \setbox0=\hbox{\unhbox1 \unskip\unpenalty\unskip \global\setbox2=\lastbox}%
   \setbox0=\vbox{\unvbox2 \setbox0=\lastbox}%
}
\newcommand{\doublehat}[1]{%
   \measurehat{#1}\dimen0=\wd0 \measurehat{\kern0pt{#1}}%
   \raise.35ex\rlap{\kern\dimexpr\dimen0-\wd0$\hat{\phantom{#1}}$}{\hat{#1}}%
}

\usepackage{dcolumn,ulem}  
\usepackage{xspace}

\begin{document}
\newcommand{\pcut}[1]{{\color{blue}{\sout{#1}}}}   
\newcommand{\Vel}{\ensuremath{V_\mathrm{el}}}
\newcommand{\SH}{\ensuremath{\mathrm{SH}}}
\newcommand{\SHplus}{\ensuremath{\mathrm{SH}^+}}
\newcommand{\il}{{\it et~al.}}

\title{Quasi-Diabatic Propagation Scheme for Simulating Polariton Chemistry}

\author{Deping Hu}
\email{deping.hu@rochester.edu}
\affiliation{Department of Chemistry, University of Rochester, 120 Trustee Road, Rochester, NY 14627, U.S.A.}
\author{Arkajit Mandal}
\affiliation{Department of Chemistry, University of Rochester, 120 Trustee Road, Rochester, NY 14627, U.S.A.}
\affiliation{Department of Chemistry, Columbia University, New York, NY 10027, U.S.A.}

\author{Braden M. Weight}
\affiliation{Department of Physics and Astronomy, University of Rochester, Rochester, NY 14627, U.S.A.}

\author{Pengfei Huo}
\email{pengfei.huo@rochester.edu}
\affiliation{Department of Chemistry, University of Rochester, 120 Trustee Road, Rochester, NY 14627, U.S.A.}
\affiliation{The Institute of Optics, Hajim School of Engineering, University of Rochester, Rochester, New York, 14627, U.S.A.}

\begin{abstract}
We generalize the quasi-diabatic (QD) propagation scheme to simulate the non-adiabatic polariton dynamics in molecule-cavity hybrid systems. The adiabatic-Fock states, which are the tensor product states of the adiabatic electronic states of the molecule and photon Fock states, are used as the {\it locally} well-defined {\it diabatic} states for the dynamics propagation. These locally well-defined diabatic states allow using any diabatic quantum dynamics methods for dynamics propagation, and the definition of these states will be updated at every nuclear time step. We use several recently developed non-adiabatic mapping approaches as the diabatic dynamics methods to simulate polariton quantum dynamics in a Shin-Metiu model coupled to an optical cavity. The results obtained from the mapping approaches provide very accurate population dynamics compared to the numerically exact method and outperform the widely used mixed quantum-classical approaches, such as the Ehrenfest dynamics and the fewest switches surface hopping approach. We envision that the generalized QD scheme developed in this work will provide a powerful tool to perform the non-adiabatic polariton simulations by allowing a direct interface between the diabatic dynamics methods and {\it ab initio} polariton information.

\end{abstract}

\maketitle

\section{Introduction}
Coupling molecules to the quantized radiation field inside an optical cavity creates a set of new photon-matter hybrid states, which are commonly referred to as polaritons,\cite{Ebbesen2016ACR,kowalewski_manipulating_2017,Flick2017pnas,Ribeiro2018,Feist2018,Mandal2019JPCL} which have been shown to facilitate new chemical reactivities.\cite{Hutchison2012ACIE,Ebbesen2016ACR,Thomas2019S,Mandal2019JPCL,Mandal2020JPCB} Theoretical investigations play a crucial role in understanding the fundamental limit and basic principles in this emerging field,\cite{Mandal2019JPCL,Feist2018,Luk2017jctc,Groenhof2018jpcl,Groenhof2019jpcl,Groenhof2019jpcl,Groenhof2021JCP} as these polariton chemical reactions often involve a rich dynamical interplay among the electronic, nuclear, and photonic degrees of freedom (DOFs). Accurately simulating polaritonic quantum dynamics remains a challenging task and is beyond the scope of photochemistry or quantum optics.\cite{kowalewski_manipulating_2017}

The trajectory-based non-adiabatic dynamics approaches\cite{Tully2012jcp,BarbattiSH,Mai2015} play an important role in simulating the non-adiabatic dynamics of the coupled electronic-nuclear DOFs. Two of the most commonly used mixed quantum-classical (MQC) methods are the Ehrenfest and fewest switches surface hopping (FSSH) approaches.\cite{Tully,tully94jcp} Both approaches describe the electronic subsystem quantum mechanically, and treat the nuclear DOFs classically. It is thus a natural idea for the theoretical chemistry community to extend these two approaches to investigate polariton chemistry by treating the electronic-photonic DOFs (or so-called polariton subsystem) quantum mechanically and the nuclear DOFs classically. Incorporating the description of the photon field into the MQC methods has become a basic strategy to simulate polariton chemistry.\cite{Luk2017jctc,Groenhof2018jpcl,Groenhof2019jpcl,Groenhof2021JCP,Fregoni2018,Fregoni2020,fregoni_strong_2020,Zhang2019jcp} The key ingredient in the MQC simulations of polariton dynamics is the expression of the nuclear gradient. Recently, we derived a rigorous expression of the nuclear gradient using the quantum electrodynamics (QED) Hamiltonian without making the usual approximations,\cite{Zhou2022} such as the rotating wave approximation. These gradient expressions, together with the corresponding MQC approaches (Ehrenfest and FSSH approaches), is valid for any any number of electronic states or Fock states at any light-matter coupling strength. However, the inherent semi-classical approximation in these approaches can lead to the break-down of detailed balance\cite{ParandekarJCTC2006} (incorrect long time population) in Ehrenfest dynamics and the creation of artificial electronic coherence\cite{subotnik2016arpc} or incorrect
chemical kinetics\cite{subotnik2016arpc} for the FSSH dynamics without invoking \textit{ad hoc} decoherence corrections.

In response to these theoretical challenges, a wide range of non-adiabatic dynamics approaches have been developed in the diabatic representation. Many of them belong to the family of non-adiabatic mapping dynamics which are based on the Meyer-Miller-Stock-Thoss (MMST) mapping formalism.\cite{MeyerJCP1979,StockPRL1997,ThossPRA1999} These methods include partial linearized density matrix\cite{HuoJCP2011,HuoMP2012} (PLDM), symmetrical quasi-classical\cite{CottonJCP2013,CottonJPCA2013} (SQC), the quantum-classical Liouville equation (QCLE) dynamics.\cite{HsiehJCP2012,HsiehJCP2013} In particular, the recently developed $\gamma$-SQC has been shown\cite{CottonJCP2019_2} to provide impressively accurate non-adiabatic photo-dissociation quantum dynamics with coupled Morse potentials through the adjusted zero-point energy (ZPE) parameter of the mapping variables, thus appearing to be a promising method to simulate on-the-fly quantum dynamics of complex molecular systems. In addition, the spin-mapping Linearized Semi-Classical approach (spin-LSC)\cite{richardson2019,richardson2020,Duncan2022JCP}, which uses generalized spin mapping representation\cite{richardson2020} for the electronic DOF as well as the Linearization approximation\cite{MillerJCP98,shi2004} for the nuclear DOF, has also shown a significant improvement of the population dynamics in the system-bath model problems (such as in spin-boson systems\cite{richardson2019} and many-state exciton Hamiltonians of light-harvesting complexes.\cite{richardson2020}) The $\gamma$-SQC approach has already demonstrated\cite{WeightJCP2021} its ability to outperform MQC approaches (Ehrenfest and FSSH) in describing the electronic non-adiabatic dynamics for {\it ab initio} on-the-fly simulations. These new mapping approaches should, in principle, also outperform the MQC methods in simulating the polaritonic non-adiabatic dynamics that happens in the electron-photon subspace coupling to the motion of the nuclei. Unfortunately, to the best of our knowledge, there are only limited studies of using mapping dynamics to investigate polariton chemistry for model systems with strict diabatic states.\cite{Mandal2019JPCL,Chowdhury2021jcp,Mandal2020JPCB}

Recently, we have developed the quasi-diabatic (QD) propagation scheme\cite{MandalJCTC2018,MandalJCP2018,MandalJPCA2019,SandovalJCP2018,ZhouJCPL2019, WeightJCP2021} as a general framework to seamlessly combine a diabatic quantum dynamics approach, such as the mapping based methods,\cite{CottonJCP2019_2,richardson2020} with the adiabatic outputs of an electronic structure method. The QD propagation scheme uses the adiabatic states at a reference nuclear geometry (the so-called ``crude adiabatic" states) as the {\it locally well-defined} diabatic states during a short-time propagation and then dynamically updates the QD basis at each consecutive nuclear propagation step. In this propagation scheme, one does not construct a {\it global diabatic representation} but instead, uses a sequence of locally diabatic representations (one for each short-time segment) to propagate the dynamics. We have both analytically\cite{MandalJCTC2018} and numerically\cite{MandalJPCA2019,SandovalJCP2018} demonstrated that the QD scheme provides exactly the same results compared to the direct diabatic quantum dynamics at the single trajectory level. 

In this work, we generalize the QD propagation scheme to simulate polariton non-adiabatic dynamics in a molecule-cavity hybrid system. In particular, we use the adiabatic-Fock state at a reference nuclear geometry as the locally well-defined diabatic basis to propagate the polariton dynamics, and dynamically update the definition of these local diabatic states between two consecutive propagation steps. These adiabatic-Fock states are tensor products of the electronic adiabatic states for the molecular system and the Fock states of the photon field inside an optical cavity. We use the Shin-Metiu (SM) model\cite{Shin1995jcp,Hoffmann2020} as the ``\textit{ab initio}" model molecular system to investigate strong and ultra-strong light-matter interactions between a molecule and an optical cavity. Through numerical simulations, we demonstrate the accuracy of using both $\gamma$-SQC\cite{CottonJCP2019_2} and spin-LSC\cite{richardson2019,richardson2020} to obtain non-adiabatic polariton dynamics, which outperforms widely used MQC approaches.

\section{Theory and Methods} \label{methods}
\subsection{The Pauli-Fierz QED Hamiltonian}
The Pauli-Fierz (PF) QED Hamiltonian for one molecule coupled to quantized radiation field inside an optical cavity can be written as 
\begin{equation}\label{total_PF_H}
\hat{H}=\hat{T}_\mathrm{n}+\hat{H}_\mathrm{en}+\hat{H}_\mathrm{p}+\hat{H}_\mathrm{enp}+\hat{H}_\mathrm{d},
\end{equation}
where $\hat{T}_\mathrm{n}$ represents the nuclear kinetic energy operator, $\hat{H}_\mathrm{en}$ is the electronic Hamiltonian that describes electron-nucleus interactions. Further, $\hat{H}_\mathrm{p}$, $\hat{H}_\mathrm{enp}$, and $\hat{H}_\mathrm{d}$ represent the photonic Hamiltonian, electronic-nuclear-photonic interactions, and the dipole self-energy (DSE) term, respectively. A full derivation of this Hamiltonian, as well as its connection with the various atomic cavity QED models can be found in the Appendix of Ref.~\citenum{Chowdhury2021jcp}.

The electronic-nuclear potential $\hat{H}_\mathrm{en}$, which describes the common molecular Hamiltonian excluding the nuclear kinetic energy is described as follows
\begin{equation}\label{eqn:Hen}
\hat{H}_\mathrm{en} = \hat{T}_\mathrm{e} + \hat{V}_\mathrm{ee} + \hat{V}_\mathrm{en} + \hat{V}_\mathrm{nn}.
\end{equation}
The above expression includes electronic kinetic energy, electron-electron interaction, electron-nucleus interaction and nucleus-nucleus interaction. The expressions of these four terms can be found in previous work.\cite{Doltsinis2002jctc,Schafer2018pra,Marxbook} Modern electronic structure theory have been developed around solving the eigenvalue problem of $\hat{H}_\mathrm{en}$, providing the following electronically adiabatic energy and its corresponding state
\begin{equation}\label{HenPsi}
\hat{H}_\mathrm{en}|\phi_\alpha(\mathbf{R}) \rangle= E_{\alpha}({\mathbf R})|\phi_\alpha(\mathbf{R}) \rangle.
\end{equation}
Here, $|\phi_\alpha(\mathbf{R}) \rangle$ represents the $\alpha_\mathrm{th}$ many-electron adiabatic state for a given molecular system, with the adiabatic energy $E_{\alpha}({\mathbf R})$.

For clarity, we restrict our discussions to the cavity with only one photonic mode, and all the formulas presented here can be easily generalized into a more realistic, many-mode cavity. The photonic Hamiltonian is written as 
\begin{equation}\label{eqn:Hp}
\hat{H}_\mathrm{p}=\frac12 \left(\hat{p}_\mathrm{c}^2+\omega_\mathrm{c}^2\hat{q}_\mathrm{c}^2\right)=\hbar\omega_\mathrm{c} \Big(\hat{a}^{\dagger}\hat{a}+\frac12 \Big),
\end{equation}
where $\hat{q}_\mathrm{c}=\sqrt{\hbar/2 \omega_\mathrm{c}}(\hat{a}^{\dagger}+\hat{a})$ and $\hat{p}_\mathrm{c}=i\sqrt{\hbar\omega_\mathrm{c}/2}(\hat{a}^{\dagger}-\hat{a})$ are photon field operators, $\hat{a}^{\dagger}$ and $\hat{a}$ are the photonic creation and annihilation operators, respectively and $\omega_\mathrm{c}$ is the photon frequency.

The light-matter coupling term (electronic-nuclear-photonic interactions) under the dipole gauge is expressed as 
\begin{equation}\label{eq:enp}
\hat{H}_\mathrm{enp}=\omega_\mathrm{c}\hat{q}_\mathrm{c} ( \boldsymbol{\lambda} \cdot {\hat{\boldsymbol \mu}})=g_\mathrm{c} \boldsymbol{\epsilon} \cdot  {\hat{\boldsymbol \mu}} (\hat{a}^{\dagger} + \hat{a}).
\end{equation}
where $\boldsymbol{\lambda}=\lambda \cdot{\boldsymbol\epsilon}$ characterizes the cavity photon field strength, ${\boldsymbol\epsilon}$ is the direction of the field polarization. The photon field strength is determined by the volume of the cavity as $\lambda=\sqrt{1/\epsilon_{0}\mathcal{V}_{0}}$, where $\epsilon_{0}$ is the permittivity inside the cavity and $\mathcal{V}_{0}$ is the effective quantization volume inside the cavity. Another way to characterize the light-matter coupling strength is using $g_\mathrm{c}=\sqrt{\hbar\omega_\mathrm{c}/2}\lambda$. Note that the common notation used in the literature,\cite{Kowalewski2016JPCL,Mandal2019JPCL} the definition of $g_\mathrm{c}$ also includes $\boldsymbol{\lambda} \cdot {\hat{\boldsymbol \mu}}$. Further, the total dipole operator of both electrons and nuclei is defined as
\begin{equation}\label{dipole}
{\hat{\boldsymbol \mu}} = -\sum_ie \hat{{\bf r}}_i +\sum_{j} Z_{j} e \hat{\bf{R}}_{j}, 
\end{equation}
where $-e$ is the charge of the electron and $Z_{j} e $ is the charge of the $j_\mathrm{th}$ nucleus.

Finally, the DSE term is expressed as
\begin{equation}\label{eqn:Hd}
\hat{H}_\mathrm{d}=\frac12 (\boldsymbol{\lambda} \cdot \hat{\boldsymbol \mu})^2=\frac{g_{\mathrm{c}}^2}{\hbar\omega_\mathrm{c}} (\boldsymbol{\epsilon} \cdot \hat{\boldsymbol \mu})^2.
\end{equation}
This is a necessary term in the PF Hamiltonian and ensures both gauge invariance of the Hamiltonian\cite{Taylor2020PRL,Mandal2020JPCB} and a bounded ground state.\cite{Rokaj2018JPB,Mandal2020JPCB,Schaefer2020AP} In this work, we do not consider the cavity loss. The cavity loss can be effectively incorporated by using Lindblad dynamics approaches with the MQC simulations.\cite{Koessler2022}

For the molecule-cavity hybrid system, a convenient basis for quantum dynamics simulations could be the photon-dressed electronic adiabatic states 
\begin{equation}\label{adia-fock}
|\psi_i({\mathbf R}) \rangle=|\phi_\alpha({\mathbf R}) \rangle \otimes |n\rangle \equiv|\phi_{\alpha}({\bf R}),n\rangle,
\end{equation}
where quantum number $i\equiv\{\alpha,n\}$ indicates both the adiabatic electronic state of the molecule and the Fock state. Note that we have introduced a shorthand notation in Eq.~\ref{adia-fock}, which will be used throughout the rest of this paper. This is one of the most straightforward choices of basis for the hybrid system because of the readily available adiabatic electronic information (\textit{e.g.,} wavefunctions, energies, and the dipole matrix) from electronic structure calculations that we need to construct the elements of Hamiltonian. 

In the MQC simulation, such as the Ehrenfest or FSSH approach, or the recently developed mapping non-adiabatic approaches, the total molecular Hamiltonian is expressed as
\begin{equation}\label{totalH}
\hat{H}=\hat{T}_\mathrm{n}+\hat{V},
\end{equation}
where $\hat{T}_\mathrm{n}$ represents the nuclear kinetic energy operator, and $\hat{V}$ represents the rest of the Hamiltonian. For a bare molecular system, $\hat{V}=\hat{H}_\mathrm{en}$ expressed in Eq.~\ref{eqn:Hen}.
For a molecule-cavity hybrid system, 
\begin{equation}\label{eqn:Henp}
\hat{V}=\hat{H}_\mathrm{en}+\hat{H}_\mathrm{p}+\hat{H}_\mathrm{enp}+\hat{H}_\mathrm{d}\equiv \hat{H}_\mathrm{pl},
\end{equation}
which is commonly referred to as the polariton Hamiltonian,\cite{Flick2017jctc,Flick2017pnas} also denoted as $\hat{H}_\mathrm{pl}$. In a similar way that electronic adiabatic states are defined in Eq.~\ref{HenPsi}, one can further define the polaritonic state\cite{Flick2017jctc,Flick2017pnas} as the eigenstate of $\hat{V}=\hat{H}_\mathrm{pl}$ (see definition in Eq.~\ref{eqn:Henp}) through the following eigenequation
\begin{equation}\label{polariton}
\hat{H}_\mathrm{pl}|\mathcal{E}_{J}({\bf R})\rangle=\mathcal{E}_{J}({\bf R})|\mathcal{E}_{J}({\bf R})\rangle,
\end{equation}
where $|\mathcal{E}_{J}({\bf R})\rangle$ is the polariton state with polariton energy $\mathcal{E}_{J}({\bf R})$. The polariton eigenstate can be expressed as 
\begin{equation}\label{pol-exp}
|\mathcal{E}_{J}({\bf R})\rangle=\sum_{\alpha,n}c^{J}_{\alpha, n}({\mathbf R})|\phi_{\alpha}({\bf R}),n\rangle,
\end{equation}
where $c^{J}_{\alpha, n}({\mathbf R})=\langle \phi_{\alpha}({\bf R}),n|\mathcal{E}_{J}({\bf R})\rangle$ and $\mathcal{E}_{J}({\bf R})$ can be obtained by diagonalizing the matrix of $\hat{V}=\hat{H}_\mathrm{pl}$ (constructed from the adiabatic-Fock state basis in Eq.~\ref{adia-fock}) as
\begin{equation}\label{eqn:diag}
\mathbf{U}^{\dagger}[V({\mathbf R})]\mathbf{U} = [\mathcal{E}({\mathbf R})],
\end{equation}
where
\begin{equation}
[V({\mathbf R})]_{ij} = \langle\psi_i({\mathbf R})| \hat{V} |\psi_j({\mathbf R}) \rangle.
\end{equation}
Note that the ${\bf R}$-dependence of $|\mathcal{E}_{J}({\bf R})\rangle$ is entirely coming from the ${\bf R}$-dependence of the adiabatic states $|\phi_\alpha({\mathbf R}) \rangle$, and the Fock state $|n\rangle$ is completely ${\bf R}$-independent. Meanwhile, the $|\mathcal{E}_{J}({\bf R})\rangle$ is the eigenstate of $\hat{V}$, whereas the adiabatic state $|\phi_\alpha({\mathbf R}) \rangle$ is only the eigenstate of $\hat{H}_\mathrm{en}$, and not for $\hat{V}$.

\subsection{Quasi-Diabatic Propagation Scheme for Molecular Cavity QED}\label{sec:QD}
The QD propagation scheme explicitly addresses the discrepancy between accurate quantum dynamics methods in the diabatic representation and the electronic structure methods in the adiabatic representation. The essential idea of the QD scheme is to use the electronic adiabatic states associated with a reference geometry as the local diabatic states during a short-time quantum propagation and dynamically updates the definition of the QD states along the time-dependent nuclear trajectory.\cite{MandalJCTC2018,MandalJCP2018,MandalJPCA2019,SandovalJCP2018,ZhouJCPL2019, WeightJCP2021}

In this work, we apply the QD propagation scheme to the case of molecular cavity QED. This requires the use of a convenient basis with a reference nuclear geometry as the $\it locally$ well-defined diabatic basis, in the sense that its character is fixed (which is automatically guaranteed because of the fixed reference geometry by construction) as well as it is a complete basis (which is only true when the geometry is close to this reference geometry). The potential candidate for this basis is the adiabatic-Fock state $|\psi_i({\mathbf R}) \rangle=|\phi_\alpha({\mathbf R}),n\rangle$ (Eq.~\ref{adia-fock}), which is not the same as the polariton states $|\mathcal{E}_{J}({\bf R})\rangle$ (Eq.~\ref{pol-exp}) except for the zero-coupling limit. In this work, we use the adiabatic-Fock state as the convenient choice due to its simplicity in terms of the polariton coupling and nuclear gradient expressions in the QD propagation framework.

Consider a short-time propagation of the nuclear DOFs during $t\in[t_0, t_1]$, where the nuclear positions evolve from ${\bf R}(t_0)$ to ${\bf R}(t_1)$, and the corresponding adiabatic-Fock basis (defined in Eq.~\ref{adia-fock}) are $\{|\psi_{i}({\bf R}(t_0))\rangle\}$ and $\{|\psi_{j}{\bf R}(t_1))\rangle\}$. We uses the basis $\{|\psi_{i}({\bf R}_0)\rangle\equiv|\phi_\alpha({\mathbf R_0}),n\rangle\}$ at the reference nuclear geometry ${\bf R}(t_0)$ as the {\it diabatic} basis during this short-time propagation, such that
\begin{equation}\label{eqn:qdidea}
|\psi_{i}({\bf R}_{0})\rangle\equiv|\psi_{i}({\bf R}(t_0))\rangle,~~\mathrm{for}~t\in[t_0,t_1].
\end{equation}
With the above QD basis defined independently of ${\bf R}(t)$ within each propagation segment, the electronic derivative couplings vanish while $\hat{V}({\bf R}(t))$ in the QD basis becomes off-diagonal. With this local diabatic basis, all of the necessary diabatic quantities can be evaluated and used to propagate quantum dynamics during $t\in[t_0,t_1]$. 

During this propagation step, the matrix element of $\hat{V}$ in the QD basis is evaluated as
\begin{equation}\label{eqn:vijt} 
V_{\alpha\beta, m n}({\bf R}(t))  = \langle \phi_\alpha ({\bf R}_0), m| \hat V ({\bf R}(t))| \phi_\beta ({\bf R}_0), n\rangle.
\end{equation}
For on-the-fly simulations, this quantity is obtained from a linear interpolation\cite{Rossky-Webster} between $V_{{\alpha}{\beta},m n} ({\bf R}_{0})$ and $V_{\alpha\beta,m n}({\bf R}(t_1))$ as follows
\begin{align}\label{eqn:interpolation}
&V_{\alpha\beta,m n}({\bf R}(t))= V_{\alpha\beta,m n}({\bf R}_{0}) \\
&~~~~~+ \frac {(t - t_{0})}{(t_{1} - t_{0})}\big[V_{\alpha\beta,mn}({\bf R}(t_{1})) - V_{\alpha\beta,mn}({\bf R}_{0})\big].\nonumber
\end{align} 
The above linear interpolation scheme can be further improved in future work and one potential choice is the recently developed norm-preserving interpolation scheme.\cite{meek2014evaluation,jain2016efficient}

It is straightforward to evaluate $V_{{\alpha}{\beta},m n} ({\bf R}_{0})$ and $V_{\alpha\beta,m n}({\bf R}(t_1))$ separately for the molecule-cavity hybrid system, as discussed below. Using electronic {\it ab initio} calculation, as well as the properties of $\hat{a}^{\dagger}$ and $\hat{a}$ for the photonic DOF, we can explicitly evaluate each term of $V_{\alpha\beta,m n}({\bf R}_{0})$ (see Eq.~\ref{eqn:Henp}) as follows
\begin{subequations}
\begin{align}
&H^\mathrm{en}_{\alpha\beta,m n}({\bf R}_{0}) = \langle \phi_\alpha({\bf R}_{0}), m | \hat{H}_\mathrm{en} ({\bf R}_{0})|\phi_\beta({\bf R}_{0}), n\rangle  \nonumber \\ 
&~~~= E_{\alpha}({\bf R}_{0}) \delta_{{\alpha},{\beta}} \delta_{m,n} \\
&H^\mathrm{p}_{\alpha\beta,m n}({\bf R}_{0}) = \langle \phi_\alpha({\bf R}_{0}), m | \hat{H}_\mathrm{p} | \phi_\beta({\bf R}_{0}), n \rangle \nonumber \\ 
&~~~= \hbar\omega_\mathrm{c} (n+\frac12)\delta_{\alpha,\beta} \delta_{m,n}\\
&H^\mathrm{enp}_{\alpha\beta,m n}({\bf R}_{0}) = \langle \phi_\alpha({\bf R}_{0}), m | \hat{H}_\mathrm{enp} ({\bf R}_{0}) | \phi_\beta({\bf R_{0}}), n \rangle \nonumber \\
&~~~= g_\mathrm{c} \boldsymbol{\epsilon} \cdot {\boldsymbol\mu}_{\alpha\beta}({\bf R}_{0}) \big(\sqrt{n}\delta_{m,n-1} + \sqrt{n+1}\delta_{m,n+1} \big) \\
&H^\mathrm{d}_{\alpha\beta,m n}({\bf R}_{0}) = \langle \phi_\alpha({\bf R}_{0}), m | \hat{H}_\mathrm{d} ({\bf R}_{0}) | \phi_\beta({\bf R}_{0}), n \rangle \nonumber\\
&~~~= \frac{g_{\mathrm{c}}^2}{\hbar\omega_\mathrm{c}} \sum_{\gamma} (\boldsymbol{\epsilon} \cdot {\boldsymbol\mu}_{\alpha\gamma} ({\bf R}_{0})) (\boldsymbol{\epsilon} \cdot {\boldsymbol\mu}_{\gamma\beta} ({\bf R}_{0})) \delta_{m,n} \nonumber\\ &
~~~\equiv D^2_{\alpha\beta}({\bf R}_{0}) \delta_{m,n},
\end{align}
\end{subequations}
where $\hat{H}_\mathrm{p}$ (see its definition in Eq.~\ref{eqn:Hp}) is an ${\bf R}$-independent operator, the sum $\sum_{\gamma}$ in the matrix element of $\hat{H}_\mathrm{d}$ runs over the diabatic states, and $D^{2}_{\alpha\beta}$ denotes the elements of DSE. Further, the matrix element of the dipole operator under the diabatic representation is expressed as
\begin{equation}\label{dipole-mat}
{\boldsymbol\mu}_{\alpha\beta} ({\bf R}_{0})\equiv\langle \phi_\alpha ({\bf R}_{0}) |\hat{\boldsymbol\mu} ({\bf R}_{0})|\phi_\beta ({\bf R}_{0}) \rangle.
\end{equation}
Similarly, at time $t_1$, the matrix element $V_{\alpha\beta, m n} ({\bf R}(t_1))=\langle \phi_\alpha ({\bf R}_{0}), m| \hat V ({\bf R}(t_1))| \phi_\beta ({\bf R}_{0}), n\rangle$ can also be written explicitly, with each term expressed as follows
\begin{subequations}
\begin{align}
&H^\mathrm{en}_{\alpha\beta,mn}({\bf R}(t_{1})) = \langle\phi_\alpha({\bf R}_{0}), m | \hat{H}_\mathrm{en} ({\bf R}(t_1)) | \phi_\beta({\bf R}_{0}), n \rangle \nonumber\\
&~~~= H^{\mathrm{en}}_{\alpha\beta}({\bf R}(t_1)) \delta_{m,n} \\
&H^\mathrm{p}_{\alpha\beta,mn}({\bf R}(t_{1})) = \langle \phi_\alpha({\bf R}_{0}), m | \hat{H}_\mathrm{p} | \phi_\beta({\bf R}_{0}),n \rangle \nonumber\\ 
&~~~= \hbar\omega_\mathrm{c} (n+\frac12)\delta_{\alpha,\beta} \delta_{m,n}\\
&H^\mathrm{enp}_{\alpha\beta,mn}({\bf R}(t_{1})) = \langle \phi_\alpha({\bf R}_{0}), m | \hat{H}_\mathrm{enp} ({\bf R}(t_1)) | \phi_\beta({\bf R}_{0}), n \rangle \nonumber\\ 
&~~~= g_\mathrm{c} \boldsymbol{\epsilon} \cdot {\boldsymbol\mu}_{\alpha\beta}({\bf R}(t_1)) \big(\sqrt{n}\delta_{m,n-1} + \sqrt{n+1}\delta_{m,n+1} \big)\\
&H^\mathrm{d}_{\alpha\beta,mn}({\bf R}(t_{1})) = \langle \phi_\alpha({\bf R}_{0}), m | \hat{H}_\mathrm{d} ({\bf R}(t_1)) | \phi_\beta({\bf R_{0}}), n \rangle \nonumber\\ 
&~~~=\frac{g_{\mathrm{c}}^2}{\hbar\omega} \sum_{\gamma} (\boldsymbol{\epsilon} \cdot {\boldsymbol\mu}_{\alpha\gamma} ({\bf R}(t_1))) (\boldsymbol{\epsilon} \cdot {\boldsymbol\mu}_{\gamma\beta} ({\bf R}(t_1))) \delta_{m,n} \nonumber\\ 
&~~~\equiv D^2_{\alpha\beta}({\bf R}(t_1)) \delta_{m,n}, 
\end{align}
\end{subequations}
where $H^{\mathrm{en}}_{\alpha\beta}({\bf R}(t_1))\equiv\langle\phi_\alpha({\bf R}_{0})| \hat{H}_\mathrm{en} ({\bf R}(t_1)) | \phi_\beta({\bf R}_{0})\rangle$, and ${\boldsymbol\mu}_{\alpha\beta} ({\bf R}(t_1))\equiv\langle \phi_\alpha ({\bf R}_{0}) |\hat{\boldsymbol\mu} ({\bf R}(t_1))|\phi_\beta ({\bf R}_{0}) \rangle$.\\

To conveniently calculate $H^{\mathrm{en}}_{\alpha\beta}({\bf R}(t_1))$ and ${\boldsymbol\mu}_{\alpha\beta} ({\bf R}(t_1))$, we use the following relations
\begin{subequations}
\begin{align}
H^{\mathrm{en}}_{\alpha\beta}({\bf R}(t_1)) & =\sum_{\lambda\nu}S_{\alpha\lambda} \tilde{H}^{\mathrm{en}}_{\lambda\nu}({\bf R}(t_1)) S^{\dagger}_{\beta\nu} \label{eqn:elect2} \\
{\boldsymbol\mu}_{\alpha\beta} ({\bf R}(t_1)) & =\sum_{\lambda\nu}S_{\alpha\lambda}\tilde{{\boldsymbol\mu}}_{\lambda\nu} ({\bf R}(t_1)) S^{\dagger}_{\beta\nu}, \label{eqn:mu2}
\end{align}
\end{subequations}
where the matrix elements at ${\bf R}(t_1)$ are expressed as
\begin{subequations}
\begin{align}
\tilde{H}^{\mathrm{en}}_{\lambda\nu}({\bf R}(t_1)) & =\langle \phi_{\lambda}({\bf R}(t_{1}))| \hat{H}_\mathrm{en} ({\bf R}(t_1))|\phi_{\nu}({\bf R}(t_{1})) \rangle \\
&=E_{\lambda}({\bf R}(t_1))\delta_{\lambda\nu} \nonumber \\
\tilde{\boldsymbol\mu}_{\lambda\nu} ({\bf R}(t_1)) & =\langle \phi_{\lambda}({\bf R}(t_{1}))| \hat{\boldsymbol\mu} ({\bf R}(t_1)) |\phi_{\nu}({\bf R}(t_{1})) \rangle,
\end{align}
\end{subequations}
and the overlap matrix between two electronic adiabatic states (with two different nuclear geometries) are  
\begin{subequations}
\begin{align}
S_{\alpha\lambda}&= \langle \phi_{\alpha}({\bf R}_0)|\phi_{\lambda}({\bf R}(t_{1}))\rangle\label{eqn:wf_overlap}\\
S^{\dagger}_{\beta\nu}&= \langle \phi_{\nu}({\bf R}(t_{1}))|\phi_{\beta}({\bf R}_0)\rangle.\label{eqn:wf_overlap_diag}
\end{align}
\end{subequations}
Using the above information, as well as Eq.~\ref{eqn:interpolation}, we can obtain each tern of $V_{\alpha\beta,m n}({\bf R}(t))$ for propagating the dynamics of the quantum subsystem that contains both electronic and photonic DOFs.

Next, we need to evaluate the nuclear gradients to propagate the dynamics of the classical subsystem, which contains the nuclear DOFs. In particular, we need to evaluate the nuclear gradients on each term of $\nabla V_{\alpha\beta,mn}({\bf R}(t_1))$. First, let us focus on the gradient term from the electron-nuclear coupling term\cite{ZhouJCPL2019} as follows
\begin{align}\label{dHen}
&{\nabla} H^\mathrm{en}_{\alpha\beta,m n}({\bf R}(t_1))\\
&= {\nabla} \langle \phi_{\alpha} ({\bf R}(t_0)),m|\hat{H}_\mathrm{en}({\bf R}(t_1))|\phi_{\beta}({\bf R}(t_0)),n\rangle \nonumber\\
&=\langle \phi_{\alpha} ({\bf R}(t_0))|{\nabla}\hat{H}_\mathrm{en}({\bf R}(t_1))|\phi_{\beta}({\bf R}(t_0))\rangle\cdot\langle m|n\rangle \nonumber\\
&=\sum_{\lambda\nu}S_{\alpha\lambda} \langle \phi_\lambda({{\bf R}({t_1})}) |\nabla{\hat{H}}_{\mathrm{en}} ({\bf R}(t_1))| \phi_\nu({{\bf R}(t_1)}) \rangle S^{\dagger}_{\beta\nu}\cdot \delta_{mn}. \nonumber
\end{align}
Here, from the first to the second line, we have used the fact that neither $\langle\phi_{\alpha}({\bf R}_0)|$ nor $\langle m|$ are ${\bf R}$-dependent, which allows ${\nabla}$ to bypass both and directly act on $\hat{V}({\bf R}(t_{1}))$. We have also used the fact that $\hat{H}_\mathrm{en}$ does not contain any photonic operators. The gradient term $\langle \phi_\lambda({{\bf R}({t_1})}) |\nabla{\hat{H}}_{\mathrm{en}} ({\bf R}(t_1))| \phi_\nu({{\bf R}(t_1)}) \rangle$ can be evaluated using the following well-known equality\cite{tully94jcp}
\begin{equation}\label{eqn:HF}
\langle \phi_{\lambda} ({\bf R}) |\nabla \hat{H}_\mathrm{en}({\bf R}) | \phi_{\nu} ({\bf R}) \rangle = 
\begin{cases}
     \nabla E_{\lambda} \quad & (\lambda=\nu) \\
      {\bf d}_{\lambda\nu} \left( E_{\nu} - E_{\lambda} \right) \quad & (\lambda \neq \nu), 
\end{cases}
\end{equation}
where the non-adiabatic coupling (NAC) vector (or so-called  derivative coupling) is 
\begin{equation}\label{elec-nac}
{\bf d}_{\lambda\nu}=\langle \phi_{\lambda} ({\bf R}) |\nabla|\phi_{\nu} ({\bf R}) \rangle.
\end{equation}

For the gradient on the matrix $H^\mathrm{p}_{\alpha\beta,m n}$, because there is no nuclear DOF in $\hat{H}_\mathrm{p}$, thus 
\begin{equation}
\nabla H^\mathrm{p}_{\alpha\beta,m n}({\bf R}({t_1}))= \nabla \big[\hbar\omega_\mathrm{c} (n+\frac12)\delta_{\alpha,\beta} \delta_{m,n}\big] = 0.
\end{equation}

For the gradient on the light-matter interaction term $H^\mathrm{enp}_{\alpha\beta,m n}$, we have

\begin{widetext}
\begin{align}\label{eqn:gradient_en}
&\nabla H^\mathrm{enp}_{\alpha\beta,m n}({\bf R}(t_1))= {\nabla}\langle \phi_{\alpha} ({\bf R}(t_0)),m|\hat{H}_\mathrm{enp}({\bf R}(t_1))|\phi_{\beta}({\bf R}(t_0)),n\rangle \\
& = \langle \phi_{\alpha} ({\bf R}(t_0))|{\nabla} \hat{\boldsymbol \mu}({\bf R}(t_1))|\phi_{\beta}({\bf R}(t_0))\rangle\cdot \boldsymbol{\epsilon}\cdot  g_\mathrm{c} \big(\sqrt{n}\delta_{m,n-1} + \sqrt{n+1}\delta_{m,n+1}\big)\nonumber\\
&=\sum_{\lambda\nu}S_{\alpha\lambda} \langle \phi_\lambda({{\bf R}({t_1})})|\nabla\hat{\boldsymbol\mu} ({\bf R}(t_1))| \phi_\nu({{\bf R}(t_1)}) \rangle S^{\dagger}_{\beta\nu}\cdot \boldsymbol{\epsilon}\cdot g_\mathrm{c} \big(\sqrt{n}\delta_{m,n-1} + \sqrt{n+1}\delta_{m,n+1}\big),\nonumber
\end{align}
\end{widetext}

where $S_{\alpha\lambda}$ and $S^{\dagger}_{\beta\nu}$ are defined in Eq.~\ref{eqn:wf_overlap} and Eq.~\ref{eqn:wf_overlap_diag}, respectively. To evaluate the term $\langle\phi_\lambda({\bf R})|\nabla\hat{\boldsymbol\mu} ({\bf R})| \phi_\nu({\bf R)}\rangle$ that appears in Eq.~\ref{eqn:gradient_en}, we use a simple relation based on the chain rule as follows
\begin{widetext}
\begin{align}\label{eq:muderive}
&\langle\phi_\lambda({\bf R})|\nabla\hat{\boldsymbol\mu} ({\bf R})| \phi_\nu({\bf R})\rangle=\nabla\langle\phi_{\lambda}({\bf R})|\hat{\boldsymbol\mu} ({\bf R})|\phi_{\nu}({\bf R})\rangle - \langle\nabla \phi_{\lambda}({\bf R})|\hat{\boldsymbol\mu} ({\bf R})|\phi_{\nu}({\bf R})\rangle-\langle \phi_{\lambda}({\bf R})|\hat{\boldsymbol\mu} ({\bf R})|\nabla\phi_{\nu}({\bf R})\rangle \nonumber \\
&=\nabla {\boldsymbol\mu}_{\lambda\nu}({\bf R}) -\sum_{\gamma}\langle\nabla \phi_{\lambda}({\bf R})|\phi_{\gamma}({\bf R})\rangle {\boldsymbol\mu}_{\gamma \nu}({\bf R})-\sum_{\gamma}{\boldsymbol\mu}_{\lambda\gamma} ({\bf R})\langle\phi_{\gamma} ({\bf R}) |\nabla\phi_{\nu}({\bf R})\rangle\nonumber\\
&=\nabla {\boldsymbol\mu}_{\lambda\nu}({\bf R})+ \sum_{\gamma}\big[{\bf d}_{\lambda \gamma}({\bf R}) {\boldsymbol\mu}_{\gamma \nu}({\bf R}) -{\boldsymbol\mu}_{\lambda\gamma}({\bf R}){\bf d}_{\gamma\nu}({\bf R})\big], 
\end{align}
\end{widetext}
where ${\bf d}_{\lambda\gamma}({\bf R})=\langle\phi_{\lambda} ({\bf R})|\nabla\phi_{\gamma}({\bf R})\rangle=-\langle\nabla\phi_{\lambda}({\bf R})|\phi_{\gamma}({\bf R})\rangle$ and ${\bf d}_{\gamma\nu}({\bf R})=\langle\phi_{\gamma} ({\bf R})|\nabla\phi_{\nu}({\bf R})\rangle=-\langle\nabla\phi_{\gamma}({\bf R})|\phi_{\nu}({\bf R})\rangle$ are the electronic derivative couplings (defined in Eq.~\ref{elec-nac}). Note that from the first line to the second line, we have inserted $\hat{\mathcal P}=\sum_{\gamma}|\phi_{\gamma}({\bf R})\rangle\langle\phi_{\gamma}({\bf R})|$, which is the resolution of identity in the electronic subspace. As one can clearly see, this term requires the evaluation of derivative coupling ${\bf d}_{\lambda\gamma}({\bf R})$ and ${\bf d}_{\gamma\nu}({\bf R})$, and the derivative on dipole matrix element $\nabla {\boldsymbol\mu}_{\lambda\nu}({\bf R})$.
For most of the electronic structure methods, the derivatives on dipole matrix elements $\nabla {\boldsymbol\mu}_{\lambda\nu}({\bf R})$ are not implemented. Nevertheless, recent theoretical development has made these quantities available.\cite{Zhang2019jcp}

Finally, the gradient from the DSE term is expressed as follows
\begin{widetext}
\begin{align} \label{eqn:gradient_d}
&\nabla H^\mathrm{d}_{\alpha\beta,m n}({\bf R}(t_1))= {\nabla} \langle \phi_{\alpha} ({\bf R}_{0}),m|\hat{H}_\mathrm{d}({\bf R}(t_1))|\phi_{\beta}({\bf R}_{0}), n\rangle \\
=& {\nabla} \Big[\sum_{\gamma} (\boldsymbol{\epsilon} \cdot \langle \phi_{\alpha} ({\bf R}_{0}) |\hat{\boldsymbol \mu}({\bf R}(t_1))|\phi_{\gamma}({\bf R}_{0})\rangle) (\boldsymbol{\epsilon} \cdot \langle \phi_{\gamma} ({\bf R}_{0}) |\hat{\boldsymbol \mu}({\bf R}(t_1))|\phi_{\beta}({\bf R}_{0})\rangle) \Big] \frac{g_{\mathrm{c}}^2}{\hbar\omega} \delta_{m,n} \nonumber \\
=& \Big[\sum_{\gamma} (\boldsymbol{\epsilon} \cdot \langle \phi_{\alpha} ({\bf R}_{0}) |{\nabla} \hat{\boldsymbol \mu}({\bf R}(t_1))|\phi_{\gamma}({\bf R}_{0})\rangle) (\boldsymbol{\epsilon} \cdot \langle \phi_{\gamma} ({\bf R}_{0}) |\hat{\boldsymbol \mu}({\bf R}(t_1))|\phi_{\beta}({\bf R}_{0})\rangle) \nonumber  \\
& + (\boldsymbol{\epsilon} \cdot \langle \phi_{\alpha} ({\bf R}_{0}) | \hat{\boldsymbol \mu}({\bf R}(t_1))|\phi_{\gamma}({\bf R}_{0})\rangle) (\boldsymbol{\epsilon} \cdot \langle \phi_{\gamma} ({\bf R}_{0}) |{\nabla} \hat{\boldsymbol \mu}({\bf R}(t_1))|\phi_{\beta}({\bf R}_{0})\rangle) \Big] \frac{g_{\mathrm{c}}^2}{\hbar\omega} \delta_{m,n},\nonumber
\end{align}
\end{widetext}
where the term $\langle \phi_{\alpha} ({\bf R}(t_0)) | \hat{\boldsymbol \mu}({\bf R}(t_1))|\phi_{\gamma}({\bf R}(t_0))\rangle$ can be evaluated using Eq.~\ref{eqn:mu2}, and the $\langle \phi_{\alpha} ({\bf R}_{0}) |{\nabla} \hat{\boldsymbol \mu}({\bf R}(t_1))|\phi_{\beta}({\bf R}_{0})\rangle$ type of derivative can be computed in the same fashion as elaborated in Eq.~\ref{eqn:gradient_en}.

Using the matrix elements $V_{\alpha\beta,m n}({\bf R}(t))$ (Eq.~\ref{eqn:interpolation}) and the nuclear gradient $\nabla V_{\alpha\beta,mn}({\bf R})$ (as outlined in Eqs.~\ref{dHen} -~\ref{eqn:gradient_d}), one can in principle use any trajectory-based approaches or wavepacket approaches with guiding trajectories\cite{IzmaylovJCP2018,Ben-Nun1998jcp,mcEhrenfest} to propagate the quantum dynamics in the time step of $t\in[t_0,t_1]$. During the next short-time propagation segment $t\in[t_1,t_2]$, the QD scheme adopts a new reference geometry ${{\bf R}'_{0}}\equiv {\bf R}(t_1)$ and new {\it diabatic} basis $|\psi_{j}({{\bf R}'_{0}})\rangle\equiv|\psi_{j}({\bf R}(t_1))\rangle=|\phi_{\beta}({\bf R}(t_{1})),m\rangle$. Between the $t\in [t_0,t_1]$ propagation and the $t \in [t_1,t_2]$ propagation segments, all of the necessary quantities will be transformed from $\{|\psi_{i}({\bf R}_{0})\rangle\}$ to the $\{|\psi_{j}({\bf R}'_{0})\rangle\}$ basis, using the relation 
\begin{equation}\label{eqn:basis}
    |\psi_{j}({\bf R}(t_{1}))\rangle=\sum_{i} \langle \psi_{i}({\bf R}(t_{0}))| \psi_{j}({\bf R}(t_{1}))\rangle|\psi_{i}({\bf R}(t_{0}))\rangle.
\end{equation}

For the model calculations in this work, these overlap integrals are evaluated as 
\begin{equation}\label{overlap}
\langle\psi_{i}({\bf R}(t_{0}))|\psi_{j}({\bf R}(t_{1}))\rangle=\langle\phi_{\alpha}({\bf R}(t_{0}))|\phi_{\beta}({\bf R}(t_{1}))\rangle\cdot\langle n|m\rangle,
\end{equation}
where the electronic adiabatic state overlaps $\langle\phi_{\alpha}({\bf R}(t_{0}))|\phi_{\beta}({\bf R}(t_{1}))\rangle$ are directly calculated using the discrete variable representation (DVR) basis, and the Fock states are orthonormal to each other $\langle n|m\rangle=\delta_{n,m}$. 
When performing the transformation in Eq.~\ref{eqn:basis} (as well as in  Eq.~\ref{map-trans2} for the non-adiabatic mapping methods), the eigenvectors maintain their mutual orthogonality subject to a very small error when they are expressed in terms of the previous basis due to the incompleteness of the basis.\cite{GranucciJCP2001,PlasserJCP2012} Nevertheless, the orthogonality remains to be well satisfied among  $\{|\psi_{i}({\bf R}(t_0))\rangle\}$ or $\{|\psi_{j}({\bf R}(t_1))\rangle\}$. This small numerical error generated from each step can, however, accumulate over many steps and cause a significant error at longer times, leading to non-unitary dynamics.\cite{GranucciJCP2001,PlasserJCP2012} This potential issue can be easily resolved by using orthonormalization procedure among the vectors of the overlap matrix composed by $\langle \psi_{i}({\bf R}(t_{0}))| \psi_{j}({\bf R}(t_{1}))\rangle$, as been done in our previous work\cite{MandalJCP2018} for simulating photo-induced charge transfer dynamics. Here, we perform the L\"owdin orthogonalization procedure\cite{LowdinJCP1950} as commonly used in the local diabatization approach\cite{GranucciJCP2001} to ensure unitary propagation.

As the nuclear geometry closely follows the reference geometry throughout the propagation, the QD basis forms a convenient and compact basis. Note that, in principle, one needs an infinite set of crude adiabatic states $\{|\psi_{i}({\bf R}_0)\rangle\}$ to represent the time-dependent electronic wavefunction because the electronic wavefunction could change rapidly with the motion of the nuclei, and the crude adiabatic basis is only convenient when the reference geometry ${\bf R}_{0}$ is close to the nuclear geometry ${\bf R}$. By dynamically updating the basis in the QD scheme, the time-dependent electronic wavefunction is expanded with the ``moving crude adiabatic basis"\cite{IzmaylovJCP2018} that explores the most relevant and important parts of the Hilbert space, thus requiring only a few states for quantum dynamics propagation. 

\subsection{Non-adiabatic Mapping Dynamics Methods}\label{subsec:mapping_method}
The Meyer-Miller-Stock-Thoss (MMST) formalism\cite{MeyerJCP1979,StockPRL1997,ThossPRA1999} maps the discrete electronic DOFs onto continuous phase space variables. In the strict diabatic basis $\{|i\rangle\}$ (in the sense that $\langle i|\nabla|j\rangle=0$ for all $|i\rangle$ and $|j\rangle$), the total Hamiltonian in Eq.~\ref{totalH} is expressed as
\begin{equation}\label{Diabatic-Hamiltonian}
\hat{H}={\frac {{\bf P}^2} {2{\bf M}}}+\sum_{i}V_{ii}(\hat{\bf R})|i\rangle\langle i|+\sum_{i\neq j}V_{ij}(\hat{\bf R})|i\rangle\langle j|,
\end{equation}
where $V_{ij}(\hat{\bf R})=\langle i|\hat{V}(\hat{\bf r},\hat {\bf R})|j\rangle$ are the matrix elements of the electronic Hamiltonian. Note that here $|i\rangle$ is used to represent the strict diabatic basis, and not to be confused with the adiabatic-Fock state $|\psi_i({\mathbf R}) \rangle=|\phi_{\alpha}({\bf R}),n\rangle$ introduced in Eq.~\ref{adia-fock}. Nevertheless, based on the QD scheme, these adiabatic-Fock states with a reference geometry ${\bf R}_{0}$ will be used as the diabatic state in the neighborhood of the reference geometries, as indicated in Eq.~\ref{eqn:qdidea}.
 
In the non-adiabatic mapping approach, the Hamiltonian operator in Eq.~\ref{Diabatic-Hamiltonian} is transformed into the following MMST Hamiltonian
\begin{equation} \label{eq:mapham} 
\mathcal{H}_\mathrm{m}={\frac {{\bf P}^2} {2{\bf M}}}+{\frac{1}{2}}\sum_{ij}V_{ij}({\bf R})\left(p_{i}p_{j}+q_{i}q_{j}-2\gamma_{j}\delta_{ij}\right),
\end{equation}
where $2\gamma_{j}$ is viewed as a parameter\cite{MillerFD2016} which specifies the ZPE of the mapping oscillators.\cite{UweJCP1999,MillerFD2016,richardson2019,richardson2020} In principle, $2\gamma_{j}$ is state-specific and trajectory-specific.\cite{CottonJCP2019_2} The MMST mapping Hamiltonian has been historically justified by Stock and Thoss using harmonic oscillator's raising and lowering operators as the mapping operator.\cite{StockPRL1997,ThossPRA1999} Recently, it has been derived using the $SU(N)$ Lie group theory or so-called generalized spin mapping approach.\cite{richardson2020}

Classical trajectories are generated based on Hamilton's equations of motion (EOM)
\begin{subequations}\label{eq:mapeqn}
\begin{eqnarray}  
    \dot q_{j} &=& \partial \mathcal{H}_\mathrm{m}/ \partial p_{j};~~\dot p_{i} = -\partial \mathcal{H}_\mathrm{m} / \partial q_{i}\\
    \dot {\bf R} &=& \partial \mathcal{H}_\mathrm{m}/ \partial {\bf P};~~ \dot {\bf P}=-\partial \mathcal{H}_\mathrm{m} / \partial {\bf R}= {\bf F}, 
\end{eqnarray}
\end{subequations}
with the nuclear force expressed as
\begin{equation}
\label{eq:force}
{\bf F}=-{\frac {1} {2}}\sum_{ij}\nabla V_{ij}({\bf R})\big(p_{i}p_{j}+q_{i}q_{j}-2\gamma_{j}\delta_{ij}\big).
\end{equation}
Overall, the MMST mapping provides a consistent classical footing for both electronic and nuclear DOFs, and the non-adiabatic transitions between electronic states are captured through the classical motion of the fictitious harmonic oscillators. The non-adiabatic dynamics obtained from this formalism have shown good performance in the {\it ab initio} on-the-fly dynamics.\cite{ZhouJCPL2019,HuJCTC2021,WeightJCP2021}

To sample the initial electronic condition and estimate the population, it is also convenient to use the action-angle variables, $\{\varepsilon_{j}, \theta_{j}\}$, which are related to the canonical mapping variables $\{p_{j}, q_{j}\}$ through 
\begin{equation}\label{action-angle}
    \varepsilon_j = \frac{1}{2}\left(p_j^2 + q_j^2 \right);~~~\theta_j =-\tan^{-1}\left( \frac{p_j}{q_j}\right),
\end{equation}
and the inverse relations
\begin{equation}\label{eqn:action_to_mapping}
    q_j = \sqrt{2 \varepsilon_j }\cos(\theta_j);~~p_j =-\sqrt{2 \varepsilon_j}\sin(\theta_j),
\end{equation}
where $\varepsilon_j$ is a positive-definite action variable that is directly proportional to the mapping variables' radius in action-space.\cite{CottonJCP2019_2} 

The SQC approach calculates the population of electronic state $|j\rangle$, which is to be evaluated as\cite{MillerFD2016}
\begin{align}\label{eqn:wignersqc}
\rho_{jj}(t)&=\mathrm{Tr}_{\bf R}\left[\hat{\rho}(0)e^{i\hat{H}t/\hbar}|j\rangle\langle j|e^{-i\hat{H}t/\hbar}\right]\\
&\approx\int d\boldsymbol{\tau} \rho_\mathrm{W}({\bf P},{\bf R})W_i({\boldsymbol \varepsilon}(0))W_j({\boldsymbol \varepsilon}(t)),\nonumber
\end{align}
where $\hat{\rho}(0)=\hat{\rho}_{\bf R} \otimes |i\rangle \langle i|$ is the initial density operator, $\rho_\mathrm{W}({\bf P},{\bf R})$ is the Wigner transform of $\hat{\rho}_{\bf R}$ operator for the nuclear DOFs, ${\boldsymbol \varepsilon} = \{\varepsilon_1,\varepsilon_2,...,\varepsilon_\mathcal{N}\}$ is the positive-definite action variable vector for $\mathcal{N}$ electronic states,\cite{CottonJCP2019_2} $W_i({\boldsymbol \varepsilon})= \delta (\varepsilon_i - (1+\gamma_{i}))\prod_{i \neq j} \delta(\varepsilon_j-\gamma_{j})$ is the Wigner transformed action variables,\cite{CottonJCP2016} and $d\boldsymbol{\tau}\equiv d{\bf P}\cdot d{\bf R}\cdot d{\boldsymbol \varepsilon} \cdot d\boldsymbol{\theta}$. For practical reasons, the above delta functions in $W_i({\boldsymbol \varepsilon})$ are broadened using a distribution function (so-called window function) that can be used to bin the resulting electronic action variables in action-space.\cite{MillerFD2016} Further, we use the $\gamma$-SQC approach,\cite{CottonJCP2019_2} which uses a {\it state-specific} and {\it trajectory-specific} $\gamma_{j}$ parameter in Eq.~\ref{eq:mapham} to correct the initial force according to the initially populated state. This method has been proven to provide very accurate non-adiabatic dynamics in model photo-dissociation problems (coupled Morse potential), as well as outperform FSSH (with decoherence correction) in {\it ab initio} on-the-fly simulations.\cite{HuJCTC2021, WeightJCP2021} The details of $\gamma$-SQC are provided in Appendix~\ref{apsec:details_of_sqc}.

For the spin-LSC approach,\cite{richardson2019,richardson2020} one chooses a universal ZPE parameter $2\gamma_{j}=\Gamma$ for all states and trajectories. The spin-LSC population dynamics is calculated as
\begin{align}\label{eqn:wignerlsc}
\rho_{jj}(t)&=\mathrm{Tr}_{\bf R}\left[\hat{\rho}_{R}\otimes|i\rangle\langle i|e^{i\hat{H}t/\hbar}|j\rangle\langle j|e^{-i\hat{H}t/\hbar}\right]\\
&\approx\int d\boldsymbol{\tau} \rho_\mathrm{W}({\bf P},{\bf R}) [|i\rangle\langle i|]_\mathrm{s}(0)\cdot[|j\rangle\langle j|]_\mathrm{\bar{s}}(t),\nonumber
\end{align}
where the population estimators are obtained from the Stratonovich-Weyl transformed electronic projection operators, with the expressions as follows\cite{richardson2020}
\begin{subequations}\label{sw-estimator}
\begin{align}
&\big[|i\rangle\langle i|\big]_\mathrm{s}=\frac{1}{2}(q^2_i+p^2_i-\Gamma)\\
&\big[|j\rangle\langle j|\big]_\mathrm{\bar s}
=\frac{\mathcal{N}+1}{2(1+\frac{\mathcal{N}\Gamma}{2})^2}\cdot (q^2_{j}+p^2_{j})-\frac{1-\frac{\Gamma}{2}}{1+\frac{\mathcal{N}\Gamma}{2}}.
\end{align}
\end{subequations}
The parameter $\Gamma$ is related to the radius of the generalized Bloch sphere $r_\mathrm{s}$ through $\Gamma=\frac{2}{\mathcal N}(r_\mathrm{s}-1)$, where $\mathrm{s}$ and $\mathrm{\bar s}$ are complementary indices in the Stratonovich-Weyl transform. Among the vast parameter space, one of the best performing choices\cite{richardson2019,richardson2020} is when $r_\mathrm{s}=r_\mathrm{\bar s}=\sqrt{\mathcal{N}+1}$, which is referred to as $\mathrm{s}=\mathrm{W}$, leading to a ZPE parameter
\begin{equation}\label{Gamma}
\Gamma=\frac{2}{\mathcal{N}}(\sqrt{\mathcal{N}+1}-1),
\end{equation}
as well as the identical expression of $[|i\rangle\langle i|]_\mathrm{s}$ and $[|j\rangle\langle j|]_\mathrm{\bar s}$ in Eq.~\ref{sw-estimator}. We further use the focused initial condition\cite{richardson2019,richardson2020} that replaces the sampling of the mapping variables in the $d\boldsymbol{\tau}$ integral of Eq.~\ref{eqn:wignerlsc} with specific values of the mapping variables, such that $\frac{1}{2}(q^2_i+p^2_i-\Gamma)=1$ for initially occupied state $|i\rangle$ and $\frac{1}{2}(q^2_j+p^2_j-\Gamma)=0$ for the initially unoccupied states $|j\rangle$. The angle variables $\{\theta_{j}\}$ (Eq.~\ref{action-angle}) are randomly sampled\cite{richardson2020} in the range of $[0,2\pi)$.

Using the QD propagation scheme, one can directly perform non-adiabatic using both $\gamma$-SQC and spin-LSC in their original diabatic formalism, with the information from the ``ab initio" polaritonic calculations of the molecule-cavity hybrid system. Using the schemes outlined in Sec.~\ref{sec:QD}, one can obtain the polariton coupling $\langle\psi_{i}({\bf R}_{0})|\hat{V}({\bf R})|\psi_{j}({\bf R}_{0})\rangle$ (see Eq.~\ref{eqn:vijt} and Eq.~\ref{eqn:interpolation}) and nuclear gradient $\nabla\langle\psi_{i}({\bf R}_{0})|\hat{V}({\bf R})|\psi_{j}({\bf R}_{0})\rangle$ (see Eq.~\ref{dHen}-Eq.~\ref{eqn:gradient_d}), which are the necessary ingredients to solve the MMST mapping EOMs in Eq.~\ref{eq:mapeqn} and Eq.~\ref{eq:force}. Between two propagation steps, the QD basis is transformed from $\{|\psi_{i}({\bf R}(t_0))\rangle\equiv|\phi_\alpha({\mathbf R}(t_0)),n\rangle\}$ to $\{|\psi_{j}({\bf R}(t_1))\rangle\equiv|\phi_\beta({\mathbf R}(t_1)),m\rangle\}$. This leads to the corresponding transform of mapping variables between the two consecutive QD bases as follows\cite{MandalJCTC2018,ZhouJCPL2019}
\begin{subequations}\label{map-trans2}
\begin{align}
    &\sum_{i} q_{i} \langle \psi_{i}({\bf R}(t_{0}))|\psi_{j}({\bf R}(t_{1}))\rangle \rightarrow q_{j}\\
    &\sum_{i} p_{i} \langle \psi_{i}({\bf R}(t_{0}))| \psi_{j}({\bf R}(t_{1}))\rangle \rightarrow p_{j},
\end{align}
\end{subequations}
where the overlaps between the two steps are evaluated using Eq.~\ref{overlap} (see discussion under that equation). More computational details for the $\gamma$-SQC and spin-LSC are provided in section \ref{sec:comp-detail}.

\section{Computational Details}
\subsection{The Model System}
\begin{figure}[ht!]
 \centering
  \begin{minipage}[t]{1.0\linewidth}
     \centering
     \includegraphics[width=\linewidth]{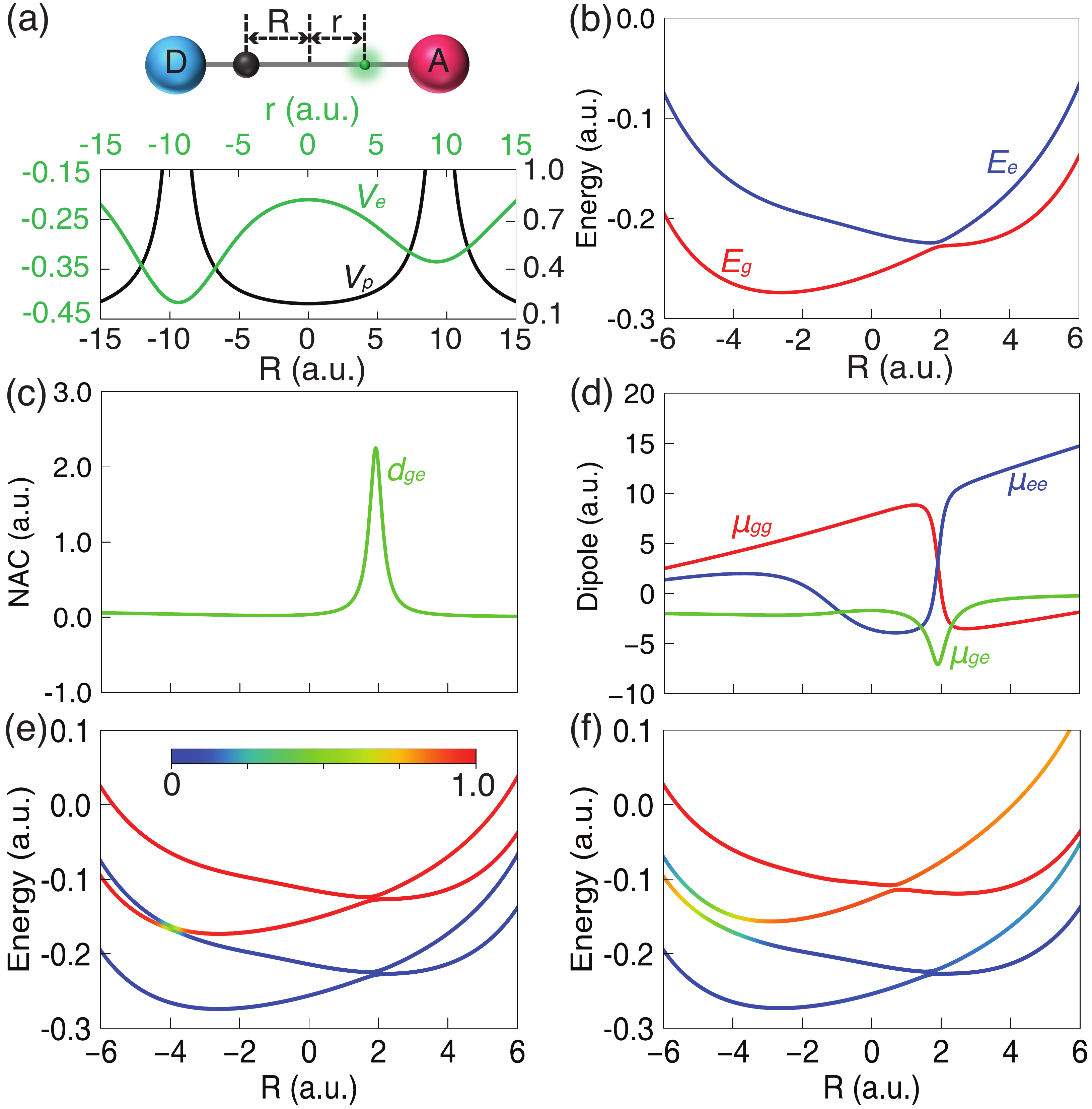}
  \end{minipage}%
   \caption{(a) The schematic illustration of the asymmetrical SM model, where one proton and one electron can transfer between two fixed ions (donor and acceptor). The distance between the donor and acceptor is 19 a.u. Here $V_p$ and $V_e$ are the potential of the transferring proton and electron, respectively. (b) The first two adiabatic electronic states, (c) NAC between these two electronic states and (d) their permanent and transition dipole moments of the SM model. The PESs of the polaritonic states inside the cavity are obtained with the light-matter coupling strength (e) $g_\mathrm{c}=0.001$ and (f) $g_\mathrm{c}=0.005$. The color used in (e) and (f) is coded according to $\langle \hat{a}^{\dagger}\hat{a}\rangle$, as shown in the upper position of panel (e).}
\label{fig:adiabatic_PES}
\end{figure}

In this work, we use the asymmetrical Shin-Metiu model\cite{Shin1995jcp,Hoffmann2020} as the ``\textit{ab initio}" model molecular system to investigate strong light-matter interactions between a molecule and an optical cavity. The model contains a transferring proton (nucleus) and an electron, as well as two fixed ions labeled as donor (D) and acceptor (A), as shown in Fig.~\ref{fig:adiabatic_PES}a. This model is usually used to describe the proton-coupled electron transfer (PCET) reaction and has been studied recently using the exact factorization approach to investigate how cavity can influence chemical reactivities.\cite{Maitra2019,Hoffmann2020,Maitra2021} The electron-nuclear interaction potential operator $\hat{H}_\mathrm{en}$ (c.f. Eq.~\ref{eqn:Hen}) is expressed as
\begin{equation}\label{Hen}
\hat{H}_\mathrm{en}=\sum_{\sigma=\pm 1} \left( \frac{1}{|R+\frac{\sigma L}{2}|}-\frac{\mathrm{erf}\Big( \frac{|r+\frac{\sigma L}{2}|}{a_{\sigma}} \Big)}{|r+\frac{\sigma L}{2}|} \right) - \frac{\mathrm{erf}\Big(\frac{|R-r|}{a_f}\Big)}{|R-r|},
\end{equation}
where the first term represents the potential of the transferring proton, the second term represents the potential of the transferring electron, and the third term represents the electron-proton coupling. We choose the same parameters used in Ref.~\citenum{Hoffmann2020}, which is $L=19 $ a.u., $a_+=3.1$ a.u., $a_{-}=4.0$ a.u., $a_f=5.0$ a.u. and the proton mass is $M=1836$ a.u. To calculate the electronic properties of the SM model, we use the Sinc discrete variable representation (DVR) basis\cite{Colbert1992} to represent the electronic adiabatic states. These adiabatic states $|\phi_{\alpha}(R)\rangle$ are computed on-the-fly for a given nuclear configuration $R$ by solving Eq.~\ref{HenPsi}. The details are provided in Appendix~\ref{Exact}.

Fig.~\ref{fig:adiabatic_PES}a also depicts the model potential in Eq.~\ref{Hen}, with the black curve depicting the proton potential (the first term in Eq.~\ref{Hen}) and the green curve depicting the electron potential (the second term in Eq.~\ref{Hen}). Fig.~\ref{fig:adiabatic_PES}b presents the two lowest adiabatic electronic states of the SM model (red and blue curves). There is an avoided crossing between the ground and the first excited state potential energy surfaces (PESs) near $R=2.0$ a.u.. Fig.~\ref{fig:adiabatic_PES}c presents the NAC between them (the green curve), which shows a strong coupling near the avoided crossing region. The matrix elements of the dipole operator under the adiabatic representation (Eq.~\ref{dipole-mat}) of the SM model are presented in Fig.~\ref{fig:adiabatic_PES}d. 

When coupling the SM molecular model with the cavity, the photon frequency of the cavity mode is chosen as $\hbar \omega_\mathrm{c}=2.721$ eV ($\approx \, 0.1 $ a.u.). Further, we assume that the cavity field polarization direction ${\boldsymbol{\epsilon}}$ is always aligned with the direction of the dipole operator $\hat{\boldsymbol\mu}$, such that ${\boldsymbol{\epsilon}}\cdot{\boldsymbol\mu}_{\gamma\nu}={\mu}_{\gamma\nu}$ (for $\{\nu,\gamma\}=\{e,g\}$) where ${\mu}_{\gamma\nu}$ is the magnitude of $\hat{\boldsymbol\mu}$. Explicitly considering the angle between ${\boldsymbol{\epsilon}}$ and $\hat{\boldsymbol\mu}$ will generate a polariton induced conical intersection (even for a diatomic molecule), which will induce geometric phase effects.\cite{Farag2021PCCP} We consider two different light-matter coupling strengths $g_\mathrm{c}=0.001$ a.u. and $g_\mathrm{c}=0.005$ a.u. in this work. The normalized coupling strength is often defined as\cite{Kockum2019rev} $\eta\equiv g_\mathrm{c}\cdot|\boldsymbol{\epsilon} \cdot  {{\boldsymbol \mu}}_{eg}|/\omega_\mathrm{c}$ where $|\boldsymbol{\epsilon} \cdot {{\boldsymbol \mu}}_{eg}|$ is the typical magnitude of the transition dipole projected along the field polarization direction. For the coupling strength considered above (and taking $\hbar\omega_\mathrm{c}\approx 2.721$ eV for the model calculation), the normalized coupling strength is $\eta = 0.06$ (for $g_\mathrm{c}=0.001$ a.u.) and $\eta = 0.3$ (for $g_\mathrm{c}=0.005$ a.u.). When $0.1<\eta<1$, the light and matter interaction achieves the ultra-strong coupling regime,\cite{Kockum2019rev,Nori2019natphys} which is difficult to achieve but still within the reach of the current experimental setup.\cite{Shegai_NC_2020,Haran2016} Thus, besides the pure theoretical value to derive the exact nuclear gradient expression, our computational results are also within the reach of the near future experimental setup.

The polaritonic PESs $\mathcal{E}_{J}(R)$ associated with polariton states $|\mathcal{E}_{J}(R)\rangle$ (see their definition in Eq.~\ref{polariton}) are presented in Fig.~\ref{fig:adiabatic_PES}e with the light-matter coupling strength $g_\mathrm{c}=0.001$ a.u. and in Fig.~\ref{fig:adiabatic_PES}f with the light-matter coupling strength $g_\mathrm{c}=0.005$. These polariton potentials are color coded (as shown in the inset of panel (e)) based on the expectation value of $\langle \hat{a}^{\dagger}\hat{a}\rangle$ indicated on top of this panel. Note that this should not be viewed as a ``photon number" operator under the dipole gauge used in the PF Hamiltonian\cite{Mandal2020JPCL,Schaefer2020AP} because the rigorous photon number operator should be obtained by applying the Power-Zienau-Woolley (PZW) Gauge transformation\cite{PZW,Cohen-Tannoudji,Taylor2020PRL} on the photon number operator $\hat{a}^{\dagger}\hat{a}$. Nevertheless, it can be viewed as an approximate estimation of the photon number when the light-matter couplings are not in the ultra-strong coupling regime.
\cite{Forn-Diaz2019rmp}

The initial state (for $t=0$) of the molecule-cavity hybrid system is
\begin{equation}\label{eq:initial-state}
|\Phi(t=0)\rangle=|e,0\rangle\otimes|\chi\rangle,
\end{equation}
which corresponds to a Franck-Condon excitation of the hybrid system to the $|e,0\rangle$ state, with $|\chi\rangle$ as the initial nuclear wavefunction. For the SM model in this work, we use $\chi(R)=\langle R|\chi\rangle\sim \exp[-M\omega_{0}(R-R_{0})^2/2\hbar]$, where $M$ is the mass of the proton (nucleus in the SM model), $R_0$ is the position with a minimum potential energy of the ground electronic state. Here, $\chi(R)$ is the vibrational ground state wavefunction on the ground electronic states, centered at $R_0$ under the harmonic approximation, with the harmonic oscillation frequency being $\omega_{0}$. We use the parameters in the original reference\cite{Hoffmann2020} for $R_{0}=-4$ and $\omega_{0}=0.000382$ a.u. To solve the exact quantum dynamics, we use the DVR basis for the nuclear DOF and the adiabatic-Fock state for the electronic-photonic subsystem. The details of the exact quantum dynamics are provided in Appendix~\ref{Exact}. 

\subsection{Details of $\gamma$-SQC and spin-LSC Dynamics}\label{sec:comp-detail}

To perform the $\gamma$-SQC dynamics, we need to sample the initial condition for the quantum subsystem. In this work, we first sample the action-angle variables $\{\varepsilon_{j}, \theta_{j}\}$ then transform them to the mapping variables$\{p_{j},q_{j}\}$ using Eq.~\ref{eqn:action_to_mapping}. Among them, the action variables $\{\varepsilon_{j}\}$ are sampled according to the window function in Eq.~\ref{EQ:TriangleWindow}, and the angle variables $\{\theta_{j}\}$ are randomly sampled from $[0,2\pi)$. The triangle window is used in this work, although the square window generates similar results. 

For the spin-LSC dynamics, we use the focused initial conditions\cite{richardson2020} as described in section \ref{subsec:mapping_method}, where the action variable $\varepsilon_{i}$ is set to be $1+\Gamma/2$ for the initially occupied state and $\Gamma/2$ for the initially unoccupied state, with $\Gamma$ expressed in Eq.~\ref{Gamma}. The angle variables $\{\theta_{j}\}$ are randomly generated between $[0,2\pi)$ as in the $\gamma$-SQC method. The canonical mapping variables are obtained from Eq.~\ref{eqn:action_to_mapping}.

The initial nuclear distribution of all trajectory-based simulations (Ehrenfest, FSSH, $\gamma$-SQC and spin-LSC) are generated by sampling the Wigner density 
\begin{equation}\label{wigner}
[\langle R|\chi\rangle]_\mathrm{w}=\frac{1}{\hbar\pi}e^{-M(P^2+\omega^2(R-R_{0})^2)/\omega\hbar},
\end{equation}
which is the Wigner transformation of the nuclear wavefunction $\chi(R)=\langle R|\chi\rangle$ in the initial state (see Eq.~\ref{eq:initial-state}). Here, $R$ and $P$ are the nuclear coordinate and momentum, respectively. The initial state for the electronic-photonic subsystem is set to $|e0\rangle$. The nuclear time step used in the QD-$\gamma$-SQC and QD-spin-LSC is $dt=0.1$ fs, with $100$ equally spaced electronic time steps for the mapping variables' integration during each nuclear time step. The equation of motion in Eq.~\ref{eq:mapeqn}-Eq.~\ref{eq:force} are integrated using a second-order symplectic integrator for the MMST variables\cite{kelly2012mapping,church2018} for a given nuclear time step, and these mapping variables are transformed based on Eq.~\ref{map-trans2} between two adjacent nuclear time steps due to the change of the QD basis. The population dynamics using all MQC and mapping methods were averaged over $5000$ trajectories, although $3000$ trajectories were enough to produce the basic trend of the polariton dynamics, see Fig. S3 in the Supplementary Material. The light-matter coupling strength $g_\mathrm{c}$ was chosen to be $0.001$ and $0.005$, according to our previous work.\cite{Zhou2022}

We also benchmark the results of non-adiabatic mapping dynamics approaches with commonly used MQC approaches, including the Ehrenfest dynamics and the FSSH method. The details of these two MQC approaches are provided in Appendix~\ref{apsec:Ehrenfest_FSSH_Details}. In particular, the Ehrenfest dynamics is equivalent to choosing $\gamma=0$ in the mapping theory (see Eq.~\ref{eq:mapham}) and an initial action-angle variables condition (see Eq.~\ref{action-angle} and Eq.~\ref{eqn:action_to_mapping}) of $\varepsilon_j=\delta_{ij}$ (for the initially occupied state $|i\rangle$) and $\theta_{j}=0$ (for all state $|j\rangle$). One can thus use the same QD scheme and the mapping equation to obtain the results of the Ehrenfest dynamics.\cite{Subotnik2016JCP} 

\begin{figure}[ht!]
 \centering
  \begin{minipage}[t]{1.0\linewidth}
     \centering
     \includegraphics[width=\linewidth]{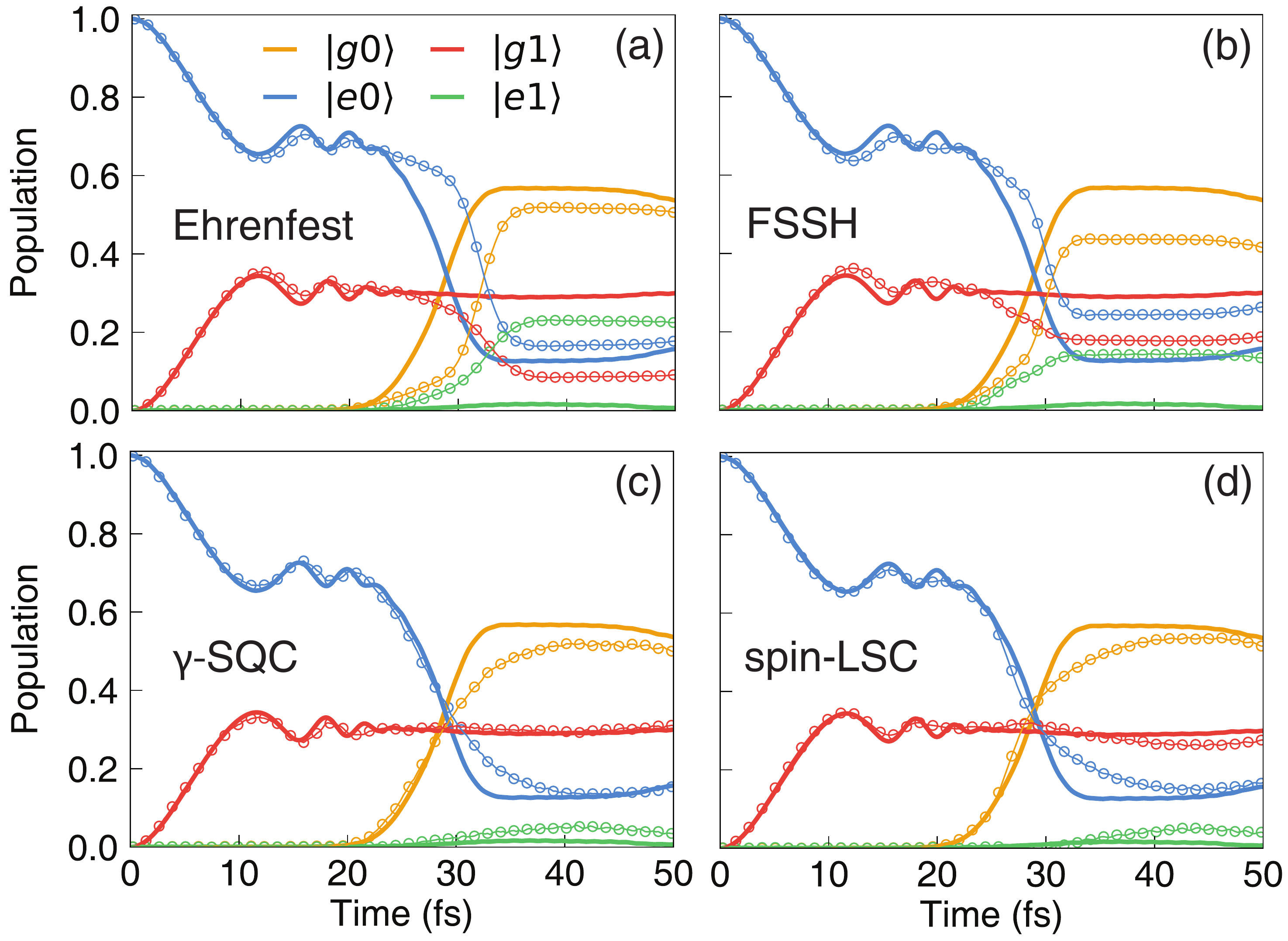}
  \end{minipage}%
   \caption{The population dynamics of the adiabatic-Fock states in Shin-Metiu-cavity model obtained from (a) Ehrenfest dynamics, (b) FSSH approach, (c) $\gamma$-SQC method, and (d) spin-LSC dynamics. The population dynamics are obtained with the approximate methods (open circles) and exact quantum propagation (solid lines). Two electronic states and two Fock states are considered in the simulation, and the light-matter coupling strength $g_\mathrm{c}=0.001$ a.u.}
\label{fig:gc001_results}
\end{figure}

\section{Results}
Fig.~\ref{fig:gc001_results} presents the population dynamics of the adiabatic-Fock states simulated using the approximate methods (open circles), including the MQC approaches (Ehrenfest and FSSH) and the mapping dynamics methods ($\gamma$-SQC and spin-LSC), compared to the numerically exact results (solid lines). The light-matter coupling strength is chosen to be $g_\mathrm{c}=0.001$ a.u. The system is initially prepared in the $|e0\rangle$ state and decays quickly into the $|g1\rangle$ state during the first $\sim12$ fs due to the large light-matter coupling strength from the large transition dipole moment ($\boldsymbol{\mu}_{eg}$) between adiabatic electronic state $|g\rangle$ and $|e\rangle$ (as shown in Fig.~\ref{fig:adiabatic_PES}d). Then, the system starts to oscillate between $|e0\rangle$ and $|g1\rangle$ until about $20$ fs. All of the MQC and mapping dynamics methods can describe the above process reasonably well compared to the exact results. After that, all the dynamics results (including the exact one) show a fast population increase of the $|g0\rangle$ state, which is due to the electronic NAC $d_{eg}$ that directly couples the $|e0\rangle$ state to $|g0\rangle$ state (gold lines). All of the approximate methods can qualitatively describe such a trend, but the MQC methods (panels a-b) are less accurate compared to the mapping-based methods (panels c-d), in terms of the rising of the $|g0\rangle$ population as well as its long time plateau. Moreover, both Ehrenfest and FSSH dynamics predict a significant population transfer from $|g1\rangle$ to $|e1\rangle$ state (panels a-b) as an {\it artifact} that is not shown in the exact dynamics results. In contrast, the $\gamma$-SQC and spin-LSC methods perform much better, where the population transfer process from $|g1\rangle$ to $|e1\rangle$ state is largely suppressed (panels c-d). Overall, the mapping methods outperform the MQC methods in this small light-matter coupling case. It is worth mentioning that the population dynamics results obtained with the FSSH method can be significantly improved if one uses the proper estimator.\cite{Landry2013JCP} We have provided details of this approach and numerical results in Appendix~\ref{apsec:Ehrenfest_FSSH_Details}. Even so, the FSSH method is still facing many challenges from the improper treatment of the quantum coherence and frustrated hop problems, which have been widely discussed.\cite{StockPRL1997,subotnik2016arpc,PrezdoSH,Granucci2007} 

\begin{figure}[ht!]
 \centering
  \begin{minipage}[t]{1.0\linewidth}
     \centering
     \includegraphics[width=\linewidth]{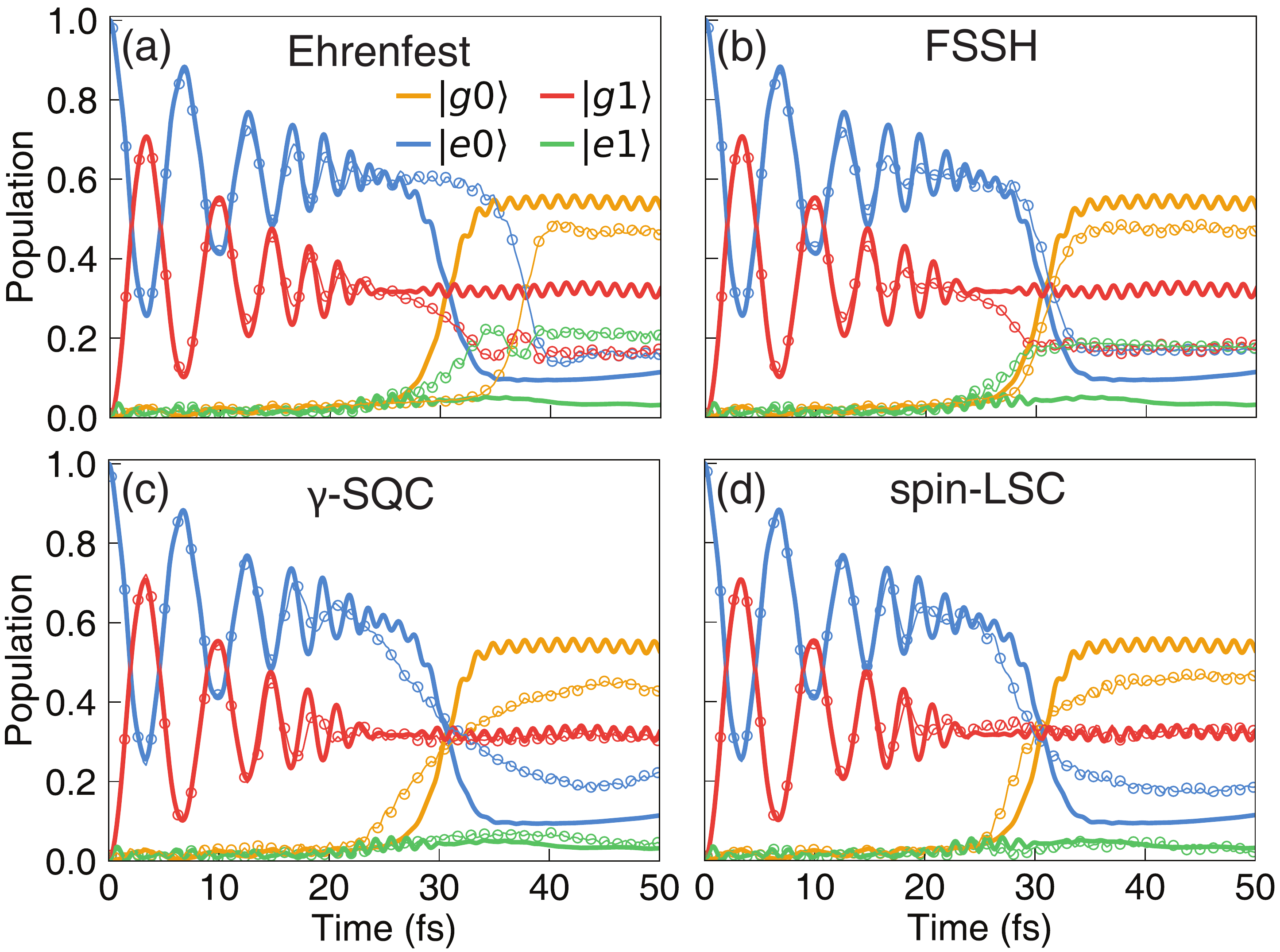}
  \end{minipage}%
   \caption{The population dynamics of the adiabatic-Fock states in Shin-Metiu-cavity model obtained from (a) Ehrenfest dynamics, (b) FSSH approach, (c) $\gamma$-SQC method, and (d) spin-LSC dynamics. The population dynamics are obtained with the approximate methods (open circles) and exact quantum propagation (solid lines). Two electronic states and two Fock states are considered in the simulation, and the light-matter coupling strength $g_\mathrm{c}=0.005$ a.u.}
\label{fig:gc005_results}
\end{figure}

Fig.~\ref{fig:gc005_results} presents the polariton population dynamics with the coupling strength $g_\mathrm{c}=0.005$ a.u. The oscillation between $|e0\rangle$ and $|g1\rangle$ state population appears much earlier and faster compared to the $g_\mathrm{c}=0.001$ results due to the larger light-matter coupling between $|e0\rangle$ and $|g1\rangle$ states (see Eq.~\ref{eq:enp}). Further, the $|g0\rangle$ and $|e1\rangle$ states are also getting populated at an earlier time, due to the permanent dipole $\mu_{gg}$ and $\mu_{ee}$ that couples $|g1\rangle$ state to $|g0\rangle$ state and $|e0\rangle$ state to $|e1\rangle$ state, respectively. Similar to the $g_\mathrm{c}=0.001$ case, all the MQC and mapping dynamics provide a reasonable accuracy for the population dynamics at a short time, while the mapping methods perform much better than the MQC methods at a longer time. In addition, the spin-LSC method outperforms $\gamma$-SQC method in the description of $|g0\rangle$ and $|e0\rangle$ state population after $t=20$ fs, as shown in Fig.~\ref{fig:gc005_results}c-d.

Until now, all of our simulations are restricted in the Hilbert subspace formed by two electronic states ($|g\rangle$ and $|e\rangle$) and two photonic Fock states ($|0\rangle$ and $|1\rangle$). The system could explore a larger Hilbert space due to the increasing light-matter coupling strength. Thus, we systematically check the polariton dynamics using the exact wavepacket dynamics method with a larger number of electronic adiabatic states and Fock states, as shown in Fig. S1 and S2 in the Supplementary Material. The results show that, for the small light-matter coupling strength case ($g_\mathrm{c}=0.001$ a.u.), truncation to the Hilbert subspace formed by two electronic states and two Fock states is enough to give an accurate description of the population dynamics for the SM model studied in this work. However, for the larger coupling strength case ($g_\mathrm{c}=0.005$ a.u.), the polariton dynamics will converge when including four adiabatic electronic states ($|g\rangle$, $|e\rangle$, $|f\rangle$, $|h\rangle$ with energies in ascending order) and four Fock states ($|0\rangle$, $|1\rangle$, $|2\rangle$, $|3\rangle$ with photon number in ascending order). 
\begin{figure}[ht!]
 \centering
  \begin{minipage}[t]{1.0\linewidth}
     \centering
     \includegraphics[width=\linewidth]{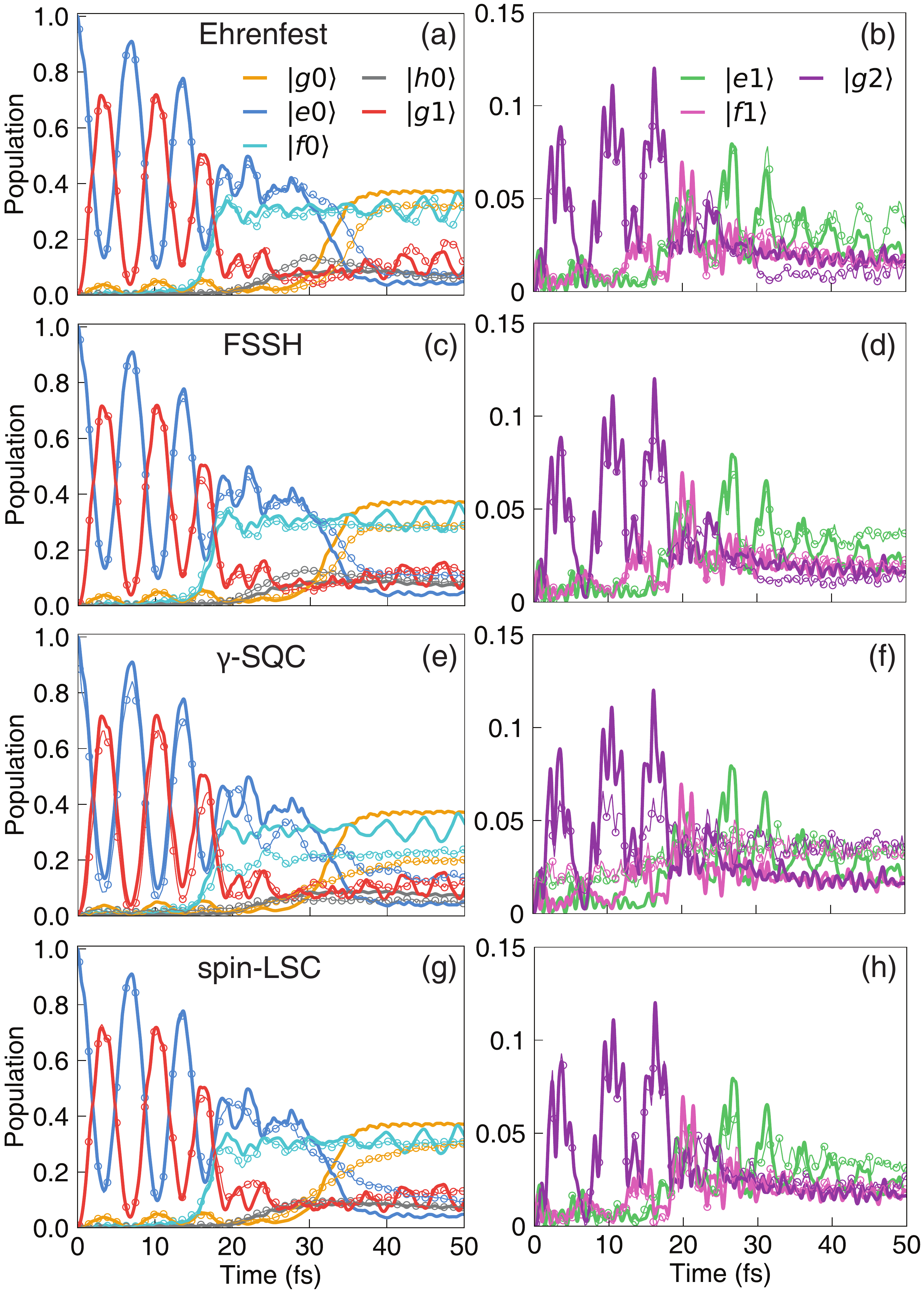}
  \end{minipage}%
   \caption{The population dynamics of the adiabatic-Fock states with (a,b) Ehrenfest dynamics, (c,d) FSSH approach, (e,f) $\gamma$-SQC method, and (g,h) spin-LSC dynamics. The population dynamics are obtained with the approximate methods (open circles) and exact quantum propagation (solid lines). Four adiabatic electronic states ($|g\rangle$, $|e\rangle$, $|f\rangle$, $|h\rangle$ with energies in ascending order) and four Fock states ($|0\rangle$, $|1\rangle$, $|2\rangle$, $|3\rangle$ with photon number in ascending order) are considered in the simulations and the light-matter coupling strength is  $g_\mathrm{c}=0.005$ a.u. Only the adiabatic-Fock states with observable populations of more than $0.01$ are plotted.}
\label{fig:more-states}
\end{figure}

To further test the performance of the mapping methods ($\gamma$-SQC and spin-LSC) as well as the MQC methods (Ehrenfest and FSSH) in such a large Hilbert subspace, which includes sixteen states formed by the tensor product of four electronic states ($|g\rangle$, $|e\rangle$, $|f\rangle$, $|h\rangle$) and four Fock states ($|0\rangle$, $|1\rangle$, $|2\rangle$, $|3\rangle$). Fig.~\ref{fig:more-states} presents the results of using Ehrenfest dynamics (a-b), FSSH (c-d), $\gamma$-SQC (e-f) and spin-LSC (g-h). Besides the adiabatic-Fock states already appear in the four-states subspace ($|g0\rangle$, $|e0\rangle$, $|g1\rangle$ and $|e1\rangle$), we can also see some other states ($|h0\rangle$, $|f0\rangle$, $|g2\rangle$ and $|f1\rangle$) are populated due to the increasing light-matter coupling strength. All of the MQC (Ehrenfest, FSSH) and mapping ($\gamma$-SQC and spin-LSC) dynamics results provide accurate agreement with the exact one in the short time ($< 20$ fs). After that, all of the methods start to generate less accurate results (especially for the $|g0\rangle$ population). Note that both $\gamma$-SQC and spin-LSC perform less accurately compared to the situation in a smaller Hilbert subspace (Fig.~\ref{fig:gc005_results}). This is because both $\gamma$-SQC and spin-LSC are sensitive to the number of states of the system. For $\gamma$-SQC, more states means less trajectory landed in the population action window.\cite{Cotton_JCP_2019_many_states} For spin-LSC, the ZPE correction $\Gamma$ (Eq.~\ref{Gamma}) explicitly depends on the number of states $\mathcal{N}$. This suggests a need for the future development of more accurate dynamics approaches. The current work, nevertheless, paves the way for those future methods to be directly used for simulating on-the-fly polariton quantum dynamics.

\section{Conclusions}
In this work, we generalize the quasi-diabatic (QD) propagation scheme\cite{MandalJCP2018,ZhouJCPL2019,MandalJPCA2019} to simulate the non-adiabatic polariton dynamics in molecule-cavity hybrid systems. The adiabatic-Fock states, which are the tensor product states of the adiabatic electronic states of the molecule and photon Fock states, are used as the {\it locally} well-defined {\it diabatic} states for the dynamics propagation.\cite{ZhouJCPL2019,MandalJPCA2019} These locally well-defined diabatic states allow using any diabatic quantum dynamics methods for dynamics propagation, and the definition of these states will be updated at every nuclear time step. The benefit of using such adiabatic-Fock states is that one can conveniently obtain the electronic adiabatic states energies, the nuclear gradient, the dipole moments and NACs between these states, which are necessary ingredients in molecular cavity QED simulations. 

We use the recently developed non-adiabatic mapping dynamics approaches, $\gamma$-SQC\cite{CottonJCP2019_2} and spin-LSC\cite{richardson2020} to investigate polariton dynamics of a Shin-Metiu model coupled to an optical cavity.\cite{Hoffmann2020,Zhou2022} To benchmark the results of the obtained polariton dynamics, we performed simulations using the Ehrenfest dynamics and the FSSH approaches, as well as the numerically exact polariton wavepacket propagation. The results show that the mapping methods can accurately describe the population dynamics of the molecule-cavity system both at a short time and a longer time when compared to exact dynamics results. In addition, the mapping methods outperform the Ehrenfest and FSSH approaches at a long time dynamics. The numerical results also demonstrate that the performance of the mapping methods ($\gamma$-SQC and spin-LSC) becomes less accurate with an increased number of states in the simulation, indicating the need for future theoretical development.

We envision that the theoretical development in this work will provide the emerging polariton chemistry field with a general theoretical tool that enables direct {\it ab initio} on-the-fly simulations of polariton photochemical processes. We also anticipate that the theoretical developments in this work will enable many recently developed diabatic quantum dynamics approaches to directly simulate polariton quantum dynamics.

\section*{Acknowledgments}
This work was supported by the National Science Foundation CAREER Award under Grant No. CHE-1845747 and by a Cottrell Scholar award (a program by Research Corporation for Science Advancement). Computing resources were provided by the Center for Integrated Research Computing (CIRC) at the University of Rochester. D.H. appreciates valuable discussions with Wanghuai Zhou. We also appreciate valuable suggestions by Prof. Yihan Shao.

\section*{Conflict of Interest}
The authors have no conflicts to disclose.

\section*{Supplementary Material}
See Supplementary Material for additional results of the convergence test for the number of trajectories and the number of adiabatic electronic states and Fock states.

\section*{Availability of Data}	
The data that support the findings of this study are available from the corresponding author upon a reasonable request.

\appendix
\section{Details of the $\gamma$-SQC Approach}\label{apsec:details_of_sqc}	
For practical reasons, the delta functions in Eq.~\ref{eqn:wignersqc} are broadened using well-explored window functions, which can be used to bin the electronic action variables in action-space. The triangle window\cite{CottonJCP2016,CottonJCP2019_2} is expressed as
\begin{equation}\label{EQ:TriangleWindow}
W_j({\boldsymbol \varepsilon}) = w_{1}(\varepsilon_j) \prod_{j'\ne j}^\mathcal{N} w_{0}(\varepsilon_j,\varepsilon_{j'}),
\end{equation}
where the window functions are defined as
\begin{equation}
    w_{1}(\varepsilon) = 
    \begin{cases} 
      (2 - \varepsilon)^{2-\mathcal{N}}, & 1 < \varepsilon < 2 \\
      0, & \mathrm{else}
   \end{cases}
\end{equation}
and 
\begin{equation}
    w_{0}(\varepsilon, \varepsilon') = 
    \begin{cases} 
      1, & \varepsilon' < 2 - \varepsilon \\
      0, & \mathrm{else},
   \end{cases}
\end{equation}
and trajectories are assigned to state $j$ at time $t$ if $\varepsilon_{j}\ge 1$ and $\varepsilon_{j'}<1$ for all $j'\neq j$. The ZPE parameter in the standard SQC method with triangle windowing is $\gamma=1/3$.

The time-dependent population of the state $|j \rangle$ is computed with Eq.~\ref{eqn:wignersqc}. Using the window function estimator, the total population is no longer properly normalized due to the fraction of trajectories that are outside of any window region at any given time.\cite{CottonJCP2013} Thus, the total population must be normalized\cite{CottonJCP2013} with the following procedure
\begin{equation}\label{eq:popNormalize}
{{\rho}_{jj}(t)}/{\sum_{i=1}^N {\rho}_{ii}(t)}\rightarrow {\rho}_{jj}(t).
\end{equation}

In the $\gamma$-SQC approach,\cite{CottonJCP2019_2} it was proposed that the mapping ZPE should be chosen in such a way as to constrain the initial force to be composed purely from the initially occupied state.\cite{CottonJCP2019_2} This new scheme has shown to provide a significant improvement for photo-dissociation problems with coupled Morse potentials\cite{CottonJCP2019_2} and has been combined with the kinematic momentum approach\cite{CottonJCP2017} to carry out on-the-fly simulations of the methaniminium cation.\cite{HuJCTC2021} The basic logic of $\gamma$-SQC is to choose an $\gamma_j$ for each state $|j\rangle$ in every given individual trajectory, such that the initial population is forced to respect the initial electronic excitation focused onto a single excited state. If the initial electronic state is $|i\rangle$, then 
\begin{equation}
     \gamma_j = \varepsilon_j - \delta_{ji},
\end{equation}
or equivalently, 
\begin{equation}\label{EQ:GAMMA}
    \delta_{ji} = \varepsilon_j - \gamma_j,
\end{equation}
where the $\{\varepsilon_j\}$ are uniformly sampled inside the window function (Eq. \ref{EQ:TriangleWindow}), and then the $\gamma_j$ are chosen to satisfy Eq. \ref{EQ:GAMMA}.

These $\gamma_j$ will be explicitly used in the EOMs in Eqs.~\ref{eq:mapeqn}-\ref{eq:force}, and in particular, the nuclear forces are now
\begin{equation}
{\bf F}=-{\frac {1} {2}}\sum_{ij}\nabla V_{ij}({\bf R})\big(p_{i}p_{j}+q_{i}q_{j}-2\gamma_{j}\delta_{ji}\big),
\end{equation}
ensuring the initial forces (at $t=0$) are simply ${\bf F}=-\nabla V_{ii}({\bf R})$. Previously, without any adjustments to $\gamma_j$, the chosen values for $\gamma_j$ were only dependent on the windowing function itself, {\it i.e.}, $\gamma_j = 0.366$ for the square Windows and $\gamma_j = 1/3$ for the triangle windows. With the above $\gamma$-correction method,\cite{CottonJCP2019_2} each individual trajectory will have its own state-specific $\gamma_j$ for state $|j\rangle$ that is completely independent of the choice of window function.

\section{Details of the Adiabatic Electronic Calculation and the Exact Quantum Dynamics Simulations}\label{Exact}
To calculate the electronic properties of the SM model, we use the Sinc DVR basis\cite{Colbert1992} to represent the electronic adiabatic wavefunction and solve Eq.~\ref{HenPsi}. The grid of DVR is uniform with spacing $\Delta x=0.147$ in the range $[-22,22]$ a.u. To test the convergence of grid points, we doubled the number of grid points and the results were identical. The matrix elements of the electronic Hamiltonian  $\hat{H}_\mathrm{el}$ in this grid basis $\{ |r_i\rangle \}$ are given by
\begin{align}\label{hel-grid}
\langle r_i | \hat{H}_\mathrm{el}|  r_j \rangle  &= \langle r_i | \hat{T}_{r} + \hat{V}_\mathrm{eN}(\hat{r},R) + \hat{V}_\mathrm{NN}(R) |r_j \rangle \\
&= \langle r_i | \hat{T}_{r} | r_j \rangle +  [\hat{V}_\mathrm{eN}(r_j,R) + \hat{V}_\mathrm{NN}(R)]\delta_{ij}, \nonumber
\end{align}
where the $\langle r_i | \hat{T}_{r} | r_j \rangle$ is given analytically\cite{Colbert1992} as follows
\begin{align}
\langle r_i | \hat{T}_{r} | r_j \rangle = & \frac{\hbar^{2}}{2m}\cdot \frac{\pi^{2}}{3(\Delta r)^{2}} \Big(1 + \frac{2}{N^{2}}\Big)\delta_{ij} \\
&+ \frac{\hbar^{2}}{2m}\cdot \frac{2 (-1)^{j-i} \pi^{2}} {(\Delta r N \sin(\frac{\pi(j-i)}{N}))^{2}} (1-\delta_{ij}). \nonumber 
\end{align}

Directly diagonalizing the matrix of $\hat{H}_\mathrm{el}$ (in Eq.~\ref{hel-grid}) at a given nuclear (proton) position $R$ in this grid basis gives the accurate adiabatic electronic states
\begin{equation}
|\phi_{\alpha}(R)\rangle=\sum_{i}c^{\alpha}_{i}(R)|r_i\rangle,
\end{equation}
where $c^{\alpha}_{i}(R)$ is the expansion coefficient, which is purely real for the adiabatic electronic states considered here.

A key ingredient for the QD propagation scheme is the overlap integral in Eq.~\ref{eqn:wf_overlap}, Eq.~\ref{eqn:wf_overlap_diag}, and  Eq.~\ref{overlap}, which involve the overlaps between two adiabatic states associated with two different reference geometries. These integrals are conveniently calculated because all of the adiabatic states are represented with the common DVR grids basis as follows
\begin{equation}
\langle\phi_{\alpha}({ R}_{0})|\phi_{\beta}({ R}_{1})\rangle=\sum_{i,j}c^{\alpha}_{i}({R}_{0})\cdot c^{\beta}_{j}({R}_{1}) \langle r_i|r_{j}\rangle.
\end{equation}
Using these bases, the matrix elements for the dipole moment operator (Eq.~\ref{dipole-mat}) are calculated as
\begin{align}\label{dipole-mat-dvr}
{\mu}_{\gamma\nu}({R})&=\sum_{ij} c^{\gamma}_{i}({R})\cdot c^{\nu}_{j}({R}) \langle r_i|(R-\hat{r})|r_{j}\rangle\\
&=\sum_{ij} c^{\gamma}_{i}({R})\cdot c^{\nu}_{j}({R})\cdot (R - r_i)\delta_{ij}.\nonumber
\end{align}

Using $|\phi_{\alpha}({R})\rangle$ and $|\phi_{\beta}({R})\rangle$ in the grid basis, we can directly evaluate the nuclear gradient $\langle \phi_{\lambda}|\nabla \hat{H}_\mathrm{en}| \phi_{\nu} \rangle$ (Eq.~\ref{eqn:HF}) as follows
\begin{equation}\label{grad-eval}
\langle \phi_{\lambda} ({R}) |\nabla \hat{H}_\mathrm{en}({R}) | \phi_{\nu} ({R}) \rangle=\sum_{ij} c^{\lambda}_{i}({R})\cdot c^{\nu}_{j}({R}) \langle r_i|\nabla \hat{H}_\mathrm{en}({R})|r_{j}\rangle,
\end{equation}
where the $\nabla \hat{H}_\mathrm{en}({R})$ is evaluated analytically using the expression of $\hat{H}_\mathrm{en}({R})$ in Eq.~\ref{Hen}. This gives the adiabatic gradient $\nabla E_{\lambda}$ (for $\lambda=\nu$) and derivative coupling (for $\lambda\neq\nu$) as $d_{\lambda\nu}=\langle \phi_{\lambda} |\nabla \hat{H}_\mathrm{en}| \phi_{\nu} \rangle/( E_{\nu} - E_{\lambda})$, as indicated in Eq.~\ref{Hen}. The nuclear gradient $\langle \phi_{\lambda}|\nabla \hat{H}_\mathrm{en}| \phi_{\nu} \rangle$ is also one of the key ingredient in the QD propagation.

Further, the nuclear gradient expression in Eq.~\ref{eq:muderive}) for a polariton system requires the derivative on the dipole matrix (Eq.~\ref{dipole-mat-dvr}). This requires the evaluation of the derivative of the expansion coefficients $\nabla c_{i}^{\gamma}(R)$ in Eq.~\ref{dipole-mat-dvr}. Instead of evaluating these derivatives, as been commonly done in electronic structure calculations,\cite{Zhang2019jcp} here, we evaluate this derivative on dipole numerically as follows
\begin{equation}
\nabla {\mu}_{\gamma\nu}(R) \approx \frac{{\mu}_{\gamma\nu}(R+\Delta R)-{\mu}_{\gamma\nu}(R-\Delta R)} {2\Delta R}.
\end{equation}

To solve the exact quantum dynamics, we represent the total wavefunction of the Hybrid system as $|\Psi_{\xi}\rangle=\sum_{i,k}c^{\xi}_{ik}|\psi_{i}(R_{k})\rangle\otimes|R_{k}\rangle$, where $\{|R_{k}\rangle\}$ is the DVR grid basis for the nucleus, and the $|\psi_{i}(R_{k})\rangle=|\phi_\alpha(R_{k}) \rangle \otimes |n\rangle$ is the adiabatic-Fock basis (Eq.~\ref{adia-fock}), where the electronic adiabatic basis $|\phi_\alpha(R_{k})\rangle$ is obtained by solving the electronic eigenequation (Eq.~\ref{HenPsi}) using the DVR basis for the electronic DOF at the nuclear configuration $R_{k}$. The coefficients for the total wavefunction $c^{\xi}_{ik}$ and the eigenvalue of the total Hamiltonian $\hat{H}$ (Eq.~\ref{total_PF_H}) will be obtained by solving the time-independent Sch\"{o}dinger equation  $\hat{H}|\Psi_{\xi}\rangle=E_{\xi}|\Psi_{\xi} \rangle$. We use the Sinc DVR basis\cite{Colbert1992} for the nuclear DOF and solve the above eigenvalue problem to obtain all the eigenvalues and eigenstates. We use finer grid points for nucleus $\Delta x = 0.016$ in the range $[-8,8]$. To test the convergence of grid points, we doubled the number of grid points and the results were identical. The time evolution dynamics is obtained by unitary evolution $|\Phi(t)\rangle = \sum_{\xi} C_\xi \exp \left(-\frac{i}{\hbar}E_{\xi} t \right) |\Psi_{\xi}\rangle$, where $C_{\xi}$ is the projection of initial total wavefunction onto the $|\Psi_{\xi}\rangle$ as  $C_\xi=\langle \Psi_\xi |\Phi(0)\rangle$, with the initial wavefunction  $|\Phi(0)\rangle$ expressed in  Eq.~\ref{eq:initial-state}. The details of the exact polariton dynamics calculation can also be found in our recent work.\cite{Zhou2022} 

\section{Details of the Ehrenfest and Surface Hopping Simulations}\label{apsec:Ehrenfest_FSSH_Details}
Besides the mapping dynamics methods ($\gamma$-SQC and spin-LSC), we also apply the commonly used Ehrenfest and Tully's FSSH \cite{Tully,tully94jcp} algorithms to run polariton dynamics. Details can be found in our previous work that develops new gradient expressions for QED simulation with the MQC methods.\cite{Zhou2022} Here, we briefly present these approaches for the completeness of this work. 

In the Ehrenfest dynamics, the wavefunction of the quantum subsystem (electronic-photonic DOFs) is written as
\begin{equation}\label{eqn:af_wf}
|\Psi({\bf r};{\bf R}(t))\rangle=\sum_{i}c_{i}(t)|\psi_{i}({\bf R}(t))\rangle,
\end{equation}
where $|\psi_{i}({\bf R}(t))\rangle=|\phi_\alpha({\bf R}(t)) \rangle \otimes |n\rangle$ is the adiabatic-Fock state basis (Eq.~\ref{adia-fock}). The quantum subsystem is described by the time-dependent Sch\"{o}dinger equation (TDSE):
\begin{equation}\label{eqn:tdse}
i\hbar \frac{\partial}{\partial t} |\Psi({\bf r};{\bf R}(t)) \rangle = \hat{V} |\Psi({\bf r};{\bf R}(t)) \rangle.
\end{equation}
The classical subsystem (nuclear DOF) is propagated using the Newton's EOM, where the nuclear force is evaluated from the time-dependent average potential (mean field) 
\begin{equation}\label{eqn:ehrenfest_force}
{\bf F} = -{\bf c}^\dagger [\nabla V] {\bf c},
\end{equation}
where ${\bf c}^\dagger $ is the transpose of the coefficient column vector ${\bf c}$ expressed as follows
\begin{equation}\label{eqn:cvector}
{\bf c}^\dagger = (c_{1}(t),c_{2}(t),\ldots,c_{N}(t)).
\end{equation}
The nuclear gradient matrix is expressed as
\begin{equation}\label{eqn:gH}
[\nabla V]\equiv\nabla [V] -[V][{\bf d}]+[{\bf d}][V],
\end{equation}
where $[V]$ and $[{\bf d}]$ are the matrix of $\hat{V}$ and derivative coupling operator in the adiabatic-Fock state basis, respectively. The full derivation of the gradient can be found in our previous work.\cite{Zhou2022}

In the FSSH dynamics, we expand the time-dependent wave function in the polaritonic basis $|\mathcal{E}_{I}({\bf R}(t)) \rangle$ (see definition in Eq.~\ref{polariton})
\begin{equation}\label{eqn:pl_wf}
|\Psi({\bf r};{\bf R}(t))\rangle = \sum_{I}c_I(t) |\mathcal{E}_{I}({\bf R}(t)) \rangle,
\end{equation}
where $c_I$ is the expansion coefficient, which will be used to compute the fewest switching probability. Here, the nuclear force comes from {\it only one} specific polariton state $|\mathcal{E}_{I}({\bf R}(t)) \rangle$ (eigenstate of $\hat{V}$, see Eq.~\ref{polariton}) as follows
\begin{equation}
{\bf F}=-\nabla \mathcal{E}_{I},
\end{equation}
where $\mathcal{E}_{I}$ is the energy of the {\it active} adiabatic polariton state, and $I$ is the active state index, which will be determined at every nuclear propagation step. The nuclear gradient is
\begin{equation} \label{grad_pola}
\nabla \mathcal{E}_{I} = \sum_{jk} \langle \mathcal{E}_I | \psi_j \rangle \langle \psi_j | \nabla V | \psi_k \rangle \langle \psi_k | \mathcal{E}_I \rangle,
\end{equation}
where the matrix element of $\nabla V$ is expressed in Eq.~\ref{eqn:gH}. The details of the nuclear gradient in the polariton basis can be found in our previous work.\cite{Zhou2022}

According to the ``fewest switches" algorithm,\cite{Tully} the probability of switching (probability flux) from the active polariton state $|\mathcal{E}_I\rangle$ to {\it any other} polariton state $|\mathcal{E}_J\rangle$ during the time interval between $t$ and $t+\delta t$ is
\begin{equation}\label{eq:hop_probability}
f_{IJ}=-\frac{2 \text{Re}[(\rho^\mathrm{pl}_{JI})^{*}\cdot{\bf{\dot{R}}}\cdot {\bf d}_{JI}({\bf R})]\,\delta t}{\rho^\mathrm{pl}_{II}},
\end{equation}
where $\rho^\mathrm{pl}_{IJ}(t)$ is the reduced density matrix element in the polariton basis expressed as follows
\begin{equation}\label{density}
\rho^\mathrm{pl}_{IJ}(t)=c_{I}(t) c_{J}^{*}(t).
\end{equation}
Since the probability should be positive definite, one sets\cite{TullyNAMD} $f_{IJ}$ to 0 if $f_{IJ}<0$. The non-adiabatic transition, {\it i.e.} stochastic switch from the currently occupied state $|\mathcal{E}_I\rangle$ to another state $|\mathcal{E}_K\rangle$, occurs if the following condition is satisfied
\begin{equation}
\sum_{J=1}^{K}{f_{IJ}} < \zeta < \sum_{J=1}^{K+1}{f_{IJ}},
\end{equation}
where $\zeta$ is a uniform randomly generated number between 0 and 1 at each nuclear time step. If the transition is accepted, the active state is set to the new adiabatic state $|\mathcal{E}_K\rangle$, while the velocities of the nuclei are rescaled along the direction of the NAC vector ${\bf d}_{IK}({\bf R})$ in order to conserve the total energy.\cite{tully94jcp} More details of performing FSSH simulation of the polariton dynamics can be found in our previous work.\cite{Zhou2022} 

For the Ehrenfest dynamics and FSSH approach, we use the fourth-order Runge-Kutta method to integrate the TDSE and the velocity Verlet algorithm to integrate Newton's EOM. The time step for the nuclear motion is $0.1$ a.u. and the sub-step for solving the TDSE of the electronic-photonic subsystem is $0.001$ a.u.. We have carefully checked that the total energy is well conserved for all the trajectories. The initial condition is described by Eq.~\ref{eq:initial-state}, where the nuclear DOF is sampled from the corresponding Wigner density described in Eq.~\ref{wigner}. For Ehrenfest dynamics, the initial coefficients $c_i(0)$ for the state $|\psi_{i}\rangle=|e0\rangle$ is set to be one, and the rest of the coefficients are set to be zero. These $\{c_j(0)\}$ can be unitary transformed into the coefficients $\{c_I(0)\}$ for each nuclear initial configuration described in Eq.~\ref{wigner}. For the FSSH simulation, one needs to choose an {\it initial active state}, and the initial state of the quantum subsystem $|e(R),0\rangle$ is not one of the eigenstates $|\mathcal{E}_{I}(R)\rangle$. We thus follow the previous work\cite{Hazra2010JPCB,Landry2013JCP} and use the  Monte-Carlo scheme to randomly choose the initial active state $|\mathcal{E}_{I}(R)\rangle$ for each trajectory, based on the {\it magnitude} of $|\langle \mathcal{E}_{I}(R)|e(R),0\rangle|^2$ for a given trajectory that has the nuclear configuration at $R$ sampled from the Wigner density (Eq.~\ref{wigner}).

When computing the population dynamics in a representation that is {\it not} the adiabatic states of $\hat{V}$, there is no unique way to calculate them in the FSSH approach.\cite{Landry2013JCP} In the main text, we present the populations of the adiabatic-Fock states using the expansion coefficients $c_{I}(t)$ in Eq.~\ref{eqn:pl_wf}. There are, of course, alternative ways to compute populations of these adiabatic-Fock states.\cite{Landry2013JCP} Below, we explore the alternative ways to compute them. 

For clarity, we denote the reduced density matrix in the adiabatic-Fock basis as $\rho^\mathrm{af}_{ij}(t)$, and  $\rho^\mathrm{pl}_{IJ}(t)$ is the reduced density matrix in the polariton basis (expressed in Eq.~\ref{density}). 
To get the adiabatic-Fock state population of the $|\psi_i\rangle$ state $\rho^\mathrm{af}_{ii}$ from the FSSH simulation, the most straightforward way (as the results presented in the main text) is through following unitary transformation   
\begin{equation}\label{eqn:pl_to_af}
[\rho^\mathrm{af}({\bf R}_{l}(t))]=\mathbf{U}[\rho^\mathrm{pl}({\bf R}_{l}(t))]\mathbf{U}^{\dagger},
\end{equation}
where $[\rho^\mathrm{pl}{\bf R}_{l}(t))]$ is the reduced density matrix in the polariton basis along a given nuclear trajectory ${\bf R}_{l}(t)$, with $l$ as the label of the trajectory, and the elements as
\begin{equation}\label{method2}
\rho^\mathrm{pl}_{IJ}({\bf R}_{l}(t))=c_{I}(t) c_{J}^{*}(t).
\end{equation}
Further, $\mathbf{U}({\bf R}_{l}(t))$ is the matrix that diagonalize the matrix $[V({\mathbf R}_{l}(t))]$ as shown in Eq.~\ref{eqn:diag}, along the same trajectory ${\bf R}_{l}(t)$. The adiabatic-Fock state population is then obtained from trajectory average as follows
\begin{equation}\label{eqn:average_pop}
P_i(t)=\frac{1}{N}\sum_{l}^{N}\left[\mathbf{U}[\rho^\mathrm{pl}({\bf R}_{l}(t))]\mathbf{U}^{\dagger}\right]_{ii},
\end{equation}
where $N$ is the total number of the trajectories. This is the estimator used in the FSSH calculation presented in the main text. 

For FSSH, there are two other commonly used choices\cite{Landry2013JCP} to calculate the populations that are not in an adiabatic representation. These methods vary on how to calculate the polaritonic state density matrix $[\rho^\mathrm{pl}({\bf R}_{l}(t))]$. The first choice is based on the active state index and ignores the polaritonic state coefficients $\{c_{I}(t)\}$ and the density matrix elements are written as
\begin{equation}\label{method1}
    \rho^\mathrm{pl}_{IJ}({\bf R}_{l}(t)) = 
    \begin{cases} 
      \delta_{IK}, & I=J \\
      0, & I \neq J,
   \end{cases}
\end{equation}
where $K$ is the active polaritonic state. This method explicitly assumes that the off-diagonal elements of the polaritonic state density matrix are zero, which is often not a good one. The adiabatic-Fock state population is then obtained from the same transformation described in Eq.~\ref{eqn:pl_to_af}, and the ensemble average over all trajectories is computed as described in Eq.~\ref{eqn:average_pop}.

The other choice\cite{Landry2013JCP} (motivated by the mixed quantum-classical Liouville approach) is to calculate the diagonal elements of $\rho^\mathrm{pl}$ using the active state index, and calculate the off-diagonal elements using the polaritonic state expansion coefficients $\{c_{I}(t)\}$
\begin{equation}\label{method3}
    \rho^\mathrm{pl}_{IJ}({\bf R}_{l}(t)) = 
    \begin{cases} 
      \delta_{IK}, & I=J \\
      c_{I} c_{J}^{*}, & I \neq J,
   \end{cases}
\end{equation}
where $K$ is the active index. The adiabatic-Fock state population is then obtained from the same transformation described in Eq.~\ref{eqn:pl_to_af}, and the ensemble average over all trajectories as described in Eq.~\ref{eqn:average_pop}.

\begin{figure}[ht!]
 \centering
  \begin{minipage}[t]{1.0\linewidth}
     \centering
     \includegraphics[width=\linewidth]{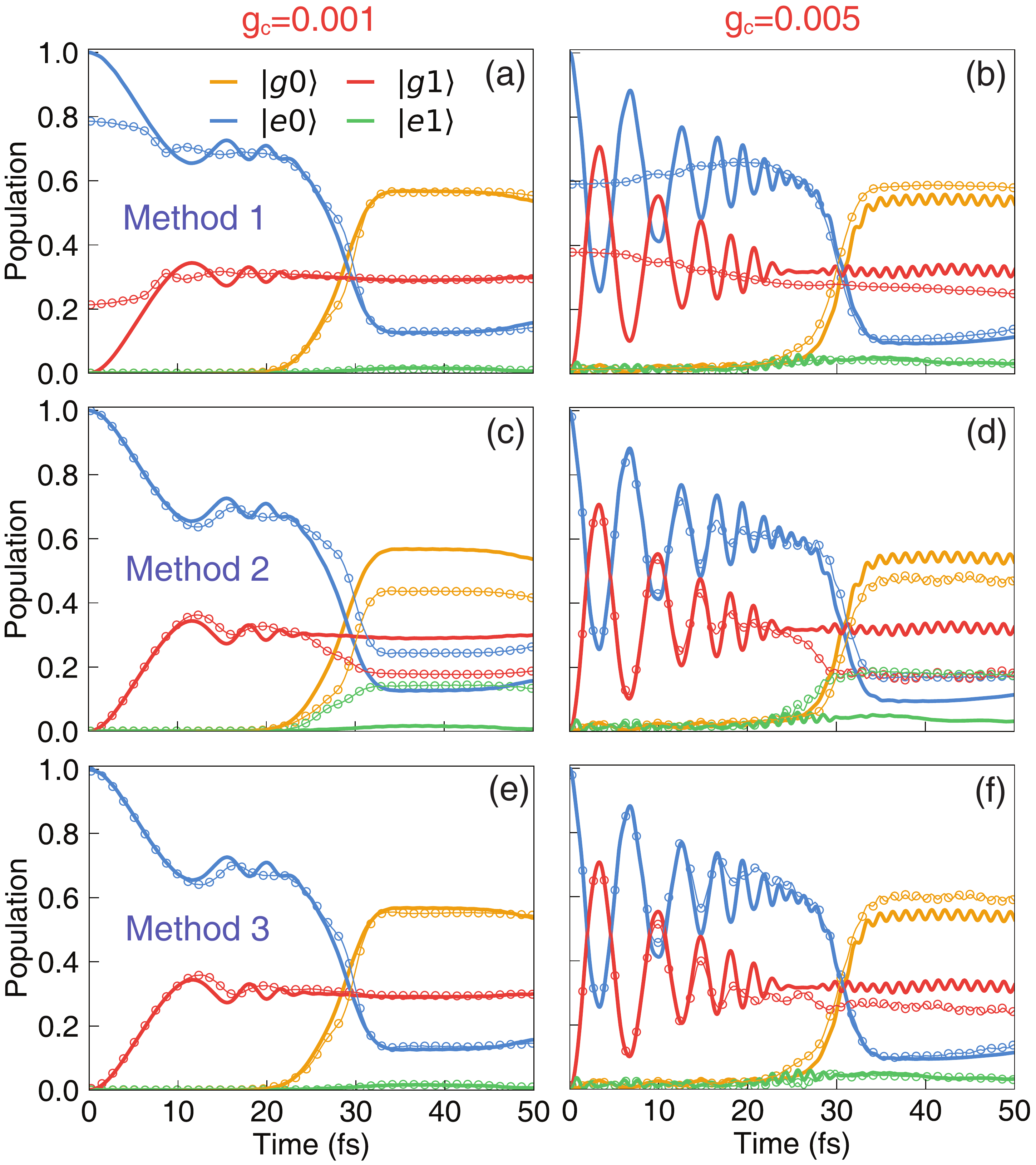}
  \end{minipage}%
   \caption{Population dynamics of the adiabatic-Fock states obtained from the FSSH method (open circles) and the exact quantum dynamics propagation (solid lines), using different population estimators.  (a,b) Method 1; (c,d) Method 2; (e,f) Method 3. The results obtained using method 2 are presented in the main text of this work. The light-matter coupling strength is set to be  (a,c,e) $g_\mathrm{c}=0.001$ a.u. and (b,d,f) $g_\mathrm{c}=0.005$ a.u.}
\label{fig:diff_estimator}
\end{figure}

Following the same notation as used in  Ref.~\citenum{Landry2013JCP}, we refer to the choice in Eq.~\ref{method1} as Method 1, the choice in Eq.~\ref{method2} as Method 2 (same as the FSSH results presented in the main text), and the choice in Eq.~\ref{method3} as Method 3. The FSSH dynamics results based on these three methods are presented in Fig.~\ref{fig:diff_estimator}. We can see that Method 3 performs much better than Method 1 and Method 2, consisting with the conclusion in Ref.~\citenum{Landry2013JCP}.

\begin{thebibliography}{93}%
\makeatletter
\providecommand \@ifxundefined [1]{%
 \@ifx{#1\undefined}
}%
\providecommand \@ifnum [1]{%
 \ifnum #1\expandafter \@firstoftwo
 \else \expandafter \@secondoftwo
 \fi
}%
\providecommand \@ifx [1]{%
 \ifx #1\expandafter \@firstoftwo
 \else \expandafter \@secondoftwo
 \fi
}%
\providecommand \natexlab [1]{#1}%
\providecommand \enquote  [1]{``#1''}%
\providecommand \bibnamefont  [1]{#1}%
\providecommand \bibfnamefont [1]{#1}%
\providecommand \citenamefont [1]{#1}%
\providecommand \href@noop [0]{\@secondoftwo}%
\providecommand \href [0]{\begingroup \@sanitize@url \@href}%
\providecommand \@href[1]{\@@startlink{#1}\@@href}%
\providecommand \@@href[1]{\endgroup#1\@@endlink}%
\providecommand \@sanitize@url [0]{\catcode `\\12\catcode `\$12\catcode
  `\&12\catcode `\#12\catcode `\^12\catcode `\_12\catcode `\%12\relax}%
\providecommand \@@startlink[1]{}%
\providecommand \@@endlink[0]{}%
\providecommand \url  [0]{\begingroup\@sanitize@url \@url }%
\providecommand \@url [1]{\endgroup\@href {#1}{\urlprefix }}%
\providecommand \urlprefix  [0]{URL }%
\providecommand \Eprint [0]{\href }%
\providecommand \doibase [0]{http://dx.doi.org/}%
\providecommand \selectlanguage [0]{\@gobble}%
\providecommand \bibinfo  [0]{\@secondoftwo}%
\providecommand \bibfield  [0]{\@secondoftwo}%
\providecommand \translation [1]{[#1]}%
\providecommand \BibitemOpen [0]{}%
\providecommand \bibitemStop [0]{}%
\providecommand \bibitemNoStop [0]{.\EOS\space}%
\providecommand \EOS [0]{\spacefactor3000\relax}%
\providecommand \BibitemShut  [1]{\csname bibitem#1\endcsname}%
\let\auto@bib@innerbib\@empty
\bibitem [{\citenamefont {Ebbesen}(2016)}]{Ebbesen2016ACR}%
  \BibitemOpen
  \bibfield  {author} {\bibinfo {author} {\bibfnamefont {T.~W.}\ \bibnamefont
  {Ebbesen}},\ }\href {\doibase 10.1021/acs.accounts.6b00295} {\bibfield
  {journal} {\bibinfo  {journal} {Acc. Chem. Res.}\ }\textbf {\bibinfo {volume}
  {49}},\ \bibinfo {pages} {2403} (\bibinfo {year} {2016})}\BibitemShut
  {NoStop}%
\bibitem [{\citenamefont {Kowalewski}\ and\ \citenamefont
  {Mukamel}(2017)}]{kowalewski_manipulating_2017}%
  \BibitemOpen
  \bibfield  {author} {\bibinfo {author} {\bibfnamefont {M.}~\bibnamefont
  {Kowalewski}}\ and\ \bibinfo {author} {\bibfnamefont {S.}~\bibnamefont
  {Mukamel}},\ }\href@noop {} {\bibfield  {journal} {\bibinfo  {journal} {Proc.
  Natl. Acad. Sci. U.S.A.}\ }\textbf {\bibinfo {volume} {114}},\ \bibinfo
  {pages} {3278} (\bibinfo {year} {2017})}\BibitemShut {NoStop}%
\bibitem [{\citenamefont {Flick}\ \emph
  {et~al.}(2017{\natexlab{a}})\citenamefont {Flick}, \citenamefont
  {Ruggenthaler}, \citenamefont {Appel},\ and\ \citenamefont
  {Rubio}}]{Flick2017pnas}%
  \BibitemOpen
  \bibfield  {author} {\bibinfo {author} {\bibfnamefont {J.}~\bibnamefont
  {Flick}}, \bibinfo {author} {\bibfnamefont {M.}~\bibnamefont {Ruggenthaler}},
  \bibinfo {author} {\bibfnamefont {H.}~\bibnamefont {Appel}}, \ and\ \bibinfo
  {author} {\bibfnamefont {A.}~\bibnamefont {Rubio}},\ }\href@noop {}
  {\bibfield  {journal} {\bibinfo  {journal} {Proc. Natl Acad. Sci. USA}\
  }\textbf {\bibinfo {volume} {114}},\ \bibinfo {pages} {3026} (\bibinfo {year}
  {2017}{\natexlab{a}})}\BibitemShut {NoStop}%
\bibitem [{\citenamefont {Ribeiro}\ \emph {et~al.}(2018)\citenamefont
  {Ribeiro}, \citenamefont {Mart\'inez-Mart\'inez}, \citenamefont {Du},
  \citenamefont {Camos-Gonzalez-Angulo},\ and\ \citenamefont
  {Yuen-Zhou}}]{Ribeiro2018}%
  \BibitemOpen
  \bibfield  {author} {\bibinfo {author} {\bibfnamefont {R.~F.}\ \bibnamefont
  {Ribeiro}}, \bibinfo {author} {\bibfnamefont {L.~A.}\ \bibnamefont
  {Mart\'inez-Mart\'inez}}, \bibinfo {author} {\bibfnamefont {M.}~\bibnamefont
  {Du}}, \bibinfo {author} {\bibfnamefont {J.}~\bibnamefont
  {Camos-Gonzalez-Angulo}}, \ and\ \bibinfo {author} {\bibfnamefont
  {J.}~\bibnamefont {Yuen-Zhou}},\ }\href@noop {} {\bibfield  {journal}
  {\bibinfo  {journal} {Chem. Sci.}\ }\textbf {\bibinfo {volume} {9}},\
  \bibinfo {pages} {6325} (\bibinfo {year} {2018})}\BibitemShut {NoStop}%
\bibitem [{\citenamefont {Feist}\ \emph {et~al.}(2018)\citenamefont {Feist},
  \citenamefont {Galego},\ and\ \citenamefont {Garcia-Vidal}}]{Feist2018}%
  \BibitemOpen
  \bibfield  {author} {\bibinfo {author} {\bibfnamefont {J.}~\bibnamefont
  {Feist}}, \bibinfo {author} {\bibfnamefont {J.}~\bibnamefont {Galego}}, \
  and\ \bibinfo {author} {\bibfnamefont {F.~J.}\ \bibnamefont {Garcia-Vidal}},\
  }\href@noop {} {\bibfield  {journal} {\bibinfo  {journal} {ACS Photonics}\
  }\textbf {\bibinfo {volume} {5}},\ \bibinfo {pages} {205} (\bibinfo {year}
  {2018})}\BibitemShut {NoStop}%
\bibitem [{\citenamefont {Mandal}\ and\ \citenamefont
  {Huo}(2019)}]{Mandal2019JPCL}%
  \BibitemOpen
  \bibfield  {author} {\bibinfo {author} {\bibfnamefont {A.}~\bibnamefont
  {Mandal}}\ and\ \bibinfo {author} {\bibfnamefont {P.}~\bibnamefont {Huo}},\
  }\href@noop {} {\bibfield  {journal} {\bibinfo  {journal} {J. Phys. Chem.
  Lett.}\ }\textbf {\bibinfo {volume} {10}},\ \bibinfo {pages} {5519} (\bibinfo
  {year} {2019})}\BibitemShut {NoStop}%
\bibitem [{\citenamefont {Hutchison}\ \emph {et~al.}(2012)\citenamefont
  {Hutchison}, \citenamefont {Schwartz}, \citenamefont {Genet}, \citenamefont
  {Devaux},\ and\ \citenamefont {Ebbesen}}]{Hutchison2012ACIE}%
  \BibitemOpen
  \bibfield  {author} {\bibinfo {author} {\bibfnamefont {J.~A.}\ \bibnamefont
  {Hutchison}}, \bibinfo {author} {\bibfnamefont {T.}~\bibnamefont {Schwartz}},
  \bibinfo {author} {\bibfnamefont {C.}~\bibnamefont {Genet}}, \bibinfo
  {author} {\bibfnamefont {E.}~\bibnamefont {Devaux}}, \ and\ \bibinfo {author}
  {\bibfnamefont {T.~W.}\ \bibnamefont {Ebbesen}},\ }\href {\doibase
  10.1002/anie.201107033} {\bibfield  {journal} {\bibinfo  {journal} {Angew.
  Chem. Int. Ed.}\ }\textbf {\bibinfo {volume} {51}},\ \bibinfo {pages} {1592}
  (\bibinfo {year} {2012})}\BibitemShut {NoStop}%
\bibitem [{\citenamefont {Thomas}\ \emph {et~al.}(2019)\citenamefont {Thomas},
  \citenamefont {Lethuillier-Karl}, \citenamefont {Nagarajan}, \citenamefont
  {Vergauwe}, \citenamefont {George}, \citenamefont {Chervy}, \citenamefont
  {Shalabney}, \citenamefont {Devaux}, \citenamefont {Genet}, \citenamefont
  {Moran},\ and\ \citenamefont {Ebbesen}}]{Thomas2019S}%
  \BibitemOpen
  \bibfield  {author} {\bibinfo {author} {\bibfnamefont {A.}~\bibnamefont
  {Thomas}}, \bibinfo {author} {\bibfnamefont {L.}~\bibnamefont
  {Lethuillier-Karl}}, \bibinfo {author} {\bibfnamefont {K.}~\bibnamefont
  {Nagarajan}}, \bibinfo {author} {\bibfnamefont {R.~M.~A.}\ \bibnamefont
  {Vergauwe}}, \bibinfo {author} {\bibfnamefont {J.}~\bibnamefont {George}},
  \bibinfo {author} {\bibfnamefont {T.}~\bibnamefont {Chervy}}, \bibinfo
  {author} {\bibfnamefont {A.}~\bibnamefont {Shalabney}}, \bibinfo {author}
  {\bibfnamefont {E.}~\bibnamefont {Devaux}}, \bibinfo {author} {\bibfnamefont
  {C.}~\bibnamefont {Genet}}, \bibinfo {author} {\bibfnamefont
  {J.}~\bibnamefont {Moran}}, \ and\ \bibinfo {author} {\bibfnamefont {T.~W.}\
  \bibnamefont {Ebbesen}},\ }\href {\doibase 10.1126/science.aau7742}
  {\bibfield  {journal} {\bibinfo  {journal} {Science}\ }\textbf {\bibinfo
  {volume} {363}},\ \bibinfo {pages} {615} (\bibinfo {year}
  {2019})}\BibitemShut {NoStop}%
\bibitem [{\citenamefont {Mandal}\ \emph
  {et~al.}(2020{\natexlab{a}})\citenamefont {Mandal}, \citenamefont {Krauss},\
  and\ \citenamefont {Huo}}]{Mandal2020JPCB}%
  \BibitemOpen
  \bibfield  {author} {\bibinfo {author} {\bibfnamefont {A.}~\bibnamefont
  {Mandal}}, \bibinfo {author} {\bibfnamefont {T.~D.}\ \bibnamefont {Krauss}},
  \ and\ \bibinfo {author} {\bibfnamefont {P.}~\bibnamefont {Huo}},\ }\href
  {\doibase 10.1021/acs.jpcb.0c03227} {\bibfield  {journal} {\bibinfo
  {journal} {J. Phys. Chem. B}\ }\textbf {\bibinfo {volume} {124}},\ \bibinfo
  {pages} {6321} (\bibinfo {year} {2020}{\natexlab{a}})}\BibitemShut {NoStop}%
\bibitem [{\citenamefont {Luk}\ \emph {et~al.}(2017)\citenamefont {Luk},
  \citenamefont {Feist}, \citenamefont {Toppari},\ and\ \citenamefont
  {Groenhof}}]{Luk2017jctc}%
  \BibitemOpen
  \bibfield  {author} {\bibinfo {author} {\bibfnamefont {H.~L.}\ \bibnamefont
  {Luk}}, \bibinfo {author} {\bibfnamefont {J.}~\bibnamefont {Feist}}, \bibinfo
  {author} {\bibfnamefont {J.~J.}\ \bibnamefont {Toppari}}, \ and\ \bibinfo
  {author} {\bibfnamefont {G.}~\bibnamefont {Groenhof}},\ }\href@noop {}
  {\bibfield  {journal} {\bibinfo  {journal} {J. Chem. Theory Comput.}\
  }\textbf {\bibinfo {volume} {13}},\ \bibinfo {pages} {4324} (\bibinfo {year}
  {2017})}\BibitemShut {NoStop}%
\bibitem [{\citenamefont {Groenhof}\ and\ \citenamefont
  {Toppari}(2018)}]{Groenhof2018jpcl}%
  \BibitemOpen
  \bibfield  {author} {\bibinfo {author} {\bibfnamefont {G.}~\bibnamefont
  {Groenhof}}\ and\ \bibinfo {author} {\bibfnamefont {J.~J.}\ \bibnamefont
  {Toppari}},\ }\href@noop {} {\bibfield  {journal} {\bibinfo  {journal} {J.
  Phys. Chem. Lett.}\ }\textbf {\bibinfo {volume} {9}},\ \bibinfo {pages}
  {4848} (\bibinfo {year} {2018})}\BibitemShut {NoStop}%
\bibitem [{\citenamefont {Groenhof}\ \emph {et~al.}(2019)\citenamefont
  {Groenhof}, \citenamefont {Climent}, \citenamefont {Feist}, \citenamefont
  {Morozov},\ and\ \citenamefont {Toppari}}]{Groenhof2019jpcl}%
  \BibitemOpen
  \bibfield  {author} {\bibinfo {author} {\bibfnamefont {G.}~\bibnamefont
  {Groenhof}}, \bibinfo {author} {\bibfnamefont {C.}~\bibnamefont {Climent}},
  \bibinfo {author} {\bibfnamefont {J.}~\bibnamefont {Feist}}, \bibinfo
  {author} {\bibfnamefont {D.}~\bibnamefont {Morozov}}, \ and\ \bibinfo
  {author} {\bibfnamefont {J.~J.}\ \bibnamefont {Toppari}},\ }\href@noop {}
  {\bibfield  {journal} {\bibinfo  {journal} {J. Phys. Chem. Lett.}\ }\textbf
  {\bibinfo {volume} {10}},\ \bibinfo {pages} {5476} (\bibinfo {year}
  {2019})}\BibitemShut {NoStop}%
\bibitem [{\citenamefont {Tichauer}\ \emph {et~al.}(2021)\citenamefont
  {Tichauer}, \citenamefont {Feist},\ and\ \citenamefont
  {Groenhof}}]{Groenhof2021JCP}%
  \BibitemOpen
  \bibfield  {author} {\bibinfo {author} {\bibfnamefont {R.~H.}\ \bibnamefont
  {Tichauer}}, \bibinfo {author} {\bibfnamefont {J.}~\bibnamefont {Feist}}, \
  and\ \bibinfo {author} {\bibfnamefont {G.}~\bibnamefont {Groenhof}},\
  }\href@noop {} {\bibfield  {journal} {\bibinfo  {journal} {J. Chem. Phys.}\
  }\textbf {\bibinfo {volume} {154}},\ \bibinfo {pages} {104112} (\bibinfo
  {year} {2021})}\BibitemShut {NoStop}%
\bibitem [{\citenamefont {Tully}(2012{\natexlab{a}})}]{Tully2012jcp}%
  \BibitemOpen
  \bibfield  {author} {\bibinfo {author} {\bibfnamefont {J.~C.}\ \bibnamefont
  {Tully}},\ }\href@noop {} {\bibfield  {journal} {\bibinfo  {journal} {J.
  Chem. Phys.}\ }\textbf {\bibinfo {volume} {137}},\ \bibinfo {pages} {22A301}
  (\bibinfo {year} {2012}{\natexlab{a}})}\BibitemShut {NoStop}%
\bibitem [{\citenamefont {Barbatti}(2011)}]{BarbattiSH}%
  \BibitemOpen
  \bibfield  {author} {\bibinfo {author} {\bibfnamefont {M.}~\bibnamefont
  {Barbatti}},\ }\href@noop {} {\bibfield  {journal} {\bibinfo  {journal}
  {Wiley Int. Rev. Comp. Mol. Sci.}\ }\textbf {\bibinfo {volume} {1}},\
  \bibinfo {pages} {620} (\bibinfo {year} {2011})}\BibitemShut {NoStop}%
\bibitem [{\citenamefont {Mai}\ \emph {et~al.}(2015)\citenamefont {Mai},
  \citenamefont {Marquetand},\ and\ \citenamefont {Gonzalez}}]{Mai2015}%
  \BibitemOpen
  \bibfield  {author} {\bibinfo {author} {\bibfnamefont {S.}~\bibnamefont
  {Mai}}, \bibinfo {author} {\bibfnamefont {P.}~\bibnamefont {Marquetand}}, \
  and\ \bibinfo {author} {\bibfnamefont {L.}~\bibnamefont {Gonzalez}},\
  }\href@noop {} {\bibfield  {journal} {\bibinfo  {journal} {Int. J. Quantum
  Chem.}\ }\textbf {\bibinfo {volume} {115}},\ \bibinfo {pages} {1215}
  (\bibinfo {year} {2015})}\BibitemShut {NoStop}%
\bibitem [{\citenamefont {Tully}(1990)}]{Tully}%
  \BibitemOpen
  \bibfield  {author} {\bibinfo {author} {\bibfnamefont {J.~C.}\ \bibnamefont
  {Tully}},\ }\href@noop {} {\bibfield  {journal} {\bibinfo  {journal} {J.
  Chem. Phys.}\ }\textbf {\bibinfo {volume} {93}},\ \bibinfo {pages} {1061}
  (\bibinfo {year} {1990})}\BibitemShut {NoStop}%
\bibitem [{\citenamefont {Hammes-Schiffer}\ and\ \citenamefont
  {Tully}(1994)}]{tully94jcp}%
  \BibitemOpen
  \bibfield  {author} {\bibinfo {author} {\bibfnamefont {S.}~\bibnamefont
  {Hammes-Schiffer}}\ and\ \bibinfo {author} {\bibfnamefont {J.~C.}\
  \bibnamefont {Tully}},\ }\href@noop {} {\bibfield  {journal} {\bibinfo
  {journal} {J. Chem. Phys.}\ }\textbf {\bibinfo {volume} {12}},\ \bibinfo
  {pages} {4657} (\bibinfo {year} {1994})}\BibitemShut {NoStop}%
\bibitem [{\citenamefont {Fregoni}\ \emph {et~al.}(2018)\citenamefont
  {Fregoni}, \citenamefont {Granucci}, \citenamefont {Coccia}, \citenamefont
  {Persico},\ and\ \citenamefont {Corni}}]{Fregoni2018}%
  \BibitemOpen
  \bibfield  {author} {\bibinfo {author} {\bibfnamefont {J.}~\bibnamefont
  {Fregoni}}, \bibinfo {author} {\bibfnamefont {G.}~\bibnamefont {Granucci}},
  \bibinfo {author} {\bibfnamefont {E.}~\bibnamefont {Coccia}}, \bibinfo
  {author} {\bibfnamefont {M.}~\bibnamefont {Persico}}, \ and\ \bibinfo
  {author} {\bibfnamefont {S.}~\bibnamefont {Corni}},\ }\href@noop {}
  {\bibfield  {journal} {\bibinfo  {journal} {Nat. Commun.}\ }\textbf {\bibinfo
  {volume} {9}},\ \bibinfo {pages} {4688} (\bibinfo {year} {2018})}\BibitemShut
  {NoStop}%
\bibitem [{\citenamefont {Fregoni}\ \emph
  {et~al.}(2020{\natexlab{a}})\citenamefont {Fregoni}, \citenamefont {Corni},
  \citenamefont {Persico},\ and\ \citenamefont {Granucci}}]{Fregoni2020}%
  \BibitemOpen
  \bibfield  {author} {\bibinfo {author} {\bibfnamefont {J.}~\bibnamefont
  {Fregoni}}, \bibinfo {author} {\bibfnamefont {S.}~\bibnamefont {Corni}},
  \bibinfo {author} {\bibfnamefont {M.}~\bibnamefont {Persico}}, \ and\
  \bibinfo {author} {\bibfnamefont {G.}~\bibnamefont {Granucci}},\ }\href@noop
  {} {\bibfield  {journal} {\bibinfo  {journal} {J. Comput. Chem.}\ }\textbf
  {\bibinfo {volume} {41}},\ \bibinfo {pages} {2033} (\bibinfo {year}
  {2020}{\natexlab{a}})}\BibitemShut {NoStop}%
\bibitem [{\citenamefont {Fregoni}\ \emph
  {et~al.}(2020{\natexlab{b}})\citenamefont {Fregoni}, \citenamefont
  {Granucci}, \citenamefont {Persico},\ and\ \citenamefont
  {Corni}}]{fregoni_strong_2020}%
  \BibitemOpen
  \bibfield  {author} {\bibinfo {author} {\bibfnamefont {J.}~\bibnamefont
  {Fregoni}}, \bibinfo {author} {\bibfnamefont {G.}~\bibnamefont {Granucci}},
  \bibinfo {author} {\bibfnamefont {M.}~\bibnamefont {Persico}}, \ and\
  \bibinfo {author} {\bibfnamefont {S.}~\bibnamefont {Corni}},\ }\href
  {\doibase 10.1016/j.chempr.2019.11.001} {\bibfield  {journal} {\bibinfo
  {journal} {Chem}\ }\textbf {\bibinfo {volume} {6}},\ \bibinfo {pages} {250}
  (\bibinfo {year} {2020}{\natexlab{b}})}\BibitemShut {NoStop}%
\bibitem [{\citenamefont {Zhang}\ \emph {et~al.}(2019)\citenamefont {Zhang},
  \citenamefont {Nelson},\ and\ \citenamefont {Tretiak}}]{Zhang2019jcp}%
  \BibitemOpen
  \bibfield  {author} {\bibinfo {author} {\bibfnamefont {Y.}~\bibnamefont
  {Zhang}}, \bibinfo {author} {\bibfnamefont {T.}~\bibnamefont {Nelson}}, \
  and\ \bibinfo {author} {\bibfnamefont {S.}~\bibnamefont {Tretiak}},\
  }\href@noop {} {\bibfield  {journal} {\bibinfo  {journal} {J. Chem. Phys.}\
  }\textbf {\bibinfo {volume} {151}},\ \bibinfo {pages} {154109} (\bibinfo
  {year} {2019})}\BibitemShut {NoStop}%
\bibitem [{\citenamefont {Zhou}\ \emph {et~al.}(2022)\citenamefont {Zhou},
  \citenamefont {Hu}, \citenamefont {Mandal},\ and\ \citenamefont
  {Huo}}]{Zhou2022}%
  \BibitemOpen
  \bibfield  {author} {\bibinfo {author} {\bibfnamefont {W.}~\bibnamefont
  {Zhou}}, \bibinfo {author} {\bibfnamefont {D.}~\bibnamefont {Hu}}, \bibinfo
  {author} {\bibfnamefont {A.}~\bibnamefont {Mandal}}, \ and\ \bibinfo {author}
  {\bibfnamefont {P.}~\bibnamefont {Huo}},\ }\href {\doibase 10.1063/5.0109395}
  {\bibfield  {journal} {\bibinfo  {journal} {J. Chem. Phys.}\ }\textbf
  {\bibinfo {volume} {157}},\ \bibinfo {pages} {104118} (\bibinfo {year}
  {2022})}\BibitemShut {NoStop}%
\bibitem [{\citenamefont {Parandekar}\ and\ \citenamefont
  {Tully}(2006)}]{ParandekarJCTC2006}%
  \BibitemOpen
  \bibfield  {author} {\bibinfo {author} {\bibfnamefont {P.~V.}\ \bibnamefont
  {Parandekar}}\ and\ \bibinfo {author} {\bibfnamefont {J.~C.}\ \bibnamefont
  {Tully}},\ }\href@noop {} {\bibfield  {journal} {\bibinfo  {journal} {J.
  Chem. Theory Comput.}\ }\textbf {\bibinfo {volume} {2}},\ \bibinfo {pages}
  {229} (\bibinfo {year} {2006})}\BibitemShut {NoStop}%
\bibitem [{\citenamefont {Subotnik}\ \emph {et~al.}(2016)\citenamefont
  {Subotnik}, \citenamefont {Jain}, \citenamefont {Landry}, \citenamefont
  {Petit}, \citenamefont {Ouyang},\ and\ \citenamefont
  {Bellonzi}}]{subotnik2016arpc}%
  \BibitemOpen
  \bibfield  {author} {\bibinfo {author} {\bibfnamefont {J.~E.}\ \bibnamefont
  {Subotnik}}, \bibinfo {author} {\bibfnamefont {A.}~\bibnamefont {Jain}},
  \bibinfo {author} {\bibfnamefont {B.}~\bibnamefont {Landry}}, \bibinfo
  {author} {\bibfnamefont {A.}~\bibnamefont {Petit}}, \bibinfo {author}
  {\bibfnamefont {W.}~\bibnamefont {Ouyang}}, \ and\ \bibinfo {author}
  {\bibfnamefont {N.}~\bibnamefont {Bellonzi}},\ }\href@noop {} {\bibfield
  {journal} {\bibinfo  {journal} {Annu. Rev. Phys. Chem.}\ }\textbf {\bibinfo
  {volume} {67}},\ \bibinfo {pages} {387} (\bibinfo {year} {2016})}\BibitemShut
  {NoStop}%
\bibitem [{\citenamefont {Meyer}\ and\ \citenamefont
  {Miller}(1979)}]{MeyerJCP1979}%
  \BibitemOpen
  \bibfield  {author} {\bibinfo {author} {\bibfnamefont {H.}~\bibnamefont
  {Meyer}}\ and\ \bibinfo {author} {\bibfnamefont {W.~H.}\ \bibnamefont
  {Miller}},\ }\href@noop {} {\bibfield  {journal} {\bibinfo  {journal} {J.
  Chem. Phys.}\ }\textbf {\bibinfo {volume} {70}},\ \bibinfo {pages} {3214}
  (\bibinfo {year} {1979})}\BibitemShut {NoStop}%
\bibitem [{\citenamefont {Stock}\ and\ \citenamefont
  {Thoss}(1997)}]{StockPRL1997}%
  \BibitemOpen
  \bibfield  {author} {\bibinfo {author} {\bibfnamefont {G.}~\bibnamefont
  {Stock}}\ and\ \bibinfo {author} {\bibfnamefont {M.}~\bibnamefont {Thoss}},\
  }\href@noop {} {\bibfield  {journal} {\bibinfo  {journal} {Phys. Rev. Lett.}\
  }\textbf {\bibinfo {volume} {78}},\ \bibinfo {pages} {578} (\bibinfo {year}
  {1997})}\BibitemShut {NoStop}%
\bibitem [{\citenamefont {Thoss}\ and\ \citenamefont
  {Stock}(1999)}]{ThossPRA1999}%
  \BibitemOpen
  \bibfield  {author} {\bibinfo {author} {\bibfnamefont {M.}~\bibnamefont
  {Thoss}}\ and\ \bibinfo {author} {\bibfnamefont {G.}~\bibnamefont {Stock}},\
  }\href@noop {} {\bibfield  {journal} {\bibinfo  {journal} {Phys. Rev. A}\
  }\textbf {\bibinfo {volume} {59}},\ \bibinfo {pages} {64} (\bibinfo {year}
  {1999})}\BibitemShut {NoStop}%
\bibitem [{\citenamefont {Huo}\ and\ \citenamefont {Coker}(2011)}]{HuoJCP2011}%
  \BibitemOpen
  \bibfield  {author} {\bibinfo {author} {\bibfnamefont {P.}~\bibnamefont
  {Huo}}\ and\ \bibinfo {author} {\bibfnamefont {D.~F.}\ \bibnamefont
  {Coker}},\ }\href@noop {} {\bibfield  {journal} {\bibinfo  {journal} {J.
  Chem. Phys.}\ }\textbf {\bibinfo {volume} {135}},\ \bibinfo {pages} {201101}
  (\bibinfo {year} {2011})}\BibitemShut {NoStop}%
\bibitem [{\citenamefont {Huo}\ and\ \citenamefont {Coker}(2012)}]{HuoMP2012}%
  \BibitemOpen
  \bibfield  {author} {\bibinfo {author} {\bibfnamefont {P.}~\bibnamefont
  {Huo}}\ and\ \bibinfo {author} {\bibfnamefont {D.~F.}\ \bibnamefont
  {Coker}},\ }\href@noop {} {\bibfield  {journal} {\bibinfo  {journal} {Mol.
  Phys.}\ }\textbf {\bibinfo {volume} {110}},\ \bibinfo {pages} {1035}
  (\bibinfo {year} {2012})}\BibitemShut {NoStop}%
\bibitem [{\citenamefont {Cotton}\ and\ \citenamefont
  {Miller}(2013{\natexlab{a}})}]{CottonJCP2013}%
  \BibitemOpen
  \bibfield  {author} {\bibinfo {author} {\bibfnamefont {S.~J.}\ \bibnamefont
  {Cotton}}\ and\ \bibinfo {author} {\bibfnamefont {W.~H.}\ \bibnamefont
  {Miller}},\ }\href@noop {} {\bibfield  {journal} {\bibinfo  {journal} {J.
  Chem. Phys.}\ }\textbf {\bibinfo {volume} {139}},\ \bibinfo {pages} {234112}
  (\bibinfo {year} {2013}{\natexlab{a}})}\BibitemShut {NoStop}%
\bibitem [{\citenamefont {Cotton}\ and\ \citenamefont
  {Miller}(2013{\natexlab{b}})}]{CottonJPCA2013}%
  \BibitemOpen
  \bibfield  {author} {\bibinfo {author} {\bibfnamefont {S.~J.}\ \bibnamefont
  {Cotton}}\ and\ \bibinfo {author} {\bibfnamefont {W.~H.}\ \bibnamefont
  {Miller}},\ }\href@noop {} {\bibfield  {journal} {\bibinfo  {journal} {J.
  Phys. Chem. A}\ }\textbf {\bibinfo {volume} {117}},\ \bibinfo {pages} {7190}
  (\bibinfo {year} {2013}{\natexlab{b}})}\BibitemShut {NoStop}%
\bibitem [{\citenamefont {Hsieh}\ and\ \citenamefont
  {Kapral}(2012)}]{HsiehJCP2012}%
  \BibitemOpen
  \bibfield  {author} {\bibinfo {author} {\bibfnamefont {C.-Y.}\ \bibnamefont
  {Hsieh}}\ and\ \bibinfo {author} {\bibfnamefont {R.}~\bibnamefont {Kapral}},\
  }\href@noop {} {\bibfield  {journal} {\bibinfo  {journal} {J. Chem. Phys.}\
  }\textbf {\bibinfo {volume} {137}},\ \bibinfo {pages} {22A507} (\bibinfo
  {year} {2012})}\BibitemShut {NoStop}%
\bibitem [{\citenamefont {Hsieh}\ and\ \citenamefont
  {Kapral}(2013)}]{HsiehJCP2013}%
  \BibitemOpen
  \bibfield  {author} {\bibinfo {author} {\bibfnamefont {C.-Y.}\ \bibnamefont
  {Hsieh}}\ and\ \bibinfo {author} {\bibfnamefont {R.}~\bibnamefont {Kapral}},\
  }\href@noop {} {\bibfield  {journal} {\bibinfo  {journal} {J. Chem. Phys.}\
  }\textbf {\bibinfo {volume} {138}},\ \bibinfo {pages} {134110} (\bibinfo
  {year} {2013})}\BibitemShut {NoStop}%
\bibitem [{\citenamefont {Cotton}\ and\ \citenamefont
  {Miller}(2019{\natexlab{a}})}]{CottonJCP2019_2}%
  \BibitemOpen
  \bibfield  {author} {\bibinfo {author} {\bibfnamefont {S.~J.}\ \bibnamefont
  {Cotton}}\ and\ \bibinfo {author} {\bibfnamefont {W.~H.}\ \bibnamefont
  {Miller}},\ }\href@noop {} {\bibfield  {journal} {\bibinfo  {journal} {J.
  Chem. Phys.}\ }\textbf {\bibinfo {volume} {150}},\ \bibinfo {pages} {194110}
  (\bibinfo {year} {2019}{\natexlab{a}})}\BibitemShut {NoStop}%
\bibitem [{\citenamefont {Runeson}\ and\ \citenamefont
  {Richardson}(2019)}]{richardson2019}%
  \BibitemOpen
  \bibfield  {author} {\bibinfo {author} {\bibfnamefont {J.~E.}\ \bibnamefont
  {Runeson}}\ and\ \bibinfo {author} {\bibfnamefont {J.~O.}\ \bibnamefont
  {Richardson}},\ }\href {\doibase 10.1063/1.5100506} {\bibfield  {journal}
  {\bibinfo  {journal} {J. Comp. Phys.}\ }\textbf {\bibinfo {volume} {151}},\
  \bibinfo {pages} {044119} (\bibinfo {year} {2019})}\BibitemShut {NoStop}%
\bibitem [{\citenamefont {Runeson}\ and\ \citenamefont
  {Richardson}(2020)}]{richardson2020}%
  \BibitemOpen
  \bibfield  {author} {\bibinfo {author} {\bibfnamefont {J.~E.}\ \bibnamefont
  {Runeson}}\ and\ \bibinfo {author} {\bibfnamefont {J.~O.}\ \bibnamefont
  {Richardson}},\ }\href {\doibase 10.1063/1.5143412} {\bibfield  {journal}
  {\bibinfo  {journal} {J. Comp. Phys.}\ }\textbf {\bibinfo {volume} {152}},\
  \bibinfo {pages} {084110} (\bibinfo {year} {2020})}\BibitemShut {NoStop}%
\bibitem [{\citenamefont {Bossion}\ \emph {et~al.}(2022)\citenamefont
  {Bossion}, \citenamefont {Ying}, \citenamefont {Chowdhury},\ and\
  \citenamefont {Huo}}]{Duncan2022JCP}%
  \BibitemOpen
  \bibfield  {author} {\bibinfo {author} {\bibfnamefont {D.}~\bibnamefont
  {Bossion}}, \bibinfo {author} {\bibfnamefont {W.}~\bibnamefont {Ying}},
  \bibinfo {author} {\bibfnamefont {S.}~\bibnamefont {Chowdhury}}, \ and\
  \bibinfo {author} {\bibfnamefont {P.}~\bibnamefont {Huo}},\ }\href@noop {}
  {\bibfield  {journal} {\bibinfo  {journal} {J. Chem. Phys.}\ }\textbf
  {\bibinfo {volume} {157}},\ \bibinfo {pages} {084105} (\bibinfo {year}
  {2022})}\BibitemShut {NoStop}%
\bibitem [{\citenamefont {Sun}\ \emph {et~al.}(1998)\citenamefont {Sun},
  \citenamefont {Wang},\ and\ \citenamefont {Miller}}]{MillerJCP98}%
  \BibitemOpen
  \bibfield  {author} {\bibinfo {author} {\bibfnamefont {X.}~\bibnamefont
  {Sun}}, \bibinfo {author} {\bibfnamefont {H.}~\bibnamefont {Wang}}, \ and\
  \bibinfo {author} {\bibfnamefont {W.~H.}\ \bibnamefont {Miller}},\
  }\href@noop {} {\bibfield  {journal} {\bibinfo  {journal} {J. Chem. Phys.}\
  }\textbf {\bibinfo {volume} {109}},\ \bibinfo {pages} {7064} (\bibinfo {year}
  {1998})}\BibitemShut {NoStop}%
\bibitem [{\citenamefont {Shi}\ and\ \citenamefont {Geva}(2004)}]{shi2004}%
  \BibitemOpen
  \bibfield  {author} {\bibinfo {author} {\bibfnamefont {Q.}~\bibnamefont
  {Shi}}\ and\ \bibinfo {author} {\bibfnamefont {E.}~\bibnamefont {Geva}},\
  }\href@noop {} {\bibfield  {journal} {\bibinfo  {journal} {J. Phys. Chem.
  A.}\ }\textbf {\bibinfo {volume} {108}} (\bibinfo {year} {2004})}\BibitemShut
  {NoStop}%
\bibitem [{\citenamefont {Weight}\ \emph {et~al.}(2021)\citenamefont {Weight},
  \citenamefont {Mandal},\ and\ \citenamefont {Huo}}]{WeightJCP2021}%
  \BibitemOpen
  \bibfield  {author} {\bibinfo {author} {\bibfnamefont {B.~M.}\ \bibnamefont
  {Weight}}, \bibinfo {author} {\bibfnamefont {A.}~\bibnamefont {Mandal}}, \
  and\ \bibinfo {author} {\bibfnamefont {P.}~\bibnamefont {Huo}},\ }\href
  {\doibase 10.1063/5.0061934} {\bibfield  {journal} {\bibinfo  {journal} {J.
  Chem. Phys.}\ }\textbf {\bibinfo {volume} {155}},\ \bibinfo {pages} {084106}
  (\bibinfo {year} {2021})}\BibitemShut {NoStop}%
\bibitem [{\citenamefont {Chowdhury}\ \emph {et~al.}(2021)\citenamefont
  {Chowdhury}, \citenamefont {Mandal},\ and\ \citenamefont
  {Huo}}]{Chowdhury2021jcp}%
  \BibitemOpen
  \bibfield  {author} {\bibinfo {author} {\bibfnamefont {S.~N.}\ \bibnamefont
  {Chowdhury}}, \bibinfo {author} {\bibfnamefont {A.}~\bibnamefont {Mandal}}, \
  and\ \bibinfo {author} {\bibfnamefont {P.}~\bibnamefont {Huo}},\ }\href@noop
  {} {\bibfield  {journal} {\bibinfo  {journal} {J. Chem. Phys.}\ }\textbf
  {\bibinfo {volume} {154}},\ \bibinfo {pages} {044109} (\bibinfo {year}
  {2021})}\BibitemShut {NoStop}%
\bibitem [{\citenamefont {Mandal}\ \emph
  {et~al.}(2018{\natexlab{a}})\citenamefont {Mandal}, \citenamefont
  {Yamijala},\ and\ \citenamefont {Huo}}]{MandalJCTC2018}%
  \BibitemOpen
  \bibfield  {author} {\bibinfo {author} {\bibfnamefont {A.}~\bibnamefont
  {Mandal}}, \bibinfo {author} {\bibfnamefont {S.~S.}\ \bibnamefont
  {Yamijala}}, \ and\ \bibinfo {author} {\bibfnamefont {P.}~\bibnamefont
  {Huo}},\ }\href {https://doi.org/10.1021/acs.jctc.7b01178} {\bibfield
  {journal} {\bibinfo  {journal} {J. Chem. Theory Comput.}\ }\textbf {\bibinfo
  {volume} {14}},\ \bibinfo {pages} {1828} (\bibinfo {year}
  {2018}{\natexlab{a}})}\BibitemShut {NoStop}%
\bibitem [{\citenamefont {Mandal}\ \emph
  {et~al.}(2018{\natexlab{b}})\citenamefont {Mandal}, \citenamefont {Shakib},\
  and\ \citenamefont {Huo}}]{MandalJCP2018}%
  \BibitemOpen
  \bibfield  {author} {\bibinfo {author} {\bibfnamefont {A.}~\bibnamefont
  {Mandal}}, \bibinfo {author} {\bibfnamefont {F.~A.}\ \bibnamefont {Shakib}},
  \ and\ \bibinfo {author} {\bibfnamefont {P.}~\bibnamefont {Huo}},\
  }\href@noop {} {\bibfield  {journal} {\bibinfo  {journal} {J. Chem. Phys.}\
  }\textbf {\bibinfo {volume} {148}},\ \bibinfo {pages} {244102} (\bibinfo
  {year} {2018}{\natexlab{b}})}\BibitemShut {NoStop}%
\bibitem [{\citenamefont {Mandal}\ \emph {et~al.}(2019)\citenamefont {Mandal},
  \citenamefont {Sandoval~C.}, \citenamefont {Shakib},\ and\ \citenamefont
  {Huo}}]{MandalJPCA2019}%
  \BibitemOpen
  \bibfield  {author} {\bibinfo {author} {\bibfnamefont {A.}~\bibnamefont
  {Mandal}}, \bibinfo {author} {\bibfnamefont {J.~S.}\ \bibnamefont
  {Sandoval~C.}}, \bibinfo {author} {\bibfnamefont {F.~A.}\ \bibnamefont
  {Shakib}}, \ and\ \bibinfo {author} {\bibfnamefont {P.}~\bibnamefont {Huo}},\
  }\href@noop {} {\bibfield  {journal} {\bibinfo  {journal} {J. Phys. Chem. A}\
  }\textbf {\bibinfo {volume} {123}},\ \bibinfo {pages} {2470} (\bibinfo {year}
  {2019})}\BibitemShut {NoStop}%
\bibitem [{\citenamefont {Sandoval}\ \emph {et~al.}(2018)\citenamefont
  {Sandoval}, \citenamefont {Mandal},\ and\ \citenamefont
  {Huo}}]{SandovalJCP2018}%
  \BibitemOpen
  \bibfield  {author} {\bibinfo {author} {\bibfnamefont {J.~S.}\ \bibnamefont
  {Sandoval}}, \bibinfo {author} {\bibfnamefont {A.}~\bibnamefont {Mandal}}, \
  and\ \bibinfo {author} {\bibfnamefont {P.}~\bibnamefont {Huo}},\ }\href
  {https://doi.org/10.1063/1.5036787} {\bibfield  {journal} {\bibinfo
  {journal} {J. Chem. Phys.}\ }\textbf {\bibinfo {volume} {149}},\ \bibinfo
  {pages} {044115} (\bibinfo {year} {2018})}\BibitemShut {NoStop}%
\bibitem [{\citenamefont {Zhou}\ \emph {et~al.}(2019)\citenamefont {Zhou},
  \citenamefont {Mandal},\ and\ \citenamefont {Huo}}]{ZhouJCPL2019}%
  \BibitemOpen
  \bibfield  {author} {\bibinfo {author} {\bibfnamefont {W.}~\bibnamefont
  {Zhou}}, \bibinfo {author} {\bibfnamefont {A.}~\bibnamefont {Mandal}}, \ and\
  \bibinfo {author} {\bibfnamefont {P.}~\bibnamefont {Huo}},\ }\href@noop {}
  {\bibfield  {journal} {\bibinfo  {journal} {J. Phys. Chem. Lett.}\ }\textbf
  {\bibinfo {volume} {10}},\ \bibinfo {pages} {7062} (\bibinfo {year}
  {2019})}\BibitemShut {NoStop}%
\bibitem [{\citenamefont {Shin}\ and\ \citenamefont
  {Metiu}(1995)}]{Shin1995jcp}%
  \BibitemOpen
  \bibfield  {author} {\bibinfo {author} {\bibfnamefont {S.}~\bibnamefont
  {Shin}}\ and\ \bibinfo {author} {\bibfnamefont {H.}~\bibnamefont {Metiu}},\
  }\href@noop {} {\bibfield  {journal} {\bibinfo  {journal} {J. Chem. Phys.}\
  }\textbf {\bibinfo {volume} {102}},\ \bibinfo {pages} {9285} (\bibinfo {year}
  {1995})}\BibitemShut {NoStop}%
\bibitem [{\citenamefont {Hoffmann}\ \emph {et~al.}(2020)\citenamefont
  {Hoffmann}, \citenamefont {Lacombe}, \citenamefont {Rubio},\ and\
  \citenamefont {Maitra}}]{Hoffmann2020}%
  \BibitemOpen
  \bibfield  {author} {\bibinfo {author} {\bibfnamefont {N.~M.}\ \bibnamefont
  {Hoffmann}}, \bibinfo {author} {\bibfnamefont {L.}~\bibnamefont {Lacombe}},
  \bibinfo {author} {\bibfnamefont {A.}~\bibnamefont {Rubio}}, \ and\ \bibinfo
  {author} {\bibfnamefont {N.~T.}\ \bibnamefont {Maitra}},\ }\href@noop {}
  {\bibfield  {journal} {\bibinfo  {journal} {J. Chem. Phys.}\ }\textbf
  {\bibinfo {volume} {153}},\ \bibinfo {pages} {104103} (\bibinfo {year}
  {2020})}\BibitemShut {NoStop}%
\bibitem [{\citenamefont {Doltsinis}\ and\ \citenamefont
  {Marx}(2002)}]{Doltsinis2002jctc}%
  \BibitemOpen
  \bibfield  {author} {\bibinfo {author} {\bibfnamefont {N.~L.}\ \bibnamefont
  {Doltsinis}}\ and\ \bibinfo {author} {\bibfnamefont {D.}~\bibnamefont
  {Marx}},\ }\href@noop {} {\bibfield  {journal} {\bibinfo  {journal} {J. Chem.
  Theory Comput.}\ }\textbf {\bibinfo {volume} {1}},\ \bibinfo {pages} {319}
  (\bibinfo {year} {2002})}\BibitemShut {NoStop}%
\bibitem [{\citenamefont {Sch\"afer}\ \emph {et~al.}(2018)\citenamefont
  {Sch\"afer}, \citenamefont {Ruggenthaler},\ and\ \citenamefont
  {Rubio}}]{Schafer2018pra}%
  \BibitemOpen
  \bibfield  {author} {\bibinfo {author} {\bibfnamefont {C.}~\bibnamefont
  {Sch\"afer}}, \bibinfo {author} {\bibfnamefont {M.}~\bibnamefont
  {Ruggenthaler}}, \ and\ \bibinfo {author} {\bibfnamefont {A.}~\bibnamefont
  {Rubio}},\ }\href@noop {} {\bibfield  {journal} {\bibinfo  {journal} {Phys.
  Rev. A}\ }\textbf {\bibinfo {volume} {98}},\ \bibinfo {pages} {043801}
  (\bibinfo {year} {2018})}\BibitemShut {NoStop}%
\bibitem [{\citenamefont {Marx}\ and\ \citenamefont {Hutter}(2009)}]{Marxbook}%
  \BibitemOpen
  \bibfield  {author} {\bibinfo {author} {\bibfnamefont {D.}~\bibnamefont
  {Marx}}\ and\ \bibinfo {author} {\bibfnamefont {J.}~\bibnamefont {Hutter}},\
  }\href@noop {} {\emph {\bibinfo {title} {Ab initio Molecular Dynamics: Basic
  Theory and Advanced Methods}}}\ (\bibinfo  {publisher} {Cambridge University
  Press},\ \bibinfo {year} {2009})\ p.~\bibinfo {pages} {11}\BibitemShut
  {NoStop}%
\bibitem [{\citenamefont {Kowalewski}\ \emph {et~al.}(2016)\citenamefont
  {Kowalewski}, \citenamefont {Bennett},\ and\ \citenamefont
  {Mukamel}}]{Kowalewski2016JPCL}%
  \BibitemOpen
  \bibfield  {author} {\bibinfo {author} {\bibfnamefont {M.}~\bibnamefont
  {Kowalewski}}, \bibinfo {author} {\bibfnamefont {K.}~\bibnamefont {Bennett}},
  \ and\ \bibinfo {author} {\bibfnamefont {S.}~\bibnamefont {Mukamel}},\
  }\href@noop {} {\bibfield  {journal} {\bibinfo  {journal} {J. Phys. Chem.
  Lett.}\ }\textbf {\bibinfo {volume} {7}},\ \bibinfo {pages} {2050} (\bibinfo
  {year} {2016})}\BibitemShut {NoStop}%
\bibitem [{\citenamefont {Taylor}\ \emph {et~al.}(2020)\citenamefont {Taylor},
  \citenamefont {Mandal}, \citenamefont {Zhou},\ and\ \citenamefont
  {Huo}}]{Taylor2020PRL}%
  \BibitemOpen
  \bibfield  {author} {\bibinfo {author} {\bibfnamefont {M.~A.~D.}\
  \bibnamefont {Taylor}}, \bibinfo {author} {\bibfnamefont {A.}~\bibnamefont
  {Mandal}}, \bibinfo {author} {\bibfnamefont {W.}~\bibnamefont {Zhou}}, \ and\
  \bibinfo {author} {\bibfnamefont {P.}~\bibnamefont {Huo}},\ }\href {\doibase
  10.1103/physrevlett.125.123602} {\bibfield  {journal} {\bibinfo  {journal}
  {Phys. Rev. Lett.}\ }\textbf {\bibinfo {volume} {125}},\ \bibinfo {pages}
  {123602} (\bibinfo {year} {2020})}\BibitemShut {NoStop}%
\bibitem [{\citenamefont {Rokaj}\ \emph {et~al.}(2018)\citenamefont {Rokaj},
  \citenamefont {Welakuh}, \citenamefont {Ruggenthaler},\ and\ \citenamefont
  {Rubio}}]{Rokaj2018JPB}%
  \BibitemOpen
  \bibfield  {author} {\bibinfo {author} {\bibfnamefont {V.}~\bibnamefont
  {Rokaj}}, \bibinfo {author} {\bibfnamefont {D.~M.}\ \bibnamefont {Welakuh}},
  \bibinfo {author} {\bibfnamefont {M.}~\bibnamefont {Ruggenthaler}}, \ and\
  \bibinfo {author} {\bibfnamefont {A.}~\bibnamefont {Rubio}},\ }\href
  {\doibase 10.1088/1361-6455/aa9c99} {\bibfield  {journal} {\bibinfo
  {journal} {J. Phys. B: At. Mol. Opt. Phys.}\ }\textbf {\bibinfo {volume}
  {51}},\ \bibinfo {pages} {034005} (\bibinfo {year} {2018})}\BibitemShut
  {NoStop}%
\bibitem [{\citenamefont {Schäfer}\ \emph {et~al.}(2020)\citenamefont
  {Schäfer}, \citenamefont {Ruggenthaler}, \citenamefont {Rokaj},\ and\
  \citenamefont {Rubio}}]{Schaefer2020AP}%
  \BibitemOpen
  \bibfield  {author} {\bibinfo {author} {\bibfnamefont {C.}~\bibnamefont
  {Schäfer}}, \bibinfo {author} {\bibfnamefont {M.}~\bibnamefont
  {Ruggenthaler}}, \bibinfo {author} {\bibfnamefont {V.}~\bibnamefont {Rokaj}},
  \ and\ \bibinfo {author} {\bibfnamefont {A.}~\bibnamefont {Rubio}},\ }\href
  {\doibase 10.1021/acsphotonics.9b01649} {\bibfield  {journal} {\bibinfo
  {journal} {{ACS} Photonics}\ }\textbf {\bibinfo {volume} {7}},\ \bibinfo
  {pages} {975} (\bibinfo {year} {2020})}\BibitemShut {NoStop}%
\bibitem [{\citenamefont {Koessler}\ \emph {et~al.}(2022)\citenamefont
  {Koessler}, \citenamefont {Mandal},\ and\ \citenamefont
  {Huo}}]{Koessler2022}%
  \BibitemOpen
  \bibfield  {author} {\bibinfo {author} {\bibfnamefont {E.~R.}\ \bibnamefont
  {Koessler}}, \bibinfo {author} {\bibfnamefont {A.}~\bibnamefont {Mandal}}, \
  and\ \bibinfo {author} {\bibfnamefont {P.}~\bibnamefont {Huo}},\ }\href@noop
  {} {\bibfield  {journal} {\bibinfo  {journal} {J. Chem. Phys.}\ }\textbf
  {\bibinfo {volume} {157}},\ \bibinfo {pages} {064101} (\bibinfo {year}
  {2022})}\BibitemShut {NoStop}%
\bibitem [{\citenamefont {Flick}\ \emph
  {et~al.}(2017{\natexlab{b}})\citenamefont {Flick}, \citenamefont {Appel},
  \citenamefont {Ruggenthaler},\ and\ \citenamefont {Rubio}}]{Flick2017jctc}%
  \BibitemOpen
  \bibfield  {author} {\bibinfo {author} {\bibfnamefont {J.}~\bibnamefont
  {Flick}}, \bibinfo {author} {\bibfnamefont {H.}~\bibnamefont {Appel}},
  \bibinfo {author} {\bibfnamefont {M.}~\bibnamefont {Ruggenthaler}}, \ and\
  \bibinfo {author} {\bibfnamefont {A.}~\bibnamefont {Rubio}},\ }\href@noop {}
  {\bibfield  {journal} {\bibinfo  {journal} {J. Chem. Theory Comput.}\
  }\textbf {\bibinfo {volume} {13}},\ \bibinfo {pages} {1616} (\bibinfo {year}
  {2017}{\natexlab{b}})}\BibitemShut {NoStop}%
\bibitem [{\citenamefont {Webster}\ \emph {et~al.}(1991)\citenamefont
  {Webster}, \citenamefont {P.J.Rossky},\ and\ \citenamefont
  {R.A.Friesner}}]{Rossky-Webster}%
  \BibitemOpen
  \bibfield  {author} {\bibinfo {author} {\bibfnamefont {F.}~\bibnamefont
  {Webster}}, \bibinfo {author} {\bibnamefont {P.J.Rossky}}, \ and\ \bibinfo
  {author} {\bibnamefont {R.A.Friesner}},\ }\href@noop {} {\bibfield  {journal}
  {\bibinfo  {journal} {Comput. Phys. Commun.}\ }\textbf {\bibinfo {volume}
  {63}},\ \bibinfo {pages} {494} (\bibinfo {year} {1991})}\BibitemShut
  {NoStop}%
\bibitem [{\citenamefont {Meek}\ and\ \citenamefont
  {Levine}(2014)}]{meek2014evaluation}%
  \BibitemOpen
  \bibfield  {author} {\bibinfo {author} {\bibfnamefont {G.~A.}\ \bibnamefont
  {Meek}}\ and\ \bibinfo {author} {\bibfnamefont {B.~G.}\ \bibnamefont
  {Levine}},\ }\href@noop {} {\bibfield  {journal} {\bibinfo  {journal} {J.
  Phys. Chem. Lett.}\ }\textbf {\bibinfo {volume} {5}},\ \bibinfo {pages}
  {2351} (\bibinfo {year} {2014})}\BibitemShut {NoStop}%
\bibitem [{\citenamefont {Jain}\ \emph {et~al.}(2016)\citenamefont {Jain},
  \citenamefont {Alguire},\ and\ \citenamefont {Subotnik}}]{jain2016efficient}%
  \BibitemOpen
  \bibfield  {author} {\bibinfo {author} {\bibfnamefont {A.}~\bibnamefont
  {Jain}}, \bibinfo {author} {\bibfnamefont {E.}~\bibnamefont {Alguire}}, \
  and\ \bibinfo {author} {\bibfnamefont {J.~E.}\ \bibnamefont {Subotnik}},\
  }\href@noop {} {\bibfield  {journal} {\bibinfo  {journal} {J. Chem. Theory
  Comput.}\ }\textbf {\bibinfo {volume} {12}},\ \bibinfo {pages} {5256}
  (\bibinfo {year} {2016})}\BibitemShut {NoStop}%
\bibitem [{\citenamefont {Joubert-Doriol}\ and\ \citenamefont
  {Izmaylov}(2018)}]{IzmaylovJCP2018}%
  \BibitemOpen
  \bibfield  {author} {\bibinfo {author} {\bibfnamefont {L.}~\bibnamefont
  {Joubert-Doriol}}\ and\ \bibinfo {author} {\bibfnamefont {A.~F.}\
  \bibnamefont {Izmaylov}},\ }\href@noop {} {\bibfield  {journal} {\bibinfo
  {journal} {J. Chem. Phys.}\ }\textbf {\bibinfo {volume} {148}},\ \bibinfo
  {pages} {114102} (\bibinfo {year} {2018})}\BibitemShut {NoStop}%
\bibitem [{\citenamefont {Ben-Nun}\ and\ \citenamefont
  {Mart\'{i}nez}(1998)}]{Ben-Nun1998jcp}%
  \BibitemOpen
  \bibfield  {author} {\bibinfo {author} {\bibfnamefont {M.}~\bibnamefont
  {Ben-Nun}}\ and\ \bibinfo {author} {\bibfnamefont {T.~J.}\ \bibnamefont
  {Mart\'{i}nez}},\ }\href@noop {} {\bibfield  {journal} {\bibinfo  {journal}
  {J. Chem. Phys.}\ }\textbf {\bibinfo {volume} {108}},\ \bibinfo {pages}
  {7244} (\bibinfo {year} {1998})}\BibitemShut {NoStop}%
\bibitem [{\citenamefont {Saita}\ and\ \citenamefont
  {Shalashilin}(2012)}]{mcEhrenfest}%
  \BibitemOpen
  \bibfield  {author} {\bibinfo {author} {\bibfnamefont {K.}~\bibnamefont
  {Saita}}\ and\ \bibinfo {author} {\bibfnamefont {D.~V.}\ \bibnamefont
  {Shalashilin}},\ }\href@noop {} {\bibfield  {journal} {\bibinfo  {journal}
  {J. Chem. Phys.}\ }\textbf {\bibinfo {volume} {137}},\ \bibinfo {pages}
  {22A506} (\bibinfo {year} {2012})}\BibitemShut {NoStop}%
\bibitem [{\citenamefont {Granucci}\ \emph {et~al.}(2001)\citenamefont
  {Granucci}, \citenamefont {Persico},\ and\ \citenamefont
  {Toniolo}}]{GranucciJCP2001}%
  \BibitemOpen
  \bibfield  {author} {\bibinfo {author} {\bibfnamefont {G.}~\bibnamefont
  {Granucci}}, \bibinfo {author} {\bibfnamefont {M.}~\bibnamefont {Persico}}, \
  and\ \bibinfo {author} {\bibfnamefont {A.}~\bibnamefont {Toniolo}},\
  }\href@noop {} {\bibfield  {journal} {\bibinfo  {journal} {J. Chem. Phys.}\
  }\textbf {\bibinfo {volume} {114}},\ \bibinfo {pages} {10608} (\bibinfo
  {year} {2001})}\BibitemShut {NoStop}%
\bibitem [{\citenamefont {Plasser}\ \emph {et~al.}(2012)\citenamefont
  {Plasser}, \citenamefont {Granucci}, \citenamefont {Pittner}, \citenamefont
  {Barbatti}, \citenamefont {Persico},\ and\ \citenamefont
  {Lischka}}]{PlasserJCP2012}%
  \BibitemOpen
  \bibfield  {author} {\bibinfo {author} {\bibfnamefont {F.}~\bibnamefont
  {Plasser}}, \bibinfo {author} {\bibfnamefont {G.}~\bibnamefont {Granucci}},
  \bibinfo {author} {\bibfnamefont {J.}~\bibnamefont {Pittner}}, \bibinfo
  {author} {\bibfnamefont {M.}~\bibnamefont {Barbatti}}, \bibinfo {author}
  {\bibfnamefont {M.}~\bibnamefont {Persico}}, \ and\ \bibinfo {author}
  {\bibfnamefont {H.}~\bibnamefont {Lischka}},\ }\href@noop {} {\bibfield
  {journal} {\bibinfo  {journal} {J. Chem. Phys.}\ }\textbf {\bibinfo {volume}
  {137}},\ \bibinfo {pages} {22A514} (\bibinfo {year} {2012})}\BibitemShut
  {NoStop}%
\bibitem [{\citenamefont {L\"owdin}(1950)}]{LowdinJCP1950}%
  \BibitemOpen
  \bibfield  {author} {\bibinfo {author} {\bibfnamefont {P.~O.}\ \bibnamefont
  {L\"owdin}},\ }\href@noop {} {\bibfield  {journal} {\bibinfo  {journal} {J.
  Chem. Phys.}\ }\textbf {\bibinfo {volume} {18}},\ \bibinfo {pages} {365}
  (\bibinfo {year} {1950})}\BibitemShut {NoStop}%
\bibitem [{\citenamefont {Miller}\ and\ \citenamefont
  {Cotton}(2016)}]{MillerFD2016}%
  \BibitemOpen
  \bibfield  {author} {\bibinfo {author} {\bibfnamefont {W.~H.}\ \bibnamefont
  {Miller}}\ and\ \bibinfo {author} {\bibfnamefont {S.~J.}\ \bibnamefont
  {Cotton}},\ }\href@noop {} {\bibfield  {journal} {\bibinfo  {journal}
  {Faraday Discuss.}\ }\textbf {\bibinfo {volume} {195}},\ \bibinfo {pages} {9}
  (\bibinfo {year} {2016})}\BibitemShut {NoStop}%
\bibitem [{\citenamefont {M\"uller}\ and\ \citenamefont
  {Stock}(1999)}]{UweJCP1999}%
  \BibitemOpen
  \bibfield  {author} {\bibinfo {author} {\bibfnamefont {U.}~\bibnamefont
  {M\"uller}}\ and\ \bibinfo {author} {\bibfnamefont {G.}~\bibnamefont
  {Stock}},\ }\href@noop {} {\bibfield  {journal} {\bibinfo  {journal} {J.
  Chem. Phys.}\ }\textbf {\bibinfo {volume} {111}},\ \bibinfo {pages} {77}
  (\bibinfo {year} {1999})}\BibitemShut {NoStop}%
\bibitem [{\citenamefont {Hu}\ \emph {et~al.}(2021)\citenamefont {Hu},
  \citenamefont {Xie}, \citenamefont {Peng},\ and\ \citenamefont
  {Lan}}]{HuJCTC2021}%
  \BibitemOpen
  \bibfield  {author} {\bibinfo {author} {\bibfnamefont {D.}~\bibnamefont
  {Hu}}, \bibinfo {author} {\bibfnamefont {Y.}~\bibnamefont {Xie}}, \bibinfo
  {author} {\bibfnamefont {J.}~\bibnamefont {Peng}}, \ and\ \bibinfo {author}
  {\bibfnamefont {Z.}~\bibnamefont {Lan}},\ }\href@noop {} {\bibfield
  {journal} {\bibinfo  {journal} {J. Chem. Theory Comput.}\ }\textbf {\bibinfo
  {volume} {17}},\ \bibinfo {pages} {3267} (\bibinfo {year}
  {2021})}\BibitemShut {NoStop}%
\bibitem [{\citenamefont {Cotton}\ and\ \citenamefont
  {Miller}(2016)}]{CottonJCP2016}%
  \BibitemOpen
  \bibfield  {author} {\bibinfo {author} {\bibfnamefont {S.~J.}\ \bibnamefont
  {Cotton}}\ and\ \bibinfo {author} {\bibfnamefont {W.~H.}\ \bibnamefont
  {Miller}},\ }\href@noop {} {\bibfield  {journal} {\bibinfo  {journal} {J.
  Chem. Phys.}\ }\textbf {\bibinfo {volume} {145}},\ \bibinfo {pages} {144108}
  (\bibinfo {year} {2016})}\BibitemShut {NoStop}%
\bibitem [{\citenamefont {Lacombe}\ \emph {et~al.}(2019)\citenamefont
  {Lacombe}, \citenamefont {Hoffmann},\ and\ \citenamefont
  {Maitra}}]{Maitra2019}%
  \BibitemOpen
  \bibfield  {author} {\bibinfo {author} {\bibfnamefont {L.}~\bibnamefont
  {Lacombe}}, \bibinfo {author} {\bibfnamefont {N.~M.}\ \bibnamefont
  {Hoffmann}}, \ and\ \bibinfo {author} {\bibfnamefont {N.~T.}\ \bibnamefont
  {Maitra}},\ }\href@noop {} {\bibfield  {journal} {\bibinfo  {journal} {Phys.
  Rev. Lett.}\ }\textbf {\bibinfo {volume} {123}},\ \bibinfo {pages} {083201}
  (\bibinfo {year} {2019})}\BibitemShut {NoStop}%
\bibitem [{\citenamefont {Martinez}\ \emph {et~al.}(2021)\citenamefont
  {Martinez}, \citenamefont {Rosenzweig}, \citenamefont {Hoffmann},
  \citenamefont {Lacombe},\ and\ \citenamefont {Maitra.}}]{Maitra2021}%
  \BibitemOpen
  \bibfield  {author} {\bibinfo {author} {\bibfnamefont {P.}~\bibnamefont
  {Martinez}}, \bibinfo {author} {\bibfnamefont {B.}~\bibnamefont
  {Rosenzweig}}, \bibinfo {author} {\bibfnamefont {N.~M.}\ \bibnamefont
  {Hoffmann}}, \bibinfo {author} {\bibfnamefont {L.}~\bibnamefont {Lacombe}}, \
  and\ \bibinfo {author} {\bibfnamefont {N.~T.}\ \bibnamefont {Maitra.}},\
  }\href@noop {} {\bibfield  {journal} {\bibinfo  {journal} {J. Chem. Phys.}\
  }\textbf {\bibinfo {volume} {154}},\ \bibinfo {pages} {014102} (\bibinfo
  {year} {2021})}\BibitemShut {NoStop}%
\bibitem [{\citenamefont {Colbert}\ and\ \citenamefont
  {Miller}(1992)}]{Colbert1992}%
  \BibitemOpen
  \bibfield  {author} {\bibinfo {author} {\bibfnamefont {D.~T.}\ \bibnamefont
  {Colbert}}\ and\ \bibinfo {author} {\bibfnamefont {W.~H.}\ \bibnamefont
  {Miller}},\ }\href@noop {} {\bibfield  {journal} {\bibinfo  {journal} {J.
  Chem. Phys.}\ }\textbf {\bibinfo {volume} {96}},\ \bibinfo {pages} {1982}
  (\bibinfo {year} {1992})}\BibitemShut {NoStop}%
\bibitem [{\citenamefont {Farag}\ \emph {et~al.}(2021)\citenamefont {Farag},
  \citenamefont {Mandal},\ and\ \citenamefont {Huo}}]{Farag2021PCCP}%
  \BibitemOpen
  \bibfield  {author} {\bibinfo {author} {\bibfnamefont {M.~H.}\ \bibnamefont
  {Farag}}, \bibinfo {author} {\bibfnamefont {A.}~\bibnamefont {Mandal}}, \
  and\ \bibinfo {author} {\bibfnamefont {P.}~\bibnamefont {Huo}},\ }\href
  {\doibase 10.1039/d1cp00943e} {\bibfield  {journal} {\bibinfo  {journal}
  {Phys. Chem. Chem. Phys.}\ }\textbf {\bibinfo {volume} {23}},\ \bibinfo
  {pages} {16868} (\bibinfo {year} {2021})}\BibitemShut {NoStop}%
\bibitem [{\citenamefont {Kockum}\ \emph {et~al.}(2019)\citenamefont {Kockum},
  \citenamefont {Miranowicz}, \citenamefont {Liberato}, \citenamefont
  {Savasta},\ and\ \citenamefont {Nori}}]{Kockum2019rev}%
  \BibitemOpen
  \bibfield  {author} {\bibinfo {author} {\bibfnamefont {A.~F.}\ \bibnamefont
  {Kockum}}, \bibinfo {author} {\bibfnamefont {A.}~\bibnamefont {Miranowicz}},
  \bibinfo {author} {\bibfnamefont {S.~D.}\ \bibnamefont {Liberato}}, \bibinfo
  {author} {\bibfnamefont {S.}~\bibnamefont {Savasta}}, \ and\ \bibinfo
  {author} {\bibfnamefont {F.}~\bibnamefont {Nori}},\ }\href@noop {} {\bibfield
   {journal} {\bibinfo  {journal} {Nature Rev. Phys.}\ }\textbf {\bibinfo
  {volume} {1}},\ \bibinfo {pages} {19} (\bibinfo {year} {2019})}\BibitemShut
  {NoStop}%
\bibitem [{\citenamefont {Stefano}\ \emph {et~al.}(2019)\citenamefont
  {Stefano}, \citenamefont {Settineri}, \citenamefont {Macri}, \citenamefont
  {Garziano}, \citenamefont {Stassi}, \citenamefont {Savasta},\ and\
  \citenamefont {Nori}}]{Nori2019natphys}%
  \BibitemOpen
  \bibfield  {author} {\bibinfo {author} {\bibfnamefont {O.~D.}\ \bibnamefont
  {Stefano}}, \bibinfo {author} {\bibfnamefont {A.}~\bibnamefont {Settineri}},
  \bibinfo {author} {\bibfnamefont {V.}~\bibnamefont {Macri}}, \bibinfo
  {author} {\bibfnamefont {L.}~\bibnamefont {Garziano}}, \bibinfo {author}
  {\bibfnamefont {R.}~\bibnamefont {Stassi}}, \bibinfo {author} {\bibfnamefont
  {S.}~\bibnamefont {Savasta}}, \ and\ \bibinfo {author} {\bibfnamefont
  {F.}~\bibnamefont {Nori}},\ }\href@noop {} {\bibfield  {journal} {\bibinfo
  {journal} {Nature Phys.}\ }\textbf {\bibinfo {volume} {15}},\ \bibinfo
  {pages} {803} (\bibinfo {year} {2019})}\BibitemShut {NoStop}%
\bibitem [{\citenamefont {Baranov}\ \emph {et~al.}(2020)\citenamefont
  {Baranov}, \citenamefont {Munkhbat}, \citenamefont {Zhukova}, \citenamefont
  {Bisht}, \citenamefont {Canales}, \citenamefont {Rousseaux}, \citenamefont
  {Johansson}, \citenamefont {Antosiewicz},\ and\ \citenamefont
  {Shegai}}]{Shegai_NC_2020}%
  \BibitemOpen
  \bibfield  {author} {\bibinfo {author} {\bibfnamefont {D.~G.}\ \bibnamefont
  {Baranov}}, \bibinfo {author} {\bibfnamefont {B.}~\bibnamefont {Munkhbat}},
  \bibinfo {author} {\bibfnamefont {E.}~\bibnamefont {Zhukova}}, \bibinfo
  {author} {\bibfnamefont {A.}~\bibnamefont {Bisht}}, \bibinfo {author}
  {\bibfnamefont {A.}~\bibnamefont {Canales}}, \bibinfo {author} {\bibfnamefont
  {B.}~\bibnamefont {Rousseaux}}, \bibinfo {author} {\bibfnamefont
  {G.}~\bibnamefont {Johansson}}, \bibinfo {author} {\bibfnamefont {T.~J.}\
  \bibnamefont {Antosiewicz}}, \ and\ \bibinfo {author} {\bibfnamefont
  {T.}~\bibnamefont {Shegai}},\ }\href@noop {} {\bibfield  {journal} {\bibinfo
  {journal} {Nat. Commun.}\ }\textbf {\bibinfo {volume} {11}},\ \bibinfo
  {pages} {1} (\bibinfo {year} {2020})}\BibitemShut {NoStop}%
\bibitem [{\citenamefont {Santhosh}\ \emph {et~al.}(2016)\citenamefont
  {Santhosh}, \citenamefont {Bitton}, \citenamefont {Chuntonov},\ and\
  \citenamefont {Haran}}]{Haran2016}%
  \BibitemOpen
  \bibfield  {author} {\bibinfo {author} {\bibfnamefont {K.}~\bibnamefont
  {Santhosh}}, \bibinfo {author} {\bibfnamefont {O.}~\bibnamefont {Bitton}},
  \bibinfo {author} {\bibfnamefont {L.}~\bibnamefont {Chuntonov}}, \ and\
  \bibinfo {author} {\bibfnamefont {G.}~\bibnamefont {Haran}},\ }\href@noop {}
  {\bibfield  {journal} {\bibinfo  {journal} {Nat. Comm.}\ }\textbf {\bibinfo
  {volume} {7}},\ \bibinfo {pages} {11823} (\bibinfo {year}
  {2016})}\BibitemShut {NoStop}%
\bibitem [{\citenamefont {Mandal}\ \emph
  {et~al.}(2020{\natexlab{b}})\citenamefont {Mandal}, \citenamefont {Vega},\
  and\ \citenamefont {Huo}}]{Mandal2020JPCL}%
  \BibitemOpen
  \bibfield  {author} {\bibinfo {author} {\bibfnamefont {A.}~\bibnamefont
  {Mandal}}, \bibinfo {author} {\bibfnamefont {S.~M.}\ \bibnamefont {Vega}}, \
  and\ \bibinfo {author} {\bibfnamefont {P.}~\bibnamefont {Huo}},\ }\href
  {\doibase 10.1021/acs.jpclett.0c02399} {\bibfield  {journal} {\bibinfo
  {journal} {J. Phys. Chem. Lett.}\ }\textbf {\bibinfo {volume} {11}},\
  \bibinfo {pages} {9215} (\bibinfo {year} {2020}{\natexlab{b}})}\BibitemShut
  {NoStop}%
\bibitem [{\citenamefont {Power}\ and\ \citenamefont {Zienau}(1959)}]{PZW}%
  \BibitemOpen
  \bibfield  {author} {\bibinfo {author} {\bibfnamefont {E.~A.}\ \bibnamefont
  {Power}}\ and\ \bibinfo {author} {\bibfnamefont {S.}~\bibnamefont {Zienau}},\
  }\href@noop {} {\bibfield  {journal} {\bibinfo  {journal} {Philos. Trans. R.
  Soc. London, Ser. A}\ }\textbf {\bibinfo {volume} {251}},\ \bibinfo {pages}
  {427} (\bibinfo {year} {1959})}\BibitemShut {NoStop}%
\bibitem [{\citenamefont {Cohen-Tannoudji}\ \emph {et~al.}(1989)\citenamefont
  {Cohen-Tannoudji}, \citenamefont {Dupont-Roc},\ and\ \citenamefont
  {Grynberg}}]{Cohen-Tannoudji}%
  \BibitemOpen
  \bibfield  {author} {\bibinfo {author} {\bibfnamefont {C.}~\bibnamefont
  {Cohen-Tannoudji}}, \bibinfo {author} {\bibfnamefont {J.}~\bibnamefont
  {Dupont-Roc}}, \ and\ \bibinfo {author} {\bibfnamefont {G.}~\bibnamefont
  {Grynberg}},\ }\href@noop {} {\bibfield  {journal} {\bibinfo  {journal} {John
  Wiley \& Sons, Inc.}\ } (\bibinfo {year} {1989})}\BibitemShut {NoStop}%
\bibitem [{\citenamefont {Forn-D\'{i}az}\ \emph {et~al.}(2019)\citenamefont
  {Forn-D\'{i}az}, \citenamefont {Lamata}, \citenamefont {Rico}, \citenamefont
  {Kono},\ and\ \citenamefont {Solano}}]{Forn-Diaz2019rmp}%
  \BibitemOpen
  \bibfield  {author} {\bibinfo {author} {\bibfnamefont {P.}~\bibnamefont
  {Forn-D\'{i}az}}, \bibinfo {author} {\bibfnamefont {L.}~\bibnamefont
  {Lamata}}, \bibinfo {author} {\bibfnamefont {E.}~\bibnamefont {Rico}},
  \bibinfo {author} {\bibfnamefont {J.}~\bibnamefont {Kono}}, \ and\ \bibinfo
  {author} {\bibfnamefont {E.}~\bibnamefont {Solano}},\ }\href@noop {}
  {\bibfield  {journal} {\bibinfo  {journal} {Rev. Mod. Phys.}\ }\textbf
  {\bibinfo {volume} {91}},\ \bibinfo {pages} {025005} (\bibinfo {year}
  {2019})}\BibitemShut {NoStop}%
\bibitem [{\citenamefont {Kelly}\ \emph {et~al.}(2012)\citenamefont {Kelly},
  \citenamefont {van Zon}, \citenamefont {Schofield},\ and\ \citenamefont
  {Kapral}}]{kelly2012mapping}%
  \BibitemOpen
  \bibfield  {author} {\bibinfo {author} {\bibfnamefont {A.}~\bibnamefont
  {Kelly}}, \bibinfo {author} {\bibfnamefont {R.}~\bibnamefont {van Zon}},
  \bibinfo {author} {\bibfnamefont {J.}~\bibnamefont {Schofield}}, \ and\
  \bibinfo {author} {\bibfnamefont {R.}~\bibnamefont {Kapral}},\ }\href@noop {}
  {\bibfield  {journal} {\bibinfo  {journal} {J. Chem. Phys.}\ }\textbf
  {\bibinfo {volume} {136}},\ \bibinfo {pages} {084101} (\bibinfo {year}
  {2012})}\BibitemShut {NoStop}%
\bibitem [{\citenamefont {Church}\ \emph {et~al.}(2018)\citenamefont {Church},
  \citenamefont {Hele}, \citenamefont {Ezra},\ and\ \citenamefont
  {Ananth}}]{church2018}%
  \BibitemOpen
  \bibfield  {author} {\bibinfo {author} {\bibfnamefont {M.~S.}\ \bibnamefont
  {Church}}, \bibinfo {author} {\bibfnamefont {T.~J.~H.}\ \bibnamefont {Hele}},
  \bibinfo {author} {\bibfnamefont {G.~S.}\ \bibnamefont {Ezra}}, \ and\
  \bibinfo {author} {\bibfnamefont {N.}~\bibnamefont {Ananth}},\ }\href@noop {}
  {\bibfield  {journal} {\bibinfo  {journal} {J. Chem. Phys.}\ }\textbf
  {\bibinfo {volume} {148}},\ \bibinfo {pages} {102326} (\bibinfo {year}
  {2018})}\BibitemShut {NoStop}%
\bibitem [{\citenamefont {Bellonzi}\ \emph {et~al.}(2016)\citenamefont
  {Bellonzi}, \citenamefont {Jain},\ and\ \citenamefont
  {Subotnik}}]{Subotnik2016JCP}%
  \BibitemOpen
  \bibfield  {author} {\bibinfo {author} {\bibfnamefont {N.}~\bibnamefont
  {Bellonzi}}, \bibinfo {author} {\bibfnamefont {A.}~\bibnamefont {Jain}}, \
  and\ \bibinfo {author} {\bibfnamefont {J.~E.}\ \bibnamefont {Subotnik}},\
  }\href {\doibase 10.1063/1.4946810} {\bibfield  {journal} {\bibinfo
  {journal} {J. Chem. Phys.}\ }\textbf {\bibinfo {volume} {144}},\ \bibinfo
  {pages} {154110} (\bibinfo {year} {2016})}\BibitemShut {NoStop}%
\bibitem [{\citenamefont {Landry}\ \emph {et~al.}(2013)\citenamefont {Landry},
  \citenamefont {Falk},\ and\ \citenamefont {Subotnik}}]{Landry2013JCP}%
  \BibitemOpen
  \bibfield  {author} {\bibinfo {author} {\bibfnamefont {B.~R.}\ \bibnamefont
  {Landry}}, \bibinfo {author} {\bibfnamefont {M.~J.}\ \bibnamefont {Falk}}, \
  and\ \bibinfo {author} {\bibfnamefont {J.~E.}\ \bibnamefont {Subotnik}},\
  }\href@noop {} {\bibfield  {journal} {\bibinfo  {journal} {J. Chem. Phys.}\
  }\textbf {\bibinfo {volume} {139}},\ \bibinfo {pages} {211101} (\bibinfo
  {year} {2013})}\BibitemShut {NoStop}%
\bibitem [{\citenamefont {Wang}\ \emph {et~al.}(2016)\citenamefont {Wang},
  \citenamefont {Akimov},\ and\ \citenamefont {Prezhdo}}]{PrezdoSH}%
  \BibitemOpen
  \bibfield  {author} {\bibinfo {author} {\bibfnamefont {L.}~\bibnamefont
  {Wang}}, \bibinfo {author} {\bibfnamefont {A.}~\bibnamefont {Akimov}}, \ and\
  \bibinfo {author} {\bibfnamefont {O.~V.}\ \bibnamefont {Prezhdo}},\
  }\href@noop {} {\bibfield  {journal} {\bibinfo  {journal} {J. Phys. Chem.
  Lett.}\ }\textbf {\bibinfo {volume} {7}},\ \bibinfo {pages} {2100} (\bibinfo
  {year} {2016})}\BibitemShut {NoStop}%
\bibitem [{\citenamefont {Granucci}\ and\ \citenamefont
  {Persico}(2007)}]{Granucci2007}%
  \BibitemOpen
  \bibfield  {author} {\bibinfo {author} {\bibfnamefont {G.}~\bibnamefont
  {Granucci}}\ and\ \bibinfo {author} {\bibfnamefont {M.}~\bibnamefont
  {Persico}},\ }\href@noop {} {\bibfield  {journal} {\bibinfo  {journal} {J.
  Chem. Phys.}\ }\textbf {\bibinfo {volume} {126}},\ \bibinfo {pages} {134114}
  (\bibinfo {year} {2007})}\BibitemShut {NoStop}%
\bibitem [{\citenamefont {Cotton}\ and\ \citenamefont
  {Miller}(2019{\natexlab{b}})}]{Cotton_JCP_2019_many_states}%
  \BibitemOpen
  \bibfield  {author} {\bibinfo {author} {\bibfnamefont {S.~J.}\ \bibnamefont
  {Cotton}}\ and\ \bibinfo {author} {\bibfnamefont {W.~H.}\ \bibnamefont
  {Miller}},\ }\href {\doibase 10.1063/1.5087160} {\bibfield  {journal}
  {\bibinfo  {journal} {J. Chem. Phys.}\ }\textbf {\bibinfo {volume} {150}},\
  \bibinfo {pages} {104101} (\bibinfo {year} {2019}{\natexlab{b}})}\BibitemShut
  {NoStop}%
\bibitem [{\citenamefont {Cotton}\ \emph {et~al.}(2017)\citenamefont {Cotton},
  \citenamefont {Liang},\ and\ \citenamefont {Miller}}]{CottonJCP2017}%
  \BibitemOpen
  \bibfield  {author} {\bibinfo {author} {\bibfnamefont {S.~J.}\ \bibnamefont
  {Cotton}}, \bibinfo {author} {\bibfnamefont {R.}~\bibnamefont {Liang}}, \
  and\ \bibinfo {author} {\bibfnamefont {W.~H.}\ \bibnamefont {Miller}},\
  }\href@noop {} {\bibfield  {journal} {\bibinfo  {journal} {J. Chem. Phys.}\
  }\textbf {\bibinfo {volume} {147}},\ \bibinfo {pages} {064112} (\bibinfo
  {year} {2017})}\BibitemShut {NoStop}%
\bibitem [{\citenamefont {Tully}(2012{\natexlab{b}})}]{TullyNAMD}%
  \BibitemOpen
  \bibfield  {author} {\bibinfo {author} {\bibfnamefont {J.~C.}\ \bibnamefont
  {Tully}},\ }\href@noop {} {\bibfield  {journal} {\bibinfo  {journal} {J.
  Chem. Phys.}\ }\textbf {\bibinfo {volume} {137}},\ \bibinfo {pages} {22A301}
  (\bibinfo {year} {2012}{\natexlab{b}})}\BibitemShut {NoStop}%
\bibitem [{\citenamefont {Hazra}\ \emph {et~al.}(2010)\citenamefont {Hazra},
  \citenamefont {Soudackov},\ and\ \citenamefont
  {Hammes-Schiffer}}]{Hazra2010JPCB}%
  \BibitemOpen
  \bibfield  {author} {\bibinfo {author} {\bibfnamefont {A.}~\bibnamefont
  {Hazra}}, \bibinfo {author} {\bibfnamefont {A.~V.}\ \bibnamefont
  {Soudackov}}, \ and\ \bibinfo {author} {\bibfnamefont {S.}~\bibnamefont
  {Hammes-Schiffer}},\ }\href@noop {} {\bibfield  {journal} {\bibinfo
  {journal} {J. Phys. Chem. B}\ }\textbf {\bibinfo {volume} {114}},\ \bibinfo
  {pages} {12319} (\bibinfo {year} {2010})}\BibitemShut {NoStop}%
\end{thebibliography}
%

\end{document}